%% file: main.tex
\documentclass[12pt]{article}

\usepackage{booktabs}
\usepackage[usestackEOL]{stackengine}
\usepackage[T1]{fontenc}
\usepackage[utf8]{inputenc}
\usepackage{subcaption}
\usepackage{color}
\usepackage{float}
\usepackage{natbib}
\usepackage[ruled,vlined]{algorithm2e}
\usepackage{listings}
\usepackage{minted}
\usepackage{enumitem}
\usepackage[most]{tcolorbox}
\usepackage{ltablex}

\usepackage{lscape}
\usepackage{longtable}
\usepackage{boldline}
\usepackage{graphicx} 
\usepackage{bbm}
\usepackage{latexsym}
\usepackage{xcolor}
\usepackage{amsmath}
\usepackage{amsfonts}
\usepackage{mathtools}
\usepackage{authblk}
\usepackage{amsthm}
\usepackage{amsfonts}
\usepackage{amssymb}
\usepackage{bm}
\allowdisplaybreaks
\usepackage{hyperref}
\usepackage[capitalise]{cleveref}
\usepackage{tikz}
\usepackage[titletoc,title]{appendix}
\usetikzlibrary{shapes.geometric, arrows.meta, positioning, fit, backgrounds}
\tikzset{
    startstop/.style={rectangle, rounded corners, minimum width=1cm, minimum height=1cm, text centered, draw=black, fill=gray!20},
    connector/.style={circle, minimum size=1cm, text centered, draw=black, fill=gray!20},
    arrow/.style={thick,-{Stealth[]}}
}

\DeclareMathOperator*{\argmin}{arg\!\min}
\DeclareUnicodeCharacter{2212}{-}

\usepackage{etoolbox} 
\apptocmd{\sloppy}{\hbadness 10000\relax}{}{} 

\newtheorem*{theorem3.2}{Theorem 3.2 (\cite{FRIEDRICH20151})}

    {\endtcolorbox}

\title{\bf Optimization of ReaxFF parameters for the \texorpdfstring{$\mathrm{Mo-S}$}{Mo-S} system using random optimization and coordinate search}

\author[*]{Arun Ravichandran}
\author[*]{Michael L. Stein}
\author[**]{Mert Y. Sengul}
\author[*]{Ying Hung}
\author[*]{Tirthankar Dasgupta}

\affil[*]{Department of Statistics, Rutgers University}
\affil[**]{Quantum Informatics, LLC}

\begin{document}

\date{}


\maketitle

\frenchspacing

\begin{abstract}
ReaxFF is a molecular dynamics method that can be considered a good approximation to quantum methods for investigating reactive molecular systems consisting of ten thousand to one hundred thousand atoms.  While ReaxFF is usually a much faster alternative to quantum methods, the force field consists of nearly 100 parameters per element, which makes the force field development a high dimensional optimization problem. In addition to the high-dimensionality, non-convexity and non-continuity make it a hard problem to optimize. We use random optimization along with coordinate search strategies to optimize efficiently and sample new parameter points that yield good molecular properties close to predefined `reference values' obtained from quantum mechanical methods for the $\mathrm{Mo-S}$ system. We also provide empirical error guaranties starting from any random sample of inputs. We discover new points for the $\mathrm{Mo-S}$ system at adjusted error levels of $13{,}000$ as compared to \citet{CLAIMED} at $70{,}000$ levels under the same loss function, registering over $80\%$ improvement. We also extend our algorithm to an out-of-sample system, $\mathrm{W-S}$, with no training data to record over $70\%$ improvement over \citet{INDEEDOPT}.
\end{abstract}
\section{Introduction}
ReaxFF is an empirical inter-atomic potential that is used to model reactive molecular systems, where the investigations require bond breaking/forming events and serves as a faster but less accurate alternative to quantum mechanics (QM). ReaxFF was originally developed to model hydrocarbon systems \citep{reaxFFHyd} but has been extended to model several other systems since. A detailed description of the ReaxFF method and how closely it models the truth (QM or experimental values) on some molecules was given by \cite{RUSSO20111549}. More recently, there is a branch of the literature that has been focusing on optimizing the force field parameters in the ReaxFF method to obtain different configurations of the system at, possibly, levels very close to reference systems, which are mostly obtained by using high level QM methods or empirical data \citep[e.g.,][]{Larsson2013GlobalOO, CLAIMED, INDEEDOPT,  JAX}.

For a given system, ReaxFF has many tuning parameters in the energy equations in ReaxFF \citep{reaxFFHyd} that produce outputs known as ``properties'' of the system. While the literature on ReaxFF uses the word ``parameters'' extensively, in this study, we refer to these parameters as inputs or tuning parameters everywhere in this text to avoid any confusion that may arise from using the term parameters, which could take on a different meaning in a statistical context. Similarly, we prefer to use  outputs for ``properties'' to avoid any confusion. Generally, the number of inputs to ReaxFF tends to be on the order of a few tens to hundreds, while the number of outputs is typically much larger, ranging from a few hundreds to thousands, depending on the system of interest. More specifically, a typical ReaxFF force field consists of approximately 100 tuning parameters per element, meaning that if we are interested in developing a force field for a molecular system that involves interactions of hydrogen, oxygen, and carbon atoms, this force field will consist of around 300 tuning parameters. However, it is common to fix some of these tuning parameters based on already optimized values from previous studies that involve a subset of similar interactions and to vary just a subset of the possible tuning parameters when trying to match ReaxFF outputs to the QM/experimental values. This is possible due to what is known as ``transferability'' of ReaxFF optimized force fields. For example, if we already have a force field for a system consisting of hydrogen, carbon, and nitrogen interactions, and if we are interested in developing a system for hydrogen, oxygen, and carbon interactions, we can use some optimized tuning parameters influencing only the common interactions. This property of ReaxFF can be very powerful when combined with a capable optimization algorithm. However, addition of a new interaction type may require changing already optimized parameters. For our purposes, it is sufficient to only consider the tuning parameters that will need to be optimized for a given system in consideration.

When considering the problem of optimization of the ReaxFF tuning parameters for a given system, the exploration of the tuning parameter landscape has been done by looking at the ReaxFF outputs as a black-box (or gray-box) method and using either surrogate models \citep{CLAIMED, INDEEDOPT} or gradient-free optimization methods such as genetic algorithms (GA) \citep{Larsson2013GlobalOO} aimed at optimizing under challenging situations of high-dimensionality, non-convexity, and infeasibility (no output) of some input points. Further, since the number of ReaxFF inputs tends to scale to about a few tens to hundreds for a given system, any surrogate model is limited in how well it can approximate the relationship between the ReaxFF inputs and the outputs of the system. Compounding this issue is the limitation on the training data size that can be obtained due to computational challenges in ReaxFF. Bigger systems may only have available a few hundred to a thousand data points to train on to optimize a $10+$ dimensional input landscape. Prior research, while having found some innovative solutions to this problem of ReaxFF parameter optimization, fails to give any form of guaranties or confidence in being able to select inputs for a given system that achieves lower error by many orders of magnitude. In our work, we uncover the gray-box nature of the ReaxFF optimization problem for the $\mathrm{Mo-S}$ system and suggest methods that successfully sample points from within an error threshold (that is not too small in some sense, as we define in the upcoming section) of the quantum reference values for the given system.

\subsection{Objective redefined}\label{sec:Chap3Objredef}
The ReaxFF typically consists of approximately 100 inputs per element considered, but since there are many optimized inputs that are transferred from other optimized systems, we will only focus on a subset of the total number of inputs to be optimized for a given system, $p$. Given any input values ($\bm x$), the ReaxFF method minimizes the total energy approximated as a linear combination of various energies \citep{reaxFFHyd} associated with each molecule of interest in the system to give a set of ``properties'' or outputs ($\bm y$). These outputs are usually a large force field training set that is composed of, but is not limited to, molecular properties (e.g., bond lengths, bond angles, charges, and energies) and simple reactions (i.e., bond breaking/formation) of reference systems, defined as reference data in \citet{Larsson2013GlobalOO}. Reference systems may be limited to individual molecules or be reactive systems that include interactions of different molecules such as distortions, bond breaking, formation, and such. To give a quick example, we study the $\mathrm{Mo-S}$ system in detail in section~\ref{sec:Mo-S}, where we explain an example of the output of a deprotonation reaction of $\mathrm{MoH_4}$. In that example, the reference systems include $\mathrm{MoH_4}$ and $\mathrm{MoH_5}$ and the molecular property that is recorded is the energy associated with the deprotonation reaction. For a more detailed explanation, see section~\ref{sec:ReaxFFexample}. For a given training set, we assume there are $q$ outputs from ReaxFF ($\bm y \in \mathbb{R}^q$). These outputs are then compared to experimental and/or quantum chemical reference values ($\bm t \in \mathbb{R}^q$) to establish whether these are a good or bad set of inputs for ReaxFF of the given system. At this point, we would like to clarify the two terms we have used so far, ``reference systems'' and quantum chemical ``reference values''. Throughout the text, ``reference systems'' are used to indicate smaller systems or molecules that are part of the bigger system under study $\mathrm{Mo-S}$. ``Reference values'' or ``gold standard values'' are values obtained by quantum chemical calculations to be compared with the outputs of ReaxFF.

The goal is to minimize the difference between the reference values and ReaxFF outputs, as it is assumed that a force field that can reproduce more accurate quantum chemical energies should be able to simulate a larger system with better predictive accuracy. Therefore, the problem of ReaxFF optimization can then be summarized mathematically as
\begin{equation}\label{eq:Chap3Obj}
  \argmin_{\bm x} \mathcal{L}(\bm y(\bm x),\bm t),  
\end{equation}
where $\bm x \in \mathbb{R}^p$ are the inputs to ReaxFF, $\bm y(\bm x) \in \mathbb{R}^q$ are the outputs of ReaxFF, $\bm t \in \mathbb{R}^q$ are the quantum reference values and $\mathcal{L}(\bm x, \bm y) = \sum_j (x_j-y_j)^2$, where the subscript $j$ is used to denote an element of a vector throughout the text, here, of $\bm x$ and $\bm y$, is an example of a loss function that is used. For even the simple choice of the loss function of squared error, the objective function landscape is generally non-convex with multiple local minima. Furthermore, it is difficult to identify minima since the lower bound of any appropriate loss function is not explicitly known and is generally strictly bounded away from $0$. This presents the additional complexity of not knowing a valid stopping time for any feasible optimization routine before we can declare a given system as having been optimized. For example, in \cite{CLAIMED}, some points with lower error than an initial sample of points were found for a squared loss function, but it was unknown if better points (with smaller error) existed or if indeed the points found were local minima. While we show in Section~\ref{sec:Chap3Results} that, indeed, there are points with much lower error than the initial sample, we believe it is better to restate the problem as a sampling problem rather than an optimization problem, as we explain below.

To approach the problem from a sampling feasibility perspective is to design a procedure such that, given an initial sample of points, we are able to find feasible points $\bm x$ in the input landscape such that the outputs $\bm y$ are within a certain error (loss) threshold from the gold standard values $\bm t$. For example, if the minimum error in the initial training data is $\epsilon$, can we find a routine that can successfully sample points with an error strictly smaller than $\epsilon$? In other words, can we find points (in terms of defined error) that are better than our training data when compared to the gold standard. In our case, we use the data from \citet{CLAIMED} as a starting point; if no such initial sample is available, devising an appropriate method to choose the initial sample would be part of the procedure.

The goal for ReaxFF optimization can then be restated as follows: to search the input landscape for values $\bm x \in \mathcal{X}$, such that $ \mathcal{L}(\bm y(\bm x),\bm t) < \epsilon$, where $\epsilon$ can be defined as the minimum error in a given training sample. Equivalently, we can rewrite this as the ability of any procedure to successfully sample $\bm x$ from a set $\mathcal{X}^{feas}$, where  
\begin{equation}\label{eq:Chap3XFeas}
\mathcal{X}^{feas} := \{\bm x \in \mathcal{X} \ \big|  \ \mathcal{L}(\bm y(\bm x),\bm t) < \epsilon\}
\end{equation}
The importance of restating the problem as above is a few. One, given an initial sample of points, it only matters whether we improve the initial sample from its original error below this threshold, the threshold signifying how close we require the points to be to the reference values. Secondly, this specification allows for a well-defined stopping criterion as opposed to stopping when a local minimum is found, which could be hard to discern given a few evaluations. In addition, one of the primary goals of ReaxFF optimization literature is to achieve several good points with similar errors that will help discover outputs close to reference values. These force fields can be tested with short large-scale MD simulations to measure which one reproduces the macro properties of a reference system of interest. Being able to sample multiple points successfully from this set, given we understand that such a set  $\mathcal{X}^{feas}$ is not a singleton for an appropriately chosen $\epsilon$, is then of utmost importance to the ReaxFF community. There are a couple of advantages to a successful procedure that can sample from $\mathcal{X}^{feas}$. First, extended local searches from $\mathcal{X}^{feas}$ may allow the discovery of further points with smaller error thresholds. Second, if such a procedure is known to work from an initial sample of points with arbitrary error ranges, the initial sample may not need to be very large to successfully sample from $\mathcal{X}^{feas}$. \citet{CLAIMED} uses an initial sample of $5{,}000$ data points while we only choose a small fraction $30$ data points to achieve the $70\%$ improvement in error.

In the next section, we introduce the system we explore throughout the rest of this paper, along with more mathematical notation for subsequent sections. The third section reveals some interesting insights into the ReaxFF problem and why some conventional methods may not be suited for it. Finally, we give a search procedure that allows us to sample successfully from a specified error threshold set $\mathcal{X}^{feas}$ as in (\ref{eq:Chap3XFeas}).

\subsection{\texorpdfstring{$\mathrm{Mo-S}$}{Mo-S} system} \label{sec:Mo-S}
In our study, we look into the $\mathrm{Mo-S}$ system used for materials science applications, which is composed of molybdenum ($\mathrm{Mo}$), sulfur ($\mathrm{S}$), hydrogen ($\mathrm{H}$) and carbon ($\mathrm{C}$) atom interactions. Reference systems for our study include geometries of various simple molecules (such as $\mathrm{MoS_2}$, $\mathrm{Mo_2S_6}$, $\mathrm{MoH_4}$, etc.) and energies associated with each. We will primarily refer to the total energy of individual simple molecules (as above) as ``energy''.  Reference systems may be limited to individual molecules or be reactive systems that include interactions of different molecules (e.g., between $\mathrm{MoS_4H_2}$ and $\mathrm{MoS_3HSH}$) such as distortions, bond breaking, formation, and such. Such reactive systems are referred to as ``properties'' or outputs ($\bm y$) in the previous section and are (usually) affine transformations of the energies of simple molecules (involved in the reactions). For instance, one of the outputs of this system refers to the energy associated with the two molecules $\mathrm{MoS_4H_2}$ and $\mathrm{MoS_3HSH}$, and output is the difference in the energies between these two molecules. We introduce the vector $\bm e$ to denote such ``intermediate'' energies, and we can use a subscript $j$ to denote the energy ($e_j$) of each individual reference system or molecule involved in producing the outputs of ReaxFF for our system. The outputs $\bm y$ of the ReaxFF system are derived from these intermediate energies $\bm e$ such that there exists an affine map $\bm y = \bm A\bm e$ such that $\bm A$ is a matrix that encodes all affine transformations that translates intermediate energies to properties. Note that $
\bm e$ can be of a different dimension than $\bm y$, which we can denote as $q'$, then, $\bm A$ is of dimension $q \times q'$.

The $\mathrm{Mo-S}$ system used in this study consists of $p=45$ inputs and $q=599$ outputs along with a vector of gold standard values $\bm t$ for the $599$ outputs. This system, selected as a model system to develop an optimization method by \citet{CLAIMED} is in comparison to other recent literature, much bigger than other systems such as $\mathrm{Ni-Cr}$ or $\mathrm{Si-OH}$ by the size (dimensionality) of either the inputs or outputs, or both. A detailed ReaxFF force field development requires incorporating as many reference systems as possible, since each output is used to capture a specific interaction that will potentially be observed during molecular dynamics applications. However, a higher number of inputs and outputs significantly complicates the parameter tuning problem. Therefore, the $\mathrm{Mo-S}$ system is a fruitful model system to develop optimization algorithms while mimicking real world applications.  In addition, very large reference systems composed of around $360$ atoms are involved, which makes it computationally more demanding compared to smaller systems, meaning we need very large runtimes to yield $\bm y(\bm x)$ for a single point, of the order of a few hours even on high performance compute nodes. While the parallelization of the core ReaxFF gray-box could help improve these times, it is still more demanding than other systems considered in the literature for the purpose of optimizing ReaxFF tuning parameters. In addition, the large reference molecules tend to accumulate potential errors in interatomic interactions, so may behave differently than small molecules, and thus, adds an extra layer of complexity to the parameter tuning problem.

It is helpful to use the $\mathrm{Mo-S}$ system to explain the ReaxFF gray-box in order to capture and communicate all aspects and complexities of the method. Thus, we give a detailed example using one of the smaller molecules, $\mathrm{MoH_4}$, which is part of this larger system $\mathrm{Mo-S}$ .

\subsection{ReaxFF method - a simple reaction example}\label{sec:ReaxFFexample}
One of the outputs used in the $\mathrm{Mo-S}$ system, $y_{j=j'} (\bm x)$, is one that monitors the ``deprotonation reaction'' between $\mathrm{MoH_4}$ and $\mathrm{MoH_5}$ through the energy difference between these individual molecules. It is known that the quantum reference values for such a reaction should be $t_{j'} = -49.85 \ \mathrm{kcal/mol}$ in our given system $\mathrm{Mo-S}$ , which shows that the hydrogen bond is broken and deprotonation occurs. Thus, the goal is to search for parameter values $\bm x$, such that, $y_{j'}(\bm x)$ is close to $t_{j'}$. For an example of parameter values, please refer to Appendix~\ref{App:labels_desc}. 

So far, we have only written the ReaxFF output as $y_{j'} (\bm x)$ when, in reality, there is a set of latent inputs that are part of the method, such as the geometries of the molecules. A known geometry of a $\mathrm{MoH_4}$ molecule is specified in the ReaxFF system as coordinates in $\mathbb{R}^3$ as shown in Table \ref{tab:tbl1}. Table \ref{tab:tbl2} gives the inter-atomic distances ($\bm r$). We use ``inter-atomic distances'' and ``geometry'' interchangeably to mean $\bm r$.  Thus, a known geometry of $\mathrm{MoH_5}$ is also specified (similar to $\mathrm{MoH_4}$ shown) in the input files. To differentiate between these two molecules, we will use $\bm r_1^0$ to refer to the known geometry of $\mathrm{MoH_4}$ and $\bm r_2^0$ to refer to $\mathrm{MoH_5}$, where a superscript $0$ is added to indicate this is the starting geometry before any ReaxFF optimization.

{
\begin{table}[!htb]
\centering
    \begin{minipage}[t]{0.45\textwidth}
    \relsize{-1}
        \centering%
        \begin{tabular}{c|r|r|r}
             Atom & x-coords & y-coords & z-coords \\\hline 
             Mo & 39.99988 & 40.00015 & 40.00000 \\
             H & 39.52332 & 41.11338 & 41.24271 \\
             H & 38.58886 & 39.42025 & 39.17105 \\
             H & 41.05202 & 40.79817 & 38.86744 \\
             H & 40.83916 & 38.66260 & 40.72336
        \end{tabular}%
        \captionsetup{justification=centering,font=small, width=1\columnwidth}
        \caption{Coordinates for $\mathrm{MoH_4}$ in $\mathbb{R}^3$}
        \par\vspace{0pt}
        \label{tab:tbl1}
    \end{minipage}%
    \hfill
    \begin{minipage}[t]{0.5\textwidth}
    \relsize{-1}
        \centering%
        \begin{tabular}{c|r|r|r|r|>{\rule{0pt}{0pt}\hspace{0.25cm}}r}
              & Mo & H & H & H & H \\\hline 
             Mo & 0& & & & \\
             H & 1.735 &  0&  & &\\
             H & 1.736 & 2.834 &0 & &\\
             H & 1.740 & 2.842 &  2.839 & 0 &\\
             H & 1.737 & 2.830 & 2.837 & 2.837 & 0
        \end{tabular}%
        \captionsetup{font=small, width=2\columnwidth}
        \caption{Inter-atomic distances (\AA) for $\mathrm{MoH_4}$}
        \par\vspace{0pt}
        \label{tab:tbl2}
    \end{minipage}
\end{table}
}

\begin{figure}[!htb]
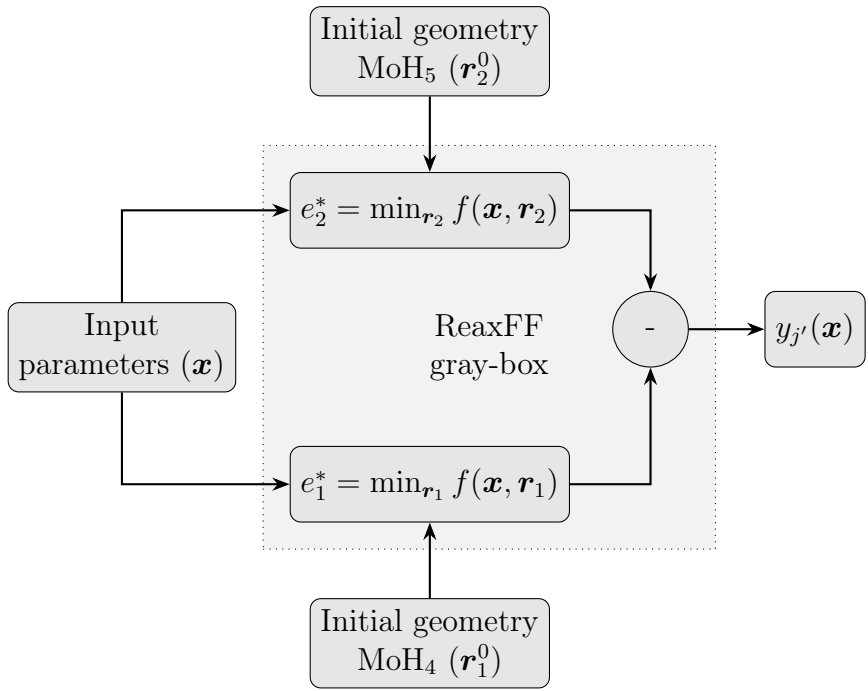

    \centering
    \include{figures/Chapter3/flowdiag}
    \caption{Pictorial representation of derviation of energies associated with the $\mathrm{MoH_4}$ de-proton  ation reaction and its associated output $y_{j'} = e_2^* - e_1^*$}
    \label{fig:deprotonMoH4}
\end{figure}

The ReaxFF gray-box, for each input $\bm x$, involves a minimization procedure for each individual molecule that allows one to identify a specific geometry $\bm r^*$ that is associated with a possibly ``lowest'' energy configuration through energy minimization methods \citep{ReaxFFEnergyMin}. For example, for the $\mathrm{MoH_4}$, for a given value of $\bm x$, ReaxFF produces $e_1^* = \min_{\bm r_1} f(\bm x, \bm r_1)$ and $\bm r_1^* = \argmin_{\bm r_1} f(\bm x,\bm r_1)$, where $f(.)$ refers to ReaxFF output. Similarly, for $\mathrm{MoH_5}$, we have $e_2^*$ and $\bm r_2^*$. Note that, for each $\bm x$, we can define $e_k^0 = f(\bm x, \bm r_k^0)$, for $k=1,2$, to be the starting condition for the minimization procedure. Then, $y_{j'}(\bm x) = e_2^* - e_1^*$ is the desired output of this deprotonation reaction, denoted as ``$\mathrm{MoH_5}-\mathrm{MoH_4}$''. Note that the deprotonation equation above omits $j'$ on the right hand side for the $e_k^*$, this is to keep notation simple and readable. A pictorial representation of the deprotonation example is shown in Figure~\ref{fig:deprotonMoH4}. Recall that this was an example for $j=j'$ which refers to this particular deprotonation reaction output $y_{j'}$.

In general, each output ($y_j$) is derived from inter-atomic distances and the total energy of one or more molecules. For every $y_j, j \in \{1,\dots, q\}$, there are a set of energies based on reference systems that are combined in a predefined manner, which can be represented by an affine map $\bm y = \bm A\bm e$, as mentioned in section~\ref{sec:Mo-S}. While we can provide further notation to show the relationship of the outputs with the intermediate variables $\bm r_j$ and $e_j$ for various molecules, it is not necessary to burden ourselves with this since we will never use that layer within the ReaxFF gray-box. However, it is important to note that the problem involves two levels of optimization, the first being the minimization over the inputs to find the minimum energy configuration $(\bm r_k^*, e_k^*)$ for each molecule obtained numerically from a known starting point $(\bm r_k^0, e_k^0)$, and the second being the optimization over the input parameter $\bm x$, to find parameter values that yield properties $y_j(\bm x)$ closer to the reference values $t_j$. Note that $k$ is used intentionally as a subscript for the energies to avoid any confusion with $j$ used for the $j$-th output $y_j$. As also mentioned in \citet{JAX}, molecules require an inner energy minimization prior to optimizing the parameters $\bm x$, to achieve a lower energy configuration that is more likely to be observed. This two-layer minimization-optimization is key to why any global methods may fail since it embeds such a local minimization procedure in the composite function that yields the ReaxFF output, $\bm y(\bm x)$. To avoid any confusion, we will always refer to the process of ``finding the minimum energy configuration'' for a single molecule as a ``minimization'' where necessary and not as optimization. We say ``optimization'' when we refer to the ``outer optimization of the parameters only''. In the later part of the paper, we refer to the outer optimization as a ``sampling procedure'' for finding input values that yield output values within a certain error threshold of the reference values as defined in (\ref{eq:Chap3XFeas}), since this makes the most sense from the problem specification point of view as outlined in Section~\ref{sec:Chap3Objredef}.

\section{Optimizing the \texorpdfstring{$\mathrm{Mo-S}$}{Mo-S} system}\label{sec:Chap3_NumExpts}
The rest of the paper predominantly focuses on results from the study of the single system $\mathrm{Mo-S}$ in great detail. 
The $\mathrm{Mo-S}$ system we use contains 45 inputs ($\bm x \in \mathbb{R}^{45}$) that we wish to optimize to obtain outputs ($\bm y \in \mathbb{R}^{599}$) that are close to the ``known'' gold standard ($\bm t \in \mathbb{R}^{599}$). In other words, we desire to sample inputs ($\bm x$) to the ReaxFF system that would give us output values ($\bm y(\bm x)$) in the $\epsilon$-neighborhood of $\bm t$. The list of inputs ($\bm x$) along with their descriptions is given in Table~\ref{tbl_param} in Appendix~\ref{App:labels_desc}.

For the loss function, we adopt the same squared error loss as in \citet{CLAIMED} with the same weights $w_j$ to ensure comparability of the methods. As noted in \citet{Larsson2013GlobalOO}, the weights could be arbitrary or could be used as normalizing constants. Thus, rewriting (\ref{eq:Chap3XFeas}) here with the loss function given by $\mathcal{L}(\bm y(\bm x),\bm t) = \sum_{j=1}^q \Big(\frac{y_j(\bm x) - t_j}{w_j}\Big)^2$, our goal is to provide a method that can successfully sample points from $\mathcal{X}^{feas}$ given by
\begin{equation}\label{eq:Chap3XFeas2}
\mathcal{X}^{feas} := \Big\{\bm x \ \Big|  \ \sum_{j=1}^q \Big(\frac{y_j(\bm x) - t_j}{w_j}\Big)^2 < \epsilon\Big\}.
\end{equation}

To make notations convenient, we call $\sum_{j=1}^q \Big(\frac{y_j(\bm x) - t_j}{w_j}\Big)^2$ the total error of the ReaxFF method for a given input $\bm x$ and denote this by $\mathcal{E}(\bm x)$. We use $N=5{,}000$ initial randomly sampled points from \citet{CLAIMED} where the minimum error of all the points was $\min_{i=1}^N \mathcal{E}(\bm x_i) = 96{,}216.427$. In their paper, their method achieves better error points reported to be of the magnitude of $80{,}000$. Our goal, hence, is to check if we can further improve and provide any error guaranties in terms of the procedure. Based on \ref{eq:Chap3XFeas}, we need to set $\epsilon = 80{,}000$, the smallest observed error (rounded below to the nearest $10{,}000$ for ease of reference) in the training data. Initial runs while learning the ReaxFF process led us to discover points with much lower error than in the training set through trial and error (down to $55{,}333$ in Section~\ref{sec:Chap3BruteSearch}). Trial and error included evaluating the ReaxFF function in a neighborhood around the lowest error points, conducting line searches along specific directions, such as the direction of maximum decrease in the ReaxFF output testing along coordinate axes of the inputs $\bm x$ at unit distance, to name a few. Thus, in Section~\ref{sec:Chap3Results}, we set the threshold $\epsilon = 50{,}000$ in our search objective in (\ref{eq:Chap3XFeas2}).

Before we outline our search procedure, we want to provide a few important insights we learned from many different numerical experiments. These findings about the ReaxFF method were key to informing our approach to a solution. We explore each by giving examples.

\subsection{Discrete jumps in the energy of molecules}\label{sec:Chap3Discontinuity}
\begin{figure}[!htb]
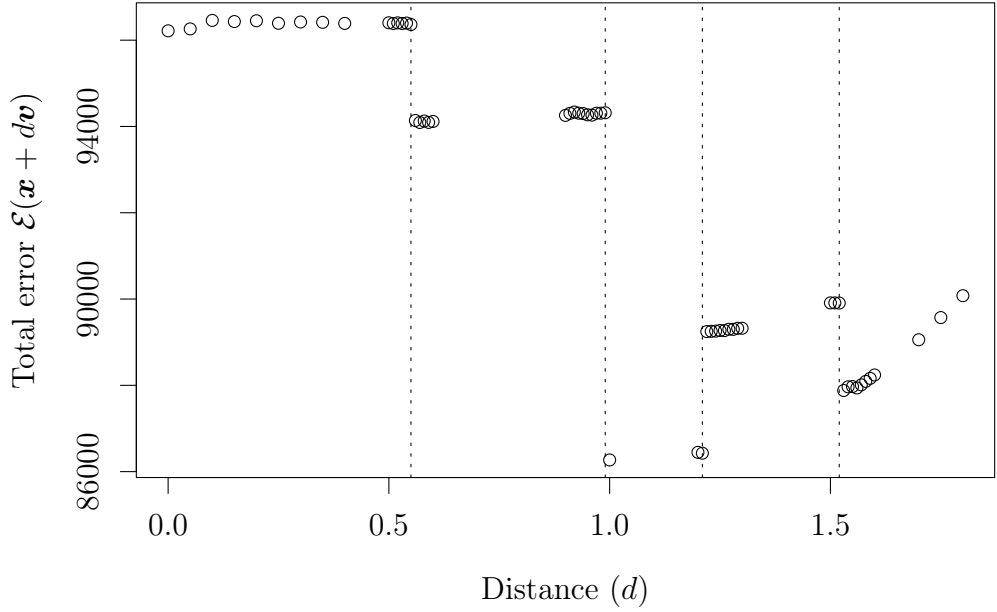

    \centering
    \include{figures/Chapter3/discrete}
    \vspace{-3\abovedisplayskip}
    \captionsetup{justification=centering,width=0.75\columnwidth}
    \caption{Plot showing discontinuity when moving along $\bm x + d\bm v$, for a known $\bm v$ given in Appendix \ref{app:discontinuity}}
    \label{fig:Chap3ReaxFFdiscontinuous}
\end{figure}

ReaxFF surface has been implicitly assumed to be continuous and differentiable in a number of papers. The use of smooth surrogate models or gradient-based descent methods, such as in \citet{CLAIMED}, \citet{INDEEDOPT}, and \citet{JAX} are some examples. We discovered that the ReaxFF surface may not be continuous everywhere, as evidenced by Figure~\ref{fig:Chap3ReaxFFdiscontinuous}. The figure shows the total error evaluations ($\mathcal{E}(\bm x)$) of the $\mathrm{Mo-S}$ system for various input configurations given by $\bm x + d\bm v$ where $\bm x$ is one of the initial sample points with error$=96{,}216.417$, $\bm v$ is a known search direction (given in Appendix~\ref{app:discontinuity}) to further minimize the error, and $d$ is the scalar step-size.

\hspace*{-5in}

The discrete jumps depicted are at the following values of $d$: $d=0.55$, $d=0.99$, $d=1.21$ and $d=1.55$. We were able to establish that these were indeed discrete jumps in the energy of specific molecules (not the same molecules for all the points of discontinuity shown) by comparing the energies of all molecules involved at points across both sides of the boundary and identifying the change as being due to a single molecule's energy. One example of such a jump is shown in Table~\ref{tab:Chap3DiscJump} at the boundary located somewhere in the interval $(0.99,1]$.

\begin{table}[!htbp]
    \centering
    \begin{tabular}{|c|c|r|}
        \hline
        Location & Molecule & Energy \\
        \hline
        Left of boundary & $\mathrm{MoS_2H_2S}$  & $-351.477$\\
        Right of boundary & $\mathrm{MoS_2H_2S}$ & $-409.087$\\
        \hline
    \end{tabular}
    \caption{Discrete jump in specific energies of a single molecule $\mathrm{MoS_2H_2S}$}
    \label{tab:Chap3DiscJump}
\end{table}

\begin{figure}[!htb]
    \centering
    \includegraphics[scale=0.5]{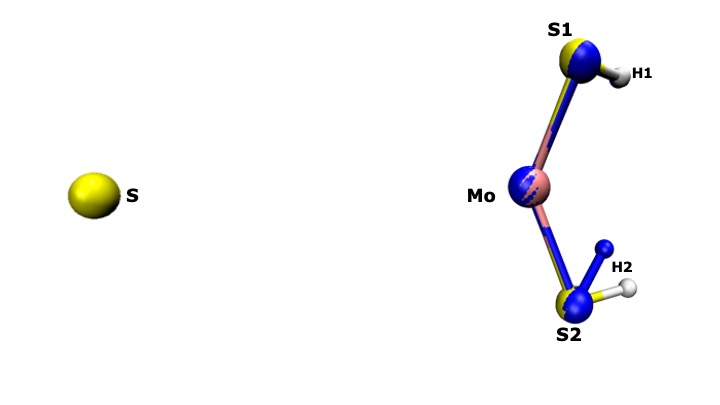}
    \caption{The $\mathrm{MoS_2H_2S}$ system before and after the error jump. The blue represents the molecule configuration before the jump.}
    \label{fig:Chap3MoS2H2S1}
\end{figure}

\begin{figure}[!htb]
    \centering
    \includegraphics[scale=0.5]{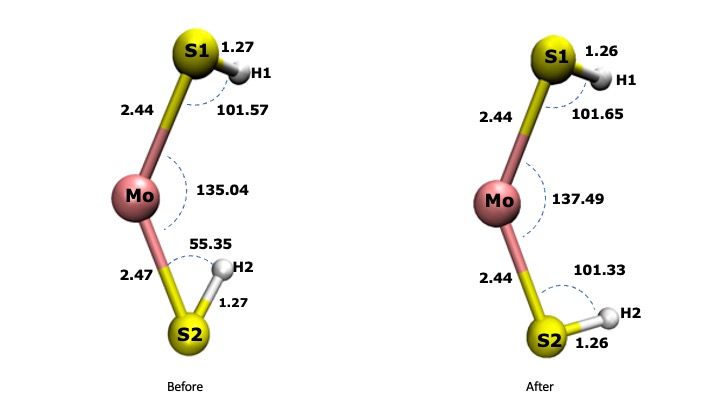}
    \caption{To improve figure visibility, the sulfur ($\mathrm{S}$) atom away from $\mathrm{MoS_2H_2}$ molecule was removed from the images to show the changes in the bond lengths and angles before and after moving a small step at the boundary at $(0.99,1]$ along direction $\bm v$.
}
    \label{fig:Chap3MoS2H2S2}
\end{figure}

Specifically, the actual geometry of the molecule switches configuration when taking a small step in the direction $\bm v$ at the boundary. Taking the case of the jump in the interval $(0.99,1]$, as can also be seen from Figure~\ref{fig:Chap3ReaxFFdiscontinuous}, a very small step in the inputs yields a discrete jump in the energies, as shown in Table~\ref{tab:Chap3DiscJump} that resulted in a change in error by about $8{,}000$ units. Inspecting the geometries of the molecule in question, we see distinctly different $\mathrm{MoS_2H_2S}$ geometries on either side of the singularity, as shown in Figures~\ref{fig:Chap3MoS2H2S1} \& ~\ref{fig:Chap3MoS2H2S2}. In Figures~\ref{fig:Chap3MoS2H2S1}, we show the full geometry of the $\mathrm{MoS_2H_2S}$ molecule with the lone sulfur atom ($\mathrm{S}$) away from the $\mathrm{MoS_2H_2}$ molecule. The blue shading represents the original geometry before it shifts to the new geometry after the jump. The changes occur in the sulfur ($\mathrm{S1}$ and $\mathrm{S2}$) and the hydrogen ($\mathrm{H1}$ and $\mathrm{H2}$) atoms.

We make this clearer in Figure~\ref{fig:Chap3MoS2H2S2} by showing only the $\mathrm{MoS_2H_2}$ molecule, removing the isolated sulfur atom for easier representation of the bond lengths and angles. From Figure~\ref{fig:Chap3MoS2H2S2}, we can see that, while the bond length between S1-Mo stays the same, the bond length between S2-Mo decreases from 2.47 (\AA) to 2.44 (\AA). The bond lengths between S1-H1 and S2-H2 decrease from 1.27 (\AA) to 1.26 (\AA), which is a negligible difference. The most affected term is the valence term; while the Mo-S1-H1 valence angle increases from 101.57 to 101.65, the Mo-S2-H2 changes the most, increasing from 55.35 to 101.33. The S1-Mo-S2 valence also slightly increases, going from 135.04 to 137.49. As a result, the most significant contribution comes from the valence angle term in the energy equation, thus creating a jump in energy.

Checking the other points of discontinuity from Figure~\ref{fig:Chap3ReaxFFdiscontinuous} revealed that different molecules make these jumps at different boundary locations given above. In the cases studied for the jumps at locations $(0.55,0.56]$, $(1.21,1.22]$ and $(1.5,1.51]$ when moving along $\bm v$, we only discovered one molecule—a different one for each location—making discrete jumps at each of the four locations identified as boundaries. Details of specific molecules for each of these jumps, along with the change in energy, are recorded in Table~\ref{tab:chapt3discontpoints} in Appendix~\ref{app:discretejumpgeo}. It is important to note that the observed discontinuities are most likely a manifestation of the inner energy minimization process. We give more details on this in Section~\ref{sec:Chap3_innermin_iter} when we look closer into the inner minimization procedure.

Discontinuity was also discovered in other directions and from other starting points, but knowing that discontinuity exists suffices to inform our approach to steer away from any methods that may use continuity of the function as an assumption. Specifically, it is worth looking into random optimization methods \cite[e.g. see][]{Baba1981ConvergenceOA, Dorea1983ExpectedNO, Nesterov2015RandomGM} in this context. In general, it is not easy to identify such discontinuity without knowing a direction and a functional evaluation at each point along the direction. With computationally expensive functional evaluations and high magnitudes of error, it is possible that discontinuity was either not suspected or fully investigated. At $\bm x$ points where the error is very high, researchers may be able to get away with a continuity assumption while still making progress in moving to better points with lower error. This suggests why there has been partial success in modeling the ReaxFF using smooth functions or using gradient descent methods to improve the error from an initial sample of points, but eventually runs into trouble due to discrete jumps becoming more noticeable at lower magnitudes of error. The ReaxFF surface may be piecewise continuous, but without additional structures such as convexity or linearity, it is unclear how any method could optimize the function successfully. We test such random optimization methods in the next sections to learn more about the response surface locally while developing our search procedure.

\subsection{Stochastic sampling in a local neighborhood}\label{sec:Chap3_StochSampling}
Initial experiments involved learning the local landscape through random moves near a given point. For example, by generating a sample of $20$ points under a normal distribution scaled to reflect the variances of each input, we can evaluate whether we are able to estimate an approximate direction for a move, given by averaging out the finite differences of the functional evaluations between each of the points in the sample and the initial point. Such experiments were informed by the literature on compressive sensing \citep{Borkar2018} and random optimization \citep{Baba1981ConvergenceOA, Dorea1983ExpectedNO, Nesterov2015RandomGM}. While estimating approximate directions to move did not always yield better points, we learned that, for almost all the points, an initial sample of $20$ points introduced at least one better point. While this heavily depends on the region of search, the notion of a trust region \citep{TrustregionColeman} can be deployed to periodically shrink the region of search when no better point is encountered. Such a method applied to $9$ randomly selected points is shown in Table~\ref{tab:Chap3randommoves} for a given point $\bm x_i$. We refer to such batch samples by $\bm x_i^k$ where $k$ denotes the index of the batch sample.

\begin{table}[!htb]
    \centering
    \begin{tabular}{|c|r|>{\rule{0pt}{0pt}\hspace{0.5cm}}r|r|}
    \hline
         $k$ & $||\bm x_i^{k}- \bm x_i^0|| $ & $\mathcal{E}(\bm x_i^{k})$ & $\mathcal{E}(\bm x_i^{k}) - \mathcal{E}(\bm x_i^{0})$\\
         \hline
        0 & $0.00 $ & $96{,}216 $ & $\phantom{-0,00}0$\\
        1 & $2.06 $ & $92{,}697 $ & $-3{,}519$\\
        2 & $4.94 $ & $95{,}980 $ & $\phantom{0}-236$\\
        3 & $6.04 $ & $86{,}271 $ & $-9{,}945$\\
        4 & $6.49 $ & $87{,}209 $ & $-9{,}007$\\
        5 & $7.81 $ & $90{,}507 $ & $-5{,}709$\\
        6 & $7.29 $ & $89{,}093 $ & $-7{,}123$\\
        7 & $8.09 $ & $95{,}843 $ & $\phantom{0}-373$\\
        8 & $8.18 $ & $94{,}395 $ & $-1{,}821$\\
        9 & $9.36 $ & $95{,}579 $ & $\phantom{0}-637$\\
        \hline
    \end{tabular}
    \caption{Total error from $9$ random moves (sampled from a normal distribution with re-scaled variances) from an initial point denoted by $k=0$.}
    \label{tab:Chap3randommoves}
\end{table}

Motivated by ideas of simultaneous perturbation methods \citep{SpallSPSA}, we also tested another method of sampling from an initial point using a mean-zero random vector that is non-normal or non-uniform, specifically a vector of Rademacher random variables ($\pm 1$ with probability $0.5$). In addition, we test moves based on random walk on $p$-directions as opposed to a batch size of $p$ evaluations, to see if we are able to move to points with smaller error that may be very far from an initial point with a very large error.

Let 
$\bm x_i^{0}$ denote the initial point. Let $\bm \Delta_i^k = (\Delta_{i1}^k,\dots,\Delta_{ip}^k)$ be defined as the random perturbation vector given by independent $\Delta_{ij}^k=\pm1$ with probability $1/2$ for every $i,j,k$. Then, for some choice of step-size $c_i^k$, we define the $k$-th move by 
\begin{equation}\label{eq:chap3SGDmove}
    \bm x_i^{k} = \bm x_i^{k-1} + c_i^k\bm \Delta_i^{k-1} \quad.
\end{equation}

\begin{table}[!htb]
    \centering
    \begin{tabular}{|c|r|r|r|r|}
    \hline
         $i$ & $\mathcal{E}(\bm x_i^0)$ & $\delta=0.01$ & $\delta=0.03$ & $\delta=0.1$ \\
         \hline
         1 & $8{,}447{,}952 $ & $-10\% $ & $-76\% $ & $-83\% $ \\
        2 & $8{,}417{,}365 $ & $0\% $ & $-69\% $ & $-52\% $ \\
        3 & $8{,}456{,}145 $ & $-12\% $ & $-30\% $ & $-61\% $ \\
        4 & $8{,}472{,}432 $ & $0\% $ & $-21\% $ & $-77\% $ \\
        5 & $8{,}479{,}381 $ & $-12\% $ & $-34\% $ & $-76\% $ \\
        6 & $8{,}531{,}646 $ & $0\% $ & $-98\% $ & $-85\% $ \\
        7 & $8{,}589{,}232 $ & $-10\% $ & $-40\% $ & $-65\% $ \\
        8 & $9{,}668{,}793 $ & $-59\% $ & $-67\% $ & $-65\% $ \\
        9 & $9{,}092{,}472 $ & $-32\% $ & $-54\% $ & $-44\% $ \\
        10 & $8{,}989{,}222 $ & $-21\% $ & $-42\% $ & $-88\% $ \\
        \hline
    \end{tabular}
    \caption{Maximum \% improvement $\Big(\min_{k=1}^{20} \mathcal{E}(\bm x_i^k)/ \mathcal{E}(\bm x_i^0) -1 \Big)$ in $20$ random moves given by (\ref{eq:chap3SGDmove}) with step-size $c_i^k = \delta ||\bm x_i^{k-1}||$}
    \label{tab:Chap3SGD}
\end{table}

Table~\ref{tab:Chap3SGD} illustrates this idea with choices on step-size ($\delta$) varying between $0.1||\bm x||$, $0.03||\bm x||$ and $0.01||\bm x||$, where $\bm x$ is the initial point. We choose such step-sizes to be able to move a fraction of the norm of the vector at each step. Looking at Table~\ref{tab:Chap3SGD}, we can see how using a bigger step-size allows moves out of the bad points with high initial error.

\begin{figure}[!htb]
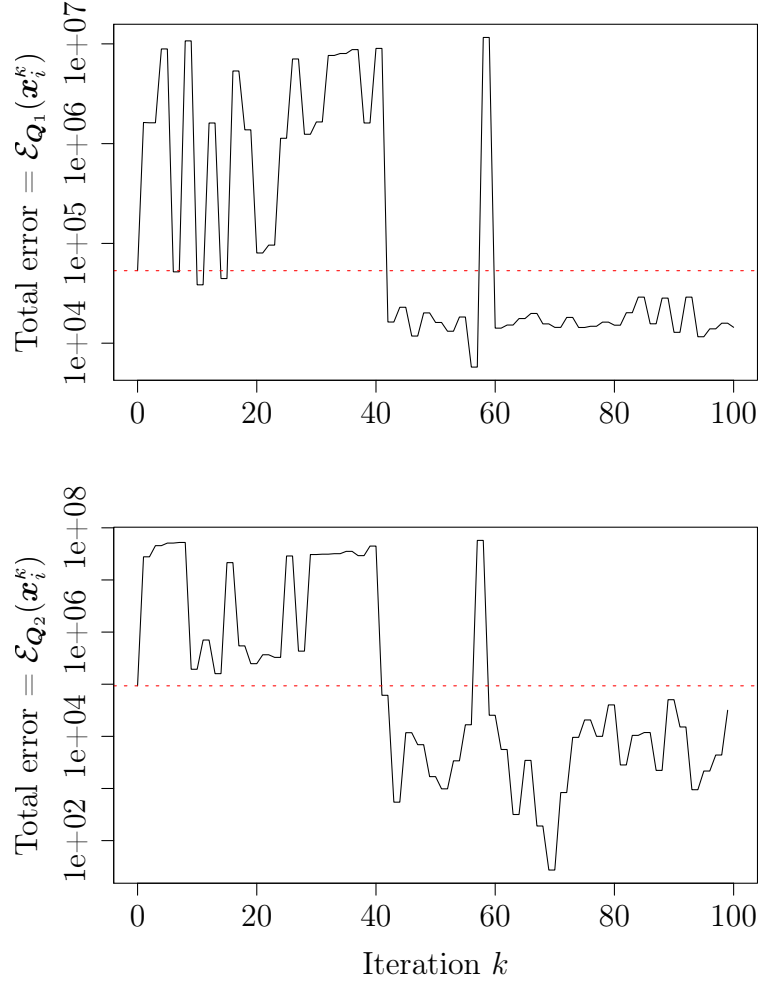

\centering
    \vspace{-1\abovedisplayskip}
    \include{figures/Chapter3/SGDwithFew-v1.tex}
    \vspace{-3\abovedisplayskip}
    \vspace{-3\belowdisplayskip}
    \include{figures/Chapter3/SGDwithFew-v2.tex}
    \vspace{-2\abovedisplayskip}
\caption{Random optimization of a subset of outputs $\bm Q_1 = \{594, 595, 596, 597, 598, 599\}$ and $\bm Q_2 = \{107, 590, 594, 595, 596, 597, 598, 599\}$ for a given point $\bm x_i$. The red dotted lines give the initial errors, $\mathcal{E}_{\bm Q_1}(\bm x_i^0)$ and $\mathcal{E}_{\bm Q_2}(\bm x_i^0)$.}
\label{fig:chap3SGDFew}
\end{figure}

\medskip
We take this one step further by introducing these moves for only a subset of the outputs to see if we are able to optimize by considering a smaller set of outputs. If we let $\bm Q \subset \{1,\dots,q\}$ refer to a subset of outputs, then we can define the total error for any given input $\bm x_i$ for just this subset using the same loss function as $\mathcal{E}_{\bm Q}(\bm x_i)$. In Figure~\ref{fig:chap3SGDFew}, we show an example of such an optimization procedure when using random optimization for only the subset of outputs given by $\bm Q_1 = \{594, 595, 596, 597, 598, 599\}$ and $\bm Q_2 = \{107, 590, 594, 595, 596, 597, 598, 599\}$ from a given point $\bm x_i$. While we tried this procedure for various subsets, $\bm Q_1$ and $\bm Q_2$ contained some of the hardest molecules to obtain stable configurations using the ReaxFF inner minimization loop (essentially not converging to a stable configuration within the chosen fixed $iter=5000$ iterations). Further, $\bm Q_2 \supset \bm Q_1$ involves outputs that are also dissimilar from the original set $\bm Q_1$, making it slightly harder than $\bm Q_1$ to converge. This is evident from the figure where $\mathcal{E}_{\bm Q_2}(\bm x_i)$ still has a high variance even at iterations $iter>60$ as compared to $\mathcal{E}_{\bm Q_1}(\bm x_i)$. Nevertheless, the fact that we are indeed able to achieve much lower error within $100$ iterations of this random optimization defined by (\ref{eq:chap3SGDmove}) allows us to use this approach to design a feasible solution for the ReaxFF problem.

\vspace{-\abovedisplayskip}

\pagebreak
\subsection{Brute search along coordinate axes}\label{sec:Chap3BruteSearch}
\begin{figure}[!htb]
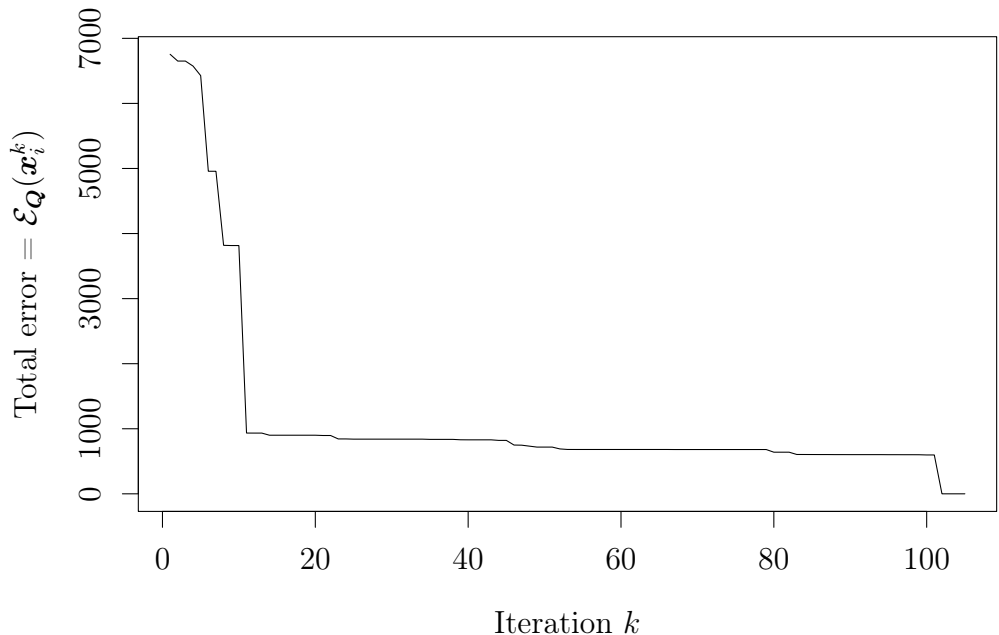

    \centering
    \include{figures/Chapter3/RandomCDv2}
    \vspace{-2\abovedisplayskip}
    \caption{Random coordinate descent of a single property $\bm Q = \{597\}$ for a given point $\bm x_i$. Note that there is insignificant change ($<10^{-3}$) in the total error at each iteration after 101 iterations and converges exactly to gold standard at 155 iterations.}
    \label{fig:Chap3RCD}
\end{figure}

Motivated by the success of random search from the previous subsection, we explored random coordinate descent for only a single randomly chosen property ($\bm y_j$) to check if some sort of convergence is indeed possible on a univariate response function that still contains challenging issues such as discontinuity, multiple local minima, and non-linearity. We picked the outputs with the highest error across all points and checked for convergence. From Figure~\ref{fig:Chap3RCD}, we can see that we are able to converge successfully to the ``gold standard'' for a chosen property within $iter=200$ iterations.

One thing to note is that coordinate descent does not always have good convergence properties, even when the underlying function is convex and nicely behaved. Being able to converge using only coordinate descent suggests that some favorable aspects exist within the ReaxFF surface that can be exploited.

\begin{table}[!htb]
    \centering
    \begin{tabular}{|c|r|r|r|r|r|}
        \hline
         $i$ & $\mathcal{E}(\bm x_i^0)$ & $\min_{l=1}^8p \mathcal{E}_l(\bm x_i^1)$ & $\min_{l=1}^8p \mathcal{E}_l(\bm x_i^2)$ & Move1 \% & Move2 \% \\
         \hline
        1 & $96{,}216 $ & $70{,}044 $ & $55{,}333 $ & $-27\% $ & $-21\% $ \\
        2 & $114{,}234 $ & $87{,}964 $ & $72{,}573 $ & $-23\% $ & $-17\% $ \\
        3 & $141{,}605 $ & $85{,}231 $ & $81{,}232 $ & $-40\% $ & $-5\% $ \\
        4 & $144{,}314 $ & $68{,}788 $ & $54{,}437 $ & $-52\% $ & $-21\% $ \\
        5 & $155{,}542 $ & $86{,}564 $ & $61{,}965 $ & $-44\% $ & $-28\% $ \\
        6 & $160{,}485 $ & $77{,}216 $ & $68{,}813 $ & $-52\% $ & $-11\% $ \\
        7 & $176{,}766 $ & $125{,}756 $ & $100{,}577 $ & $-29\% $ & $-20\% $ \\
        8 & $186{,}987 $ & $145{,}680 $ & $127{,}551 $ & $-22\% $ & $-12\% $ \\
        9 & $185{,}044 $ & $123{,}471 $ & $98{,}899 $ & $-33\% $ & $-20\% $ \\
        10 & $195{,}670 $ & $132{,}936 $ & $109{,}207 $ & $-32\% $ & $-18\% $ \\
        11 & $545{,}271 $ & $347{,}794 $ & $311{,}182 $ & $-36\% $ & $-11\% $ \\
        12 & $568{,}393 $ & $564{,}622 $ & $454{,}063 $ & $-1\% $ & $-20\% $ \\
        13 & $550{,}618 $ & $525{,}972 $ & $477{,}414 $ & $-4\% $ & $-9\% $ \\
        14 & $571{,}719 $ & $537{,}458 $ & $387{,}734 $ & $-6\% $ & $-28\% $ \\
        15 & $577{,}424 $ & $443{,}923 $ & $407{,}365 $ & $-23\% $ & $-8\% $ \\
        16 & $591{,}723 $ & $247{,}201 $ & $205{,}085 $ & $-58\% $ & $-17\% $ \\
        17 & $598{,}153 $ & $60{,}875 $ & $53{,}800 $ & $-90\% $ & $-12\% $ \\
        18 & $573{,}984 $ & $391{,}503 $ & $354{,}026 $ & $-32\% $ & $-10\% $ \\
        19 & $583{,}290 $ & $577{,}014 $ & $506{,}609 $ & $-1\% $ & $-12\% $ \\
        20 & $606{,}100 $ & $553{,}933 $ & $434{,}168 $ & $-9\% $ & $-22\% $ \\
        21 & $8{,}447{,}952 $ & $8{,}265{,}399 $ & $4{,}319{,}497 $ & $-2\% $ & $-48\% $ \\
        22 & $8{,}417{,}365 $ & $1{,}278{,}750 $ & $525{,}237 $ & $-85\% $ & $-59\% $ \\
        23 & $8{,}456{,}145 $ & $5{,}561{,}033 $ & $1{,}391{,}146 $ & $-34\% $ & $-75\% $ \\
        24 & $8{,}467{,}763 $ & $683{,}006 $ & $590{,}452 $ & $-92\% $ & $-14\% $ \\
        25 & $8{,}479{,}381 $ & $7{,}155{,}222 $ & $3{,}624{,}531 $ & $-16\% $ & $-49\% $ \\
        26 & $8{,}531{,}646 $ & $256{,}457 $ & $205{,}035 $ & $-97\% $ & $-20\% $ \\
        27 & $8{,}589{,}232 $ & $3{,}728{,}397 $ & $2{,}996{,}232 $ & $-57\% $ & $-20\% $ \\
        28 & $9{,}668{,}793 $ & $1{,}608{,}955 $ & $714{,}919 $ & $-83\% $ & $-56\% $ \\
        29 & $9{,}092{,}472 $ & $1{,}079{,}649 $ & $817{,}772 $ & $-88\% $ & $-24\% $ \\
        30 & $8{,}989{,}222 $ & $3{,}894{,}272 $ & $996{,}321 $ & $-57\% $ & $-74\% $ \\
        \hline
    \end{tabular}
    \caption{Brute search along $p=45$ dimensions with step-sizes = $\{\pm 1,\pm 0.3,\pm 0.1,\pm 0.01\}$}
    \label{tab:Chap3BS}
\end{table}

\begin{figure}[!htb]
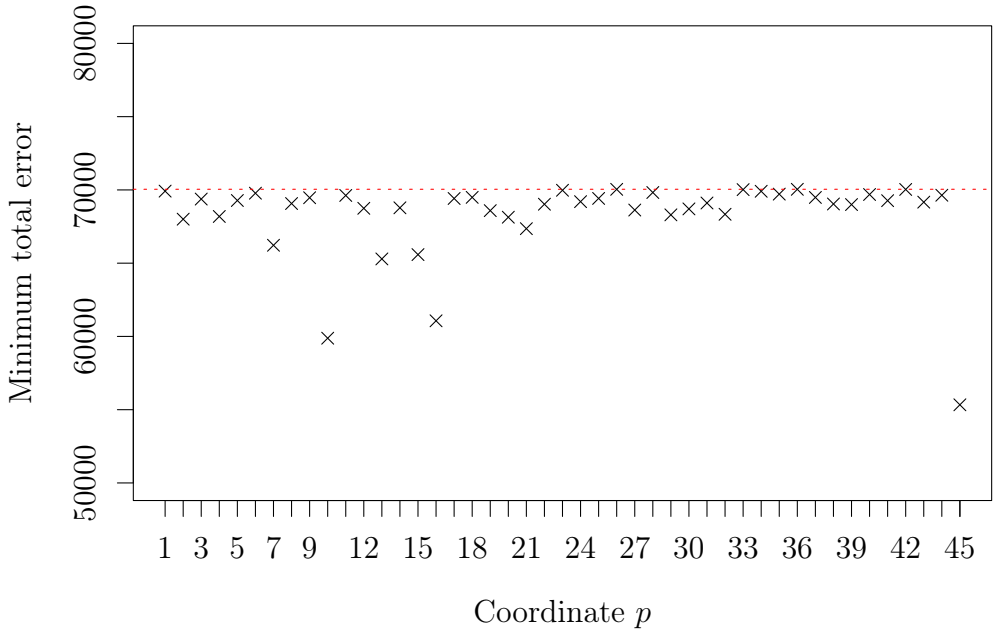

    \centering
    \include{figures/Chapter3/BruteSearch-Box}
    \vspace{-3\abovedisplayskip}
    \caption{Plot of minimum total error moving $\delta \in \{\pm 1,\pm 0.3,\pm 0.1,\pm 0.01\}$ in each coordinate direction from an initial starting point $\bm x_1^1$ in Table~\ref{tab:Chap3BS}. The red line denotes $\mathcal{E}_{p^{*}}(\bm x_1^0)= 70{,}044$.}
    \label{fig:Chap3BruteSearchBox}
\end{figure}

While the coordinate descent worked well for single property evaluations, it did not work as well when we tried to do this by cycling through all ($q=599$) outputs (randomly) to try to optimize the total error $\mathcal{E}(\bm x_i)$. Instead, we considered looking in all the coordinate directions at varying step-sizes to learn the local landscape, i.e., at $\delta \in \{\pm 1,\pm 0.3,\pm 0.1,\pm 0.01\}$. We evaluated the function for a given point $\bm x_i$ at $8$ locations along each coordinate axis given by $x_{ij} + \delta$, for $j=1,\dots,p$ and denoted the error for each evaluation as $\mathcal{E}_l(\bm x_i^k)$, where $l$ gives the index for $\delta$ and $k$ gives the iteration of the brute search. The goal was to check if by searching along such coordinates, we could find any local moves that lowered the total error $\mathcal{E}(\bm x_i^0)$. Indeed, such a brute force search seemed to do very well, especially when the total error was already low ($<200k$). The first two brute searches are shown in Table~\ref{tab:Chap3BS} for the random sample of $30$ points, along with the minimum error in each iteration of the search. The order of improvement in error is consistent in the first ten points when the error $\mathcal{E}(\bm x_i^0) < 100{,}000$. In fact, this approach led to our first discovery of an input point that breached the error threshold ($\epsilon$) of $60{,}000$. This result seemed very promising given that it worked well across the entire random sample. We also show a plot of the minimum total error in Figure~\ref{fig:Chap3BruteSearchBox} for all directions $j=1,\dots,p$ using brute search from a single point $\bm x_1$ in Table~\ref{tab:Chap3BS}. Overall, we found similar behavior locally, in that we found points with lower error when sampling along the coordinate directions from other points as well. This suggests the high likelihood of finding a point with error better (smaller) than the previous step total error $\mathcal{E}(\bm x_i^{k-1})$.

In some ways, the nice aspects we discovered through the conjugate gradient method seemed to have manifested in this brute search. Interactions often appear to be fairly weak, so that moving along coordinate axes can work well. We investigated this further for a single point $\bm x_1$ from Table~\ref{tab:Chap3BS} by conducting a full grid search on two of the directions with the biggest improvements found from the brute search during the first two moves. Let us denote the best directional move for step $k$ by $p^* = \argmin_{l} \mathcal{E}_l(\bm x_i^{k-1})$ (we suppress the dependence of $p^*$ on $k$ for ease of readability). Then, for $k=1$, $p_* = 22$ with $\mathcal{E}_{p^*}(\bm x_1^0)=70{,}044$. For $k=1$, from Figure~\ref{fig:Chap3BruteSearchBox}, we see for $p=45$ and $p=10$, the minimum total error is $55{,}333$ and $59{,}880$, respectively. Thus, we choose $p \in \{22,10,45\}$ to do a grid search across the first and second moves to understand the underlying contour maps.

\begin{figure}
    \centering
    \include{figures/Chapter3/Contour1}
    \vspace{-3\abovedisplayskip}
    \vspace{-3\belowdisplayskip}
    \include{figures/Chapter3/Contour2}
    \caption{Contour plots over a grid showing the first $k=2$ moves using brute search}
    \label{fig:Chap3Contours}
\end{figure}

The contour plots show us the optimal moves with respect to the directions $p \in \{22,10,45\}$. The first move is always along $p=22$ with a $\delta \in [1,1.3]$. The second move along $p=9$ could have been in one of two regions, our search ended up in the second local minimum on the right. For a second move along $p=45$, we can see that using $\delta \ge 0.8$ would have gotten us to the lowest levels of error. Looking at the contour plots, we can see why any attempt to use gradient information may fail, at least in the first plot. But more interestingly, the contour plots seem to reveal weak interactions between these inputs. So, we test some additive models for the two grid evaluations. The results of a vanilla generalized additive model from the mgcv package, fitted to each grid, are shown in Tables~\ref{tab:Chap3_GAM1} \& \ref{tab:Chap3_GAM2} in Appendix~\ref{sec:Chap3_GAM}. We can see that, using non-linear transformations, the ReaxFF surface can be approximated quite well with just an additive function for the two contours shown in Figure~\ref{fig:Chap3Contours}. 

Thus, based on the ReaxFF equations in \citet{reaxFFHyd}, it may be possible and helpful to learn additive relations between inputs in a local region. However, there may be situations where the interaction between two parameters may change depending on the physical interactions in the molecules. For example, the non-bonded interactions between two atoms may be enabled (or disabled) depending on the distance between these atoms, and two parameters influencing such non-bonded terms may start (or stop) interacting after a threshold value. Such unforeseen interactions are one of the reasons why force field optimization is a very complex problem, as inner energy minimization and parameter tuning interoperate, thus making this a highly local feature of the overall landscape. 

\subsection{Varying the inner minimization convergence quality}\label{sec:Chap3_innermin_iter}
Finally, some of the literature \citep{JAX} has focused on delinking the inner minimization from the outer optimization process in determining better input values for ReaxFF. We test whether we can suitably use lower values of the inner minimization iteration threshold to leverage lower computational times and possibly better functional attributes, such as continuity or differentiability of the underlying function. \citet{CLAIMED} uses $T=5000$ iterations to allow nearly all reference systems to converge successfully in the inner minimization step. We found that using any other lower number of iterations resulted in a subset of reference systems not converging fully in the inner minimization step. We tested various numbers from $T=10, 20, 50, 100$ and $500$ to see if there was any relationship we could leverage by running fewer iterations, such as monotonicity of error, where error associated with larger $T$ is lower than the error for smaller $T$. Overall, no relationship was discovered, and we found that, in general, fewer iterations could not serve as a proxy for the final converged configuration of a reference system if the system did indeed take $T=5000$ to converge.

\begin{figure}[!htb]
    \centering
    \vspace{-1.2\abovedisplayskip}
    \includegraphics[scale=0.7]{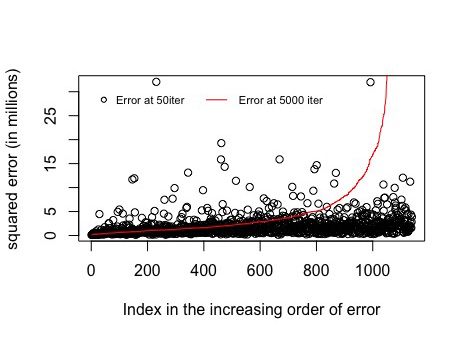}
    \vspace{-1\abovedisplayskip}
    \caption{Comparing the error at $T_1=50$ iterations to $T_2=5000$ iterations when ranked in order of error from $T_2$ iterations. Includes all points from \cite{CLAIMED} with an output. On the x-axis, index = 1 denotes the smallest error of the ReaxFF output after $T_2=5000$ iterations.}
    \label{fig:Chap3error50vs5000-1}
\end{figure}

\clearpage
\begin{figure}
    \centering
    \vspace{-1\abovedisplayskip}
    \includegraphics[scale=0.75]{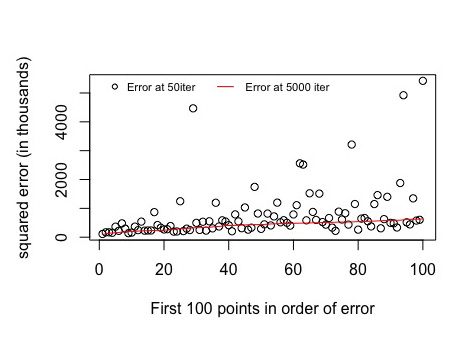}
    \vspace{-1\abovedisplayskip}
    \caption{Comparing the error at $T_1=50$ iterations to $T_2=5000$ iterations when ranked in order of error from $T_2$ iterations. Includes only the first 100 points with smallest error.}
    \label{fig:Chap3error50vs5000-2}
\end{figure}

We show in Figures~\ref{fig:Chap3error50vs5000-1} \& \ref{fig:Chap3error50vs5000-2}, the plots of the total error $\mathcal{E}(\bm x_i)$ for each point $\bm x_i$ run at inner minimization iterations of $T=50$ and $T=5000$. We will denote $\mathcal{E}(\bm x_i)^{T=5000}$ to represent the total error of the point $\bm x_i$ when the inner minimization is $T=5000$, and similarly $\mathcal{E}(\bm x_i)^{T=50}$ for when $T=50$.
Figure \ref{fig:Chap3error50vs5000-1} compares the errors of the minimization process at $T=50$ vs $T=5000$. All points above the red line are points where the error at $T=50$ iterations is larger than at $T=5000$ iterations. Given that we are more interested in how well we approximate points with low error of the order of $100{,}000$ or less, we plot in Figure~\ref{fig:Chap3error50vs5000-2}, only the first $100$ points in the increasing order of the total error $\mathcal{E}(\bm x_i)^{T=5000}$. It is evident that using a lower iteration threshold may sometimes yield points with lower $\mathcal{E}(\bm x_i)^{T=5000}$ and we may very well end up with an unstable configuration of a reference system by using a lower $T$. 

\begin{figure}[!htb]
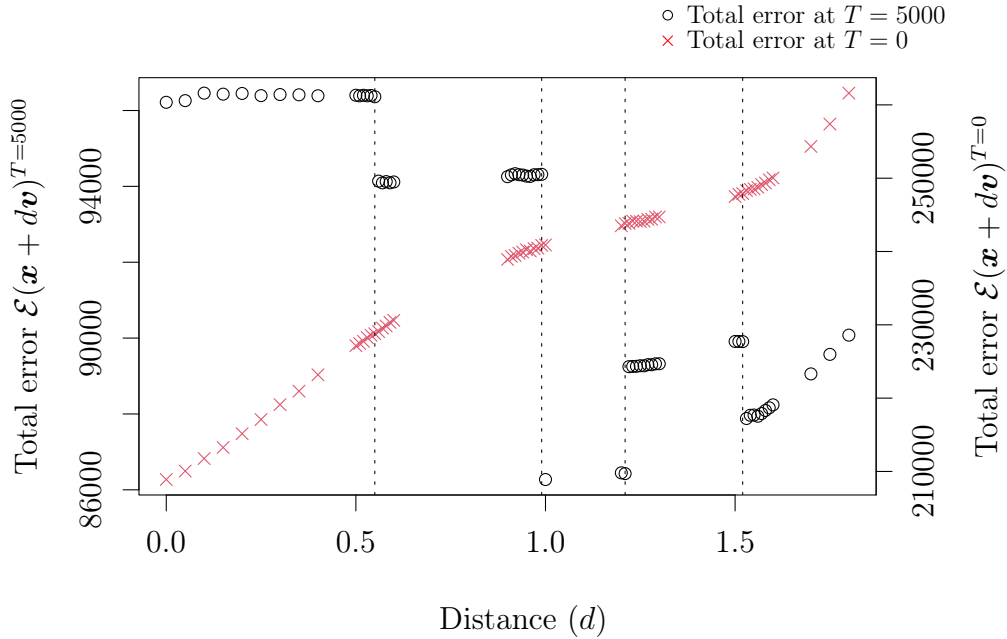

    \centering
    \vspace{-3\abovedisplayskip}
    \include{figures/Chapter3/discrete2}
    \vspace{-3\abovedisplayskip}
    \caption{Plotting ReaxFF outputs for $T=0$ and $T=5000$ for the same points with discontinuity in Figure~\ref{fig:Chap3ReaxFFdiscontinuous}}
    \label{fig:Chap3_ContinuousatT=0}
\end{figure}

Revisiting the problem of discontinuity from Section~\ref{sec:Chap3Discontinuity}, we wanted to test if the ReaxFF function was continuous if the inner minimization was not run at all. Basically, we wanted to know if $e_j^0 = f(x,r_j^0)$ is continuous. This way, we can know whether the discontinuities in Figure~\ref{fig:Chap3ReaxFFdiscontinuous} are due to the inner minimization routine. We plot the same data as in Figure~\ref{fig:Chap3ReaxFFdiscontinuous} but with an additional secondary axis that shows the plot of the total error if the inner minimization was totally neglected. As seen from Figure~\ref{fig:Chap3_ContinuousatT=0}, the function looks to be continuous at $T=0$, and all the discontinuity is introduced due to the varying convergence properties of different molecules involved. It is perhaps possible to vary $T$ over a range from $T=0$ to $T=5000$ to discover when the discontinuity starts to show considerably, but it is likely that this happens when some molecules start to converge to stable configurations prior to the stopping time $T$ specified, thus leading to discontinuous jumps when changing the input $\bm x$. In this light, using continuity and differentiability in \cite{JAX} may be justified since they essentially de-link the two routines and only apply differentiation when $T=0$. But it is apparent from Figure~\ref{fig:Chap3_ContinuousatT=0}, how optimizing over $\bm x$ at $T=0$ may not really help with optimizing the function at $T=5000$.

\hspace*{-5in}

In general, we do not see how any procedure can give a solution to the problem of optimizing the ReaxFF inputs by delinking the inner minimization from the outer optimization. As to what is a sufficient $T$ beyond which there exists some concordance in error values of points, it may very well depend on each system. For our system $\mathrm{Mo-S}$, we adopt $T=5000$ to ensure we do not have any issues with the inner minimization process that can invalidate the points discovered by any procedure we recommend.

\section{Our solution to the ReaxFF problem}
The previous section gave an in-depth pre-cursor to introducing our method for sampling successfully from a feasible set given by (\ref{eq:Chap3XFeas2}). Motivated by many of the numerical experiments run on the ReaxFF gray-box, we put together a procedure that incorporates elements from various concepts such as trust region methods \citep{TrustregionColeman}, random optimization \citep[see e.g.,]{Baba1981ConvergenceOA, Dorea1983ExpectedNO, Nesterov2015RandomGM} and simultaneous perturbation methods \citep{SpallSPSA}. 

\citet{Baba1981ConvergenceOA} gives convergence guaranties under very general conditions of non-convexity but do require continuous functions with bounded support. The main critique of random optimization methods is the extremely slow convergence rate and poor performance in high dimensional problems \citep{Sarma1990}. \citet{Nesterov2015RandomGM} proposes the use of gaussian smoothing methods to random optimization that they test on piece-wise linear but convex functions and on another setting with non-convex but smooth functions of Chebyshev polynomials. Simultaneous perturbation \citep{SpallSPSA} introduced the idea of estimating the gradient by a small number of perturbations of a mean-zero random vector with independent components (but not gaussian or uniform). Specifically, we deployed a vector of Rademacher variables in Section~\ref{sec:Chap3_StochSampling} to successfully find better moves, but without batch evaluations. So, to improve the likelihood of finding a better point at each move, we incorporate batch evaluations of the function at every iteration. Finally, using the notion of a trust region \citep{TrustregionColeman}, we can set our search within a specific radius that we can continue to shrink if batches of searches are unsuccessful at finding a better point. Thus, combining approaches of random optimization, simultaneous perturbation, and trust region methods, we devise a stochastic search procedure as one half of the solution. 

From all our numerical experiments, we gathered that any stochastic search was only good for sampling points above a certain error threshold for our given system. With our current system $\mathrm{Mo-S}$, we learned that points with very high error (on the order of millions) performed extremely well with stochastic searches to move to points with error under a million (or much smaller) in a few iterations. In contrast, points with error lower than $100{,}000$ could hardly improve by $5-10\%$ from previous error even after many iterations. Motivated by the success of the brute searches along coordinate axes from points with lower error, we combine the brute search with stochastic search so that we can, one, continue to sample better points below a certain error threshold, and two, help use a random perturbation of the initial point in a stochastic search if none of the trust region windows yield a better move. There can be improvements in how we can adaptively search all $p$ directions without having to search all $p$ directions every time, but with parallel processing, we can reduce evaluation times to a fraction of the time and cost of serial computation. Thus, we present our procedure below and the results from deploying it on the $\mathrm{Mo-S}$ and $\mathrm{W-S}$ systems in the next section.

We revisit some notation here to help with the readability of the algorithm.
Given the $\mathrm{Mo-S}$ system with $p=45$ inputs ($\bm x \in \mathbb{R}^{45}$), reintroducing (\ref{eq:Chap3XFeas2}) here for ease of reference, we wish to define a procedure to successfully sample points from the following set,
\begin{equation*}
\mathcal{X}^{feas} := \Big\{\bm x \ \Big|  \ \sum_{j=1}^q \Big(\frac{y_j(\bm x) - t_j}{w_j}\Big)^2 < \epsilon\Big\} \quad .
\end{equation*}

Since we will work with an initial random sample of points $n$, we will index the input points by $\bm x_i, \ i=1,\dots,n$. The total error for input $\bm x_i$ is denoted by $\mathcal{E}(\bm x_i)$. Define $\bm e_j \in \mathbb{R}^p$ to be the canonical basis vector such that $\forall j=1,\dots,p, \ e_{jk} = \mathbbm{1}_{j=k}$. Define error thresholds $\mathcal{E}_{th3} \ge \mathcal{E}_{th2} > \mathcal{E}_{th1} > 0$ for a stopping rule. Then, the pseudo-code for our procedure is given in Figure~\ref{fig:chap3pseudocode}. A more detailed version of our procedure is given in Appendix~\ref{app:fullproc} with details around choices of pre-determined values and constants that yielded the results in section~\ref{sec:Chap3Results}.

\begin{figure}[!htb]
    \centering
    \begin{tcolorbox}[width=15cm]
        \begin{itemize}
            \item Given $\bm x_i$. Compute $\mathcal{E}(\bm x_i)$.
            \item while ($\mathcal{E}(\bm x_i) > \mathcal{E}_{th3})$ do:
            \begin{itemize}
                \item \textit{\color{gray} Simultaneous perturbation}
                
                while ($\mathcal{E}(\bm x_i) \ge \mathcal{E}_{th1}$)  do:
                \begin{itemize}
                    \item Generate $\bm \Delta_m = (\delta_1,\dots,\delta_p), \ \delta_j = \pm 1 \ w.p. \ 0.5 , \ m=1,\dots,M$, for some pre-determined value of $M (\le 2^p$).
                    \item Evaluate ReaxFF total error for $\bm u_i^m = \bm x_i + c_i\bm \Delta_m$ for pre-determined $c_i$ (refer~\ref{eq:chap3SGDmove}).
                    \item Let $u_i^* = \argmin_{u_i^m} \mathcal{E}(\bm u_i^m)$
                    \item If ($\mathcal{E}(\bm u_i^*) < \mathcal{E}(\bm x_i)$) then: $\bm x_i = \bm u_i^*$, else: \textit{Shrink $c_i$}
                \end{itemize}
                \item \textit{\color{gray} Brute search}
                
                while ($\mathcal{E}(\bm x_i) \ge \mathcal{E}_{th2}$) do:
                \begin{itemize}
                    \item Generate $u_i^j = \bm x_i \pm \delta'\bm e_j,\ for \ j= 1,\dots,p , \delta'>0$, \(e_j = (0, 0, \dots, 1, \dots, 0)^T\), with the 1 residing in the j-th position.
                    \item Compute $u_i^* = \argmin_{u_i^j} \mathcal{E}_l(\bm u_i^j)$.
                    \item If ($\mathcal{E}(\bm u_i^*) < \mathcal{E}(\bm x_i)$) then: $\bm x_i = \bm u_i^*$, else: \textit{Shrink $\delta'$}
                \end{itemize}
            \end{itemize}
        \end{itemize}
    \end{tcolorbox}
    \vspace{-1\abovedisplayskip}
    \caption{Pseudo-code for our search procedure to sample successfully from $\mathcal{X}^{feas}$}
    \label{fig:chap3pseudocode}
\end{figure}

\section{Results}\label{sec:Chap3Results}
It is important to document the results of our procedure from two perspectives. The first involves testing whether our procedure can truly sample from within the specified error threshold that has never been reported for the $\mathrm{Mo-S}$ system, even from our own numerical experiments. The second is to check if our method can indeed discover some new input points for the $\mathrm{Mo-S}$ system with very low error. While it may be hard to establish if any of these newly discovered lowest error points are potentially local or global minima, the absence of better points (with relatively significant reduction in error) in the local searches of our procedure gives some confidence in the quality of the lowest error points found.

In addition, we present the results from applying our procedure to a completely new system that was not used in studying or developing any part of the procedure to see if our method generalizes well to arbitrary systems that may or may not have been studied already. We see this as a very important achievement since it will give rise to the rapid study of new systems that yield very stable configurations, in turn leading to expanding horizons for the study of large molecular systems and simulations. 

\subsection{Mo-S system}
\subsubsection{Random points reaching error threshold of 50k}
Based on our procedure given in Figure~\ref{fig:chap3pseudocode}, we were able to successfully sample from the set $\mathcal{X}^{feas}$ given in (\ref{eq:Chap3XFeas2}) starting from any random point. Since we were able to find a point with total error of $55{,}333$ from our numerical experiments in Section~\ref{sec:Chap3BruteSearch}, we choose $\epsilon=50{,}000$ in (\ref{eq:Chap3XFeas2}) to challenge our procedure. 

\begin{figure}[!htb]
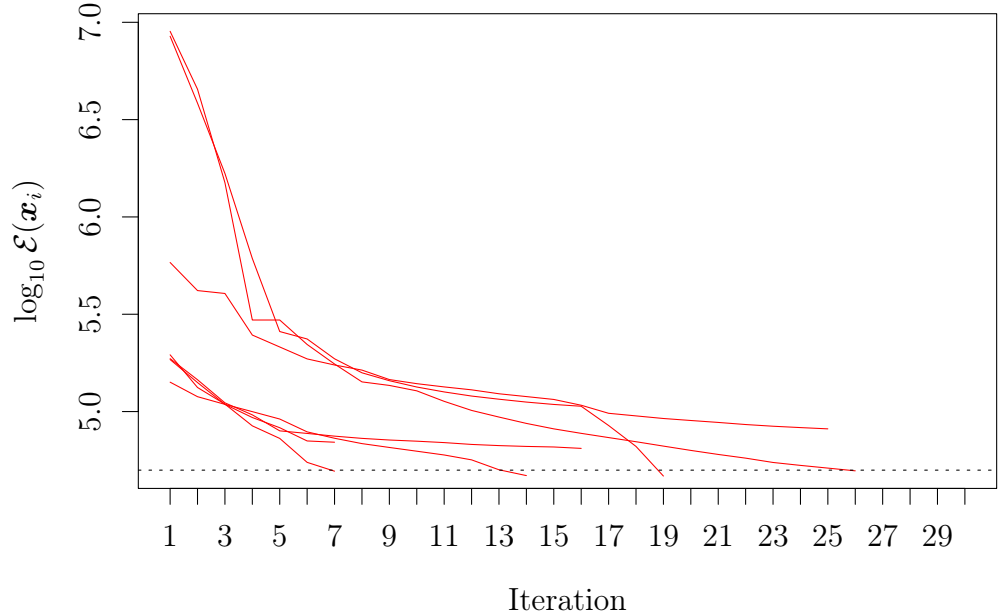

    \centering
    \include{figures/Chapter3/Results1}
    \vspace{-3\abovedisplayskip}
    \caption{Results of our procedure on random initial points. Dotted line denotes the threshold $\epsilon=50{,}000$}
    \label{fig:chap3results}
\end{figure}

Figure~\ref{fig:chap3results} shows the results of our procedure applied to $7$ randomly chosen points (from the sample of $30$ points in Section~\ref{sec:Chap3_NumExpts}). We are able to successfully sample from $\mathcal{X}^{feas}$ with $\epsilon=50{,}000$. It is interesting to note that the combined search using stochastic search and brute search was able to converge to our set $\mathcal{X}^{feas}$ in about 20-30 iterations.  

It is important to note that Figure~\ref{fig:chap3results} shows our procedure works for points with arbitrary error as the initial starting point. This reiterates the point mentioned earlier in Section~\ref{sec:Chap3Objredef}, that one of the important aspects of a successful procedure is achieving results from any starting point. In this way, we could have chosen any random starting points, and we would be able to converge to our desired set. Thus, there is no need for a huge initial training sample, and possibly even undercutting the need for any design of the initial points. We next explore the maximum improvement in error we obtained using our method in the next section.

\subsubsection{Best points with lowest error discovered}
Being able to successfully sample points with almost half the error as reported in prior literature, we wanted to test how small an $\epsilon$ we can sample points from. The lowest error points we discovered by letting our procedure run for up to $150$ iterations are shown in Table~\ref{tab:Chap3resultslowest}. 

\begin{table}[!htb]
    \centering
    \begin{minipage}{0.3\linewidth}
        \begin{tabular}{|c|r|r|}
    \hline
        $\bm x_i^*$ & $\mathcal{E}(\bm x_i^*)$ \\
        \hline
        $\bm x_1^* $ & $23{,}066$  \\
        $\bm x_2^* $ & $24{,}737$ \\
        $\bm x_3^* $ & $24{,}811$  \\
        $\bm x_4^* $ & $25{,}605$ \\
        $\bm x_5^* $ & $26{,}329$ \\
        \hline
    \end{tabular}
    \end{minipage}
    \begin{minipage}{0.45\linewidth}
    \begin{tabular}{|c|r|r|r|r|r|r|}
    \hline
         & $\bm x_1^*$ & $\bm x_2^*$ &$ \bm x_3^*$ & $\bm x_4^*$\\
        \hline
        $\bm x_2^* $ & $78.53$ &  &   &     \\
        $\bm x_3^* $ & $82.70$ & $30.18$ & &     \\
        $\bm x_4^* $ & $228.30$ & $261.60$ & $262.43$ &  \\
        $\bm x_5^* $ & $239.07$ & $280.70$ & $280.95$ & $70.79$\\
        \hline
    \end{tabular}
    \end{minipage}
    \caption{Best points discovered by running our procedure for up to 150 iterations giving the total error (left) and the distance between the best points found (right)}
    \label{tab:Chap3resultslowest}
\end{table}

A few things are worthy of being pointed out at this stage. First, this level of error was practically unknown for a system as complex as $\mathrm{Mo-S}$. Just to understand what configurations and geometries allowed such low levels of error is already a very interesting proposition for the problem at hand. In addition, of the $q=599$ outputs involved, there were about $q'=269$ outputs ($45\%$ of the total set of outputs) that were transferred from other optimized systems, only included to test their stability (optimum) from varying the $p$ inputs. Given that they were transferred from other optimized systems, they are mostly constant and contribute a constant error to the loss function considered. To be able to compare our results with the literature, we did not remove these properties when calculating the total error $\mathcal{E}(x_i)$. But, for all other purposes, if we remove the error from these ``constant'' outputs, under the loss function and chosen weights as in \ref{eq:Chap3XFeas2}, our reported error goes down by $10{,}045.848$ units. What this implies is that we have essentially optimized the $\mathrm{Mo-S}$ to similar levels as other simpler systems that have been established as optimized systems from which the outputs were transferred. Thus, in reality, our lowest error point discovered by $\bm x_1^*$ has an adjusted error of $13{,}020$ for the $\mathrm{Mo-S}$ system. Making a similar adjustment to \citet{CLAIMED} results, we find that we improved upon their best point with adjusted error levels of $70{,}000$ by over $80\%$. More importantly, we also give multiple input points with such low levels of error, as is one of the key requirements for the solution of the ReaxFF optimization problem. 

\subsection{$\mathrm{W-S}$ system}
The $\mathrm{W-S}$ is a smaller system than the $\mathrm{Mo-S}$ system we  studied in great detail until now. The setup is similar with the inputs and outputs, and in this case, we have 68 parameters to optimize across 291 properties, with quantities defined analogously to those described in section~\ref{sec:Mo-S} for the Mo-S system. The $\mathrm{W-S}$ dataset comprises interactions among tungsten ($\mathrm{W}$), selenium ($\mathrm{S}$), hydrogen ($\mathrm{H}$), oxygen ($\mathrm{O}$) and carbon ($\mathrm{C}$) atoms. The reference systems include geometries of a range of simple molecules (e.g., $\mathrm{WS_4H_2}$, $\mathrm{WCO_3H_2S_3}$), as well as reactive systems such as $\mathrm{WCO_2H_2S_4}$ → $\mathrm{WCO_2H_2S_3}$ + $\mathrm{H_2S}$ together with the associated energetics. Full details of the system specifications are given in the supplemental materials to allow for readability and concise reporting of important results. 

\subsubsection{Cold start to new systems}
A primary advantage of running our procedure on a new system is that it does not require any training data, as in other previous literature methods \citep[see e.g.,][]{CLAIMED, INDEEDOPT}. Any random initial point that yields a valid output is sufficient for the system to start suggesting stable configurations at very low error when compared to ``gold standard'' quantum mechanical values . It is, however, necessary to provide feasible regions to sample from for each parameter, as this is part of the configuration settings of every system. Note that while this may be restrictive in terms of application to completely unknown systems, this procedure only requires one point that has a finite output in terms of the property values from which it can iterate to give good configurations. 

\subsubsection{Achieving significant reduction in total error magnitudes}
Without any prior training on this specific system, we were able to achieve a significant reduction in total error (as defined in section \ref{sec:Chap3_NumExpts} by $\mathcal{E}(\bm x)$, but with the appropriate terms for this system). An initial random point chosen (details in supplementary materials) was recorded with a total error of $20{,}161{,}772$. Within 2 batch updates (or 100 iterations each) of running this system through our procedure, we were able to achieve a significant reduction of $99\%$ to a total error of $2{,}584$. 

\begin{table}[!htb]
    \centering
    \begin{tabular}{|c|r|r|r|}\hline
         S.No.&  Initial Error&  Final Error& 
\% change\\\hline
         1&     24,832,966 &     2,107 & 99.99%
\\\hline
         2&     87,356,776 &     9,348 & 99.99%
\\\hline
         3&     89,625,335 &     25,776 & 99.97%
\\\hline
         4&     16,279,512 &     24,114 & 99.85%
\\\hline
         5&     16,340,776 &     3,944 & 99.98%
\\\hline
         6&     272,469 &     21,327 & 92.17%
\\\hline
         7&     43,297 &     3,944 & 90.89%
\\\hline
         8&     26,881 &     1,700 & 93.68%
\\\hline
         9&     27,952 &     2,642 & 90.55%
\\\hline
 10&   7,850 &   4,272 &45.58%
\\\hline
 11&   2,889 &   2,253 &22.02%
\\\hline
 12&   2,313 &   2,102 &9.14%
 \\ \hline
 \end{tabular}
    \caption{Error reductions from restarts of varying error magnitudes}
    \label{tab:ErrorReductionWse}
\end{table}
Subsequently, we picked random points from the trajectory of the previous optimization as restart points for discovering additional points. The results are summarized in Table~\ref{tab:ErrorReductionWse}. Note that all starts from extremely bad points ($\approx10^6$ error) achieve significant reductions of atleast $99.85\%$ in total magnitude. This is in line with our observation for the $\mathrm{Mo-S}$ system that stochastic search is most effective above a certain threshold. Points with error magnitudes of $10^5$ or $10^6$ still achieved significant reductions of atleast $90\%$. The last 3 points have a very low error in comparison to any of the other points and were deliberately included to test if the procedure can continue to find better points even from very low error. This is where the brute search dominates the stochastic search; hence, much slower yet certain progress is made.

\subsubsection{Configurations of best point discovered}
The lowest error configuration recorded through our searches was $1{,}533.97$. To compare with previous attempts to optimize the $\mathrm{W-S}$ force field, \citet{INDEEDOPT} achieves optimized parameter values for the $\mathrm{W-S}$ system with errors reported at $5{,}313.4$ using IndeedOPT conventional method and $5{,}250.3$ using IndeedOPT with MED algorithm. This, in comparison to our best point at error of $1{,}533.97$, shows remarkable improvement (over $70\%$) over previous results, on a system that essentially acted as an out-of-sample system with no training set to benefit from in terms of fine-tuning model parameters. The optimized force field also yields very feasible configurations as shown in the figures below.

Figure~\ref{fig:WS1} shows the bond lengths and angles for the $\mathrm{WS4H4}$ molecule obtained using the optimum parameters. As can be seen, all $\mathrm{W-S}$ bonds are equal and have a length of $2.33$\AA, where the gold standard obtained by quantum chemical calculations shows $2.4$\AA \ with all equal length. Similarly, $\mathrm{S-W-S}$ angles were obtained as $109.6$, $109.3$, and $109.2$, which are close to the gold standard values of $109.6$, $109.4$, and $109.3$.

\begin{figure}[!htb]
\begin{minipage}[b]{0.45\textwidth}
    \centering
    \includegraphics[scale=0.3]{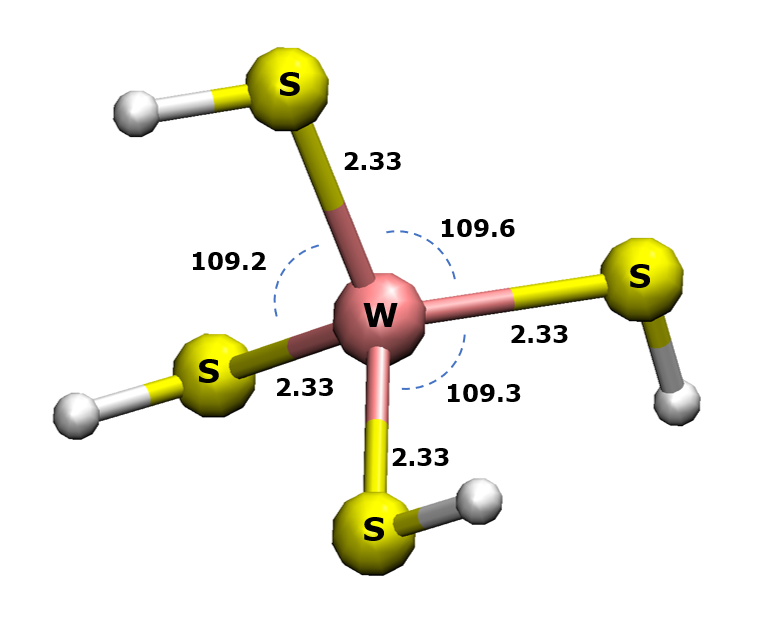}
    \caption{Final configuration of $\mathrm{WS4H4}$ using optimized parameters}
    \label{fig:WS1}
\end{minipage}
\hfill
\begin{minipage}[b]{0.48\textwidth}    
    \centering
    \includegraphics[scale=0.3]{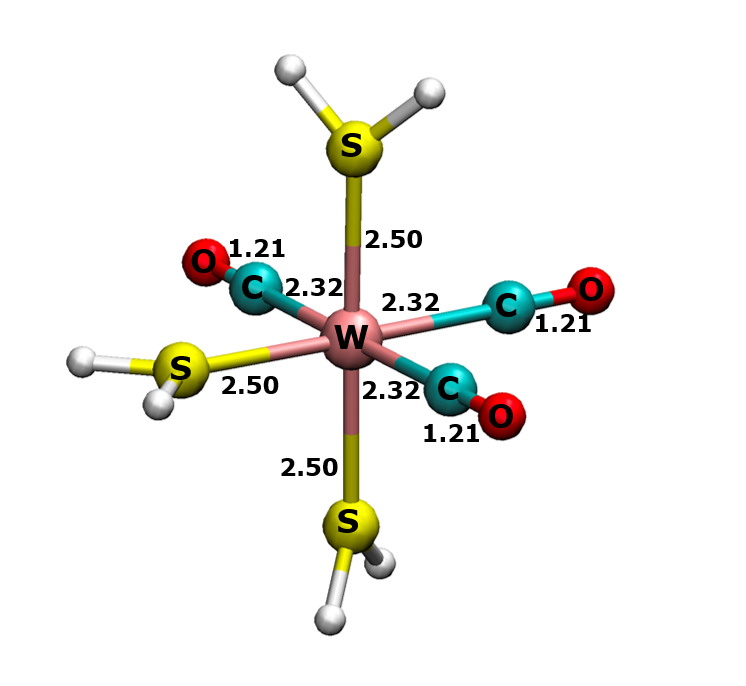}
    \caption{Final configuration of $\mathrm{WS3C3O3}$ using optimized parameters}
    \label{fig:WS2}
\end{minipage}
\end{figure}
Figure~\ref{fig:WS2} shows the bond lengths for the $\mathrm{WS3C3O3}$ molecule obtained using the optimum parameters. The optimized molecule has $\mathrm{W-S}$ bond lengths of $2.5$\AA, $\mathrm{W-C}$ bond length of $2.32$\AA, and $\mathrm{C-O}$ bond length of $1.21$\AA. All similar bond types were optimized at equal lengths. The gold standard bond lengths obtained from quantum chemical calculations are $2.54$\AA, $2.31$\AA, and $1.22$\AA \ for $\mathrm{W-S}$, $\mathrm{W-C}$, and $\mathrm{C-O}$ bonds, respectively, concluding that the optimum parameters yield molecule geometry very close to the geometries obtained by more advanced calculations.

\section{Discussion}
In our work, we revisit the problem of optimizing the tuning parameters of ReaxFF for a given system $\mathrm{Mo-S}$, conduct various numerical experiments to explore the function landscape, establish some characteristics along the way, such as discontinuity, that were not previously widely reported, and eventually provide a sampling procedure that successfully improves on arbitrary starting points to points within a specified threshold of error. A notable achievement is the discovery of tuning parameters with adjusted error as low as $13{,}000$ that can spur new interest in the $\mathrm{Mo-S}$ system, as well as in optimizing other systems with larger input-output sets. Further, it may not be unreasonable to try to optimize some other already optimized systems with our procedure to compare and contrast performances with other previous results in the literature.

While we achieve successful results from the point of view of the ReaxFF objectives for the given $\mathrm{Mo-S}$ system, there are still many unknown characteristics of ReaxFF in general. For instance, we do not elaborate on the problem of infeasibility of certain input points since, in a batch update, we have other points to choose from. In the event that a whole batch of samples is infeasible, shrinking the size of the step-size acts like a reset. With a variety of ways to ensure we can move far away from bad points, infeasible points were not a point of contention when using our methods. However, infeasibility is an important characteristic of ReaxFF that was a central issue in other work in the literature \citep{CLAIMED}CC1. Similarly, discontinuities that were discovered in a few directions were mainly a consequence of the inner minimization routine, as we have reported. There could be other causes of discontinuity within ReaxFF such as conditional clauses within specific routines that may have physical justification for their existence. One other issue along similar lines that we encountered was the issue of numerical precision. Specifically, for any input, a numerical precision below $4$ decimal places did not yield coherent outputs, showing possible discontinuous jumps that may not be due to the inner minimization routine. Further, there exists an aspect of randomness within the ReaxFF control set that adds an unknown random quantity to the inner minimization routine each time it is called upon. While this can be controlled to be ``on'' or ``off'', the inner minimization seems to produce sensible outputs only when this randomness feature is enabled. The shortcoming of such randomness is the inability to reproduce the exact results every time. While we maintain reproducibility of our outputs for the most part through known random seeds set at each evaluation, the randomness aspect is nested into many layers within ReaxFF that can still produce somewhat different results in consecutive runs. For example, running the inner minimization routines through a different order of the latent variables of the reference systems can change the output due to the randomness involved. We created structures to ensure that every output we report is reproducible to the extent possible, but it may be worthwhile to revisit the inner minimization routine with a view to the problem at hand, optimization of the tuning parameters. Modifying the inner minimization routine with a view to improving ReaxFF functional attributes such as continuity or reproducibility can help leverage some more advanced methods in the optimization literature. Other characteristics worth exploring include learning any sparsity structure that may exist locally for each input, leveraging the correlational structure of the outputs of ReaxFF, and establishing any additivity relationship among input coordinates (even if only locally) that can allow for adopting sophisticated optimization procedures that work well in lower dimensions.

Overall, we were able to exploit some characteristics of ReaxFF that we learned through the numerical experiments and propose a procedure that is system-agnostic in terms of achieving extremely low error optimized configurations. While we only provide extensive testing and reporting on the one chosen system, $\mathrm{Mo-S}$, we were also able to test our method on another system, $\mathrm{W-S}$, to provide significant improvement in error over optimized force fields compared to previous literature. For bigger systems than $\mathrm{Mo-S}$, challenges of an even higher dimensionality in the inputs may become a limiting factor in our procedure, even with the parallelization of different processes, but it is worth putting to the test as a future scope. 

\section{Acknowledgements}
The authors acknowledge the Office of Advanced Research Computing (OARC) at Rutgers, The State University of New Jersey for providing access to the Amarel cluster and associated research computing resources that have contributed to the results reported here. URL: https://it.rutgers.edu/oarc

\pagebreak
\bibliographystyle{apalike}
\bibliography{ref}

\pagebreak
\begin{appendices}
\newpage
\setcounter{page}{1}
\begin{center}
{\bf \Large  Supplementary Material\\for\\``Optimization of ReaxFF parameters using random optimization and coordinate search''}
\end{center}
    \section{$\mathrm{Mo-S}$ system specifications}\label{App:labels_desc}

    \paragraph{Parameter Descriptions for $\mathrm{Mo-S}$ system}
{\small
\begin{tabularx}{\textwidth}{c|c|p{3cm}|c|X}
    No.&Name&Desc&Details&Description\\
    \hline
    1&gamma &EEM shielding parameter&Mo&Shielding term for Coulombic interactions of Molybdenum\\
    2&chiEEM &EEM electronegativity&Mo&Electronegativity term for Coulombic interactions of Molybdenum\\
    3&etaEEM &EEM hardness &Mo&Shielding term for Coulombic interactions of Molybdenum\\
    4&De(sigma) &Sigma-bond dissociation energy &Mo-S&Dissociation energy of sigma bond between Molybdenum and Sulfur\\
    5&De(pi) &Pi-bond dissociation energy &Mo-S&Dissociation energy of pi bond between Molybdenum and Sulfur\\
    6&p(be1)  &Bond energy parameter&Mo-S&Coefficient related to bond energy between Molybdenum and Sulfur \\
    7&p(ovun1)&Overcoordination penalty parameter &Mo-S&Overcoordination penalty term for Molybdenum and Sulfur bond\\
    8&p(be2)  &Bond energy parameter &Mo-S&Coefficient related to bond energy between Molybdenum and Sulfur \\
    9&p(bo3)  &Pi bond order&Mo-S&Bond order related term for pi bond between Molybdenum and Sulfur \\
    10&p(bo4)  &Pi bond order&Mo-S&Bond order related term for pi bond between Molybdenum and Sulfur \\
    11&p(bo1)  &Sigma bond order&Mo-S&Bond order related term for sigma bond between Molybdenum and Sulfur \\
    12&p(bo2)  &Sigma bond order&Mo-S&Bond order related term for sigma bond between Molybdenum and Sulfur \\
    13&Dij&&Mo-S&Coefficient related to off-diagonal terms between Molybdenum and Sulfur \\
    14&RvdW&&Mo-S&Coefficient related to off-diagonal terms between Molybdenum and Sulfur \\
    15&alfa&&Mo-S&Coefficient related to off-diagonal terms between Molybdenum and Sulfur \\
    16&ro(sigma)&&Mo-S&Coefficient related to off-diagonal terms between Molybdenum and Sulfur \\
    17&ro(pi)&&Mo-S&Coefficient related to off-diagonal terms between Molybdenum and Sulfur \\
    18&Thetao,o&180 – equilibrium angle &S-Mo-S&Equilibrium angle of S-Mo-S compound\\
    19&p(val1)  &Valency energy (force constant) &S-Mo-S&Force constant for valency energy of S-Mo-S compound\\
    20&p(val2)  &Valency energy (force constant) &S-Mo-S&Force constant for valency energy of S-Mo-S compound\\
    21&p(val7)  &Valence energy (undercoord) parameter &S-Mo-S&Undercoordination term for valency energy of S-Mo-S compound\\
    22&p(val4)  &Valence angle energy parameter &S-Mo-S&Energy parameter for valence angle of S-Mo-S compound\\
    23&Thetao,o&180 – equilibrium angle &Mo-S-Mo&Equilibrium angle of Mo-S-Mo compound\\
    24&p(val1)  &Valency energy (force constant) &Mo-S-Mo&Force constant for valency energy of Mo-S-Mo compound\\
    25&p(val2)  &Valency energy (force constant) &Mo-S-Mo&Force constant for valency energy of Mo-S-Mo compound\\
    26&p(val7)  &Valence energy (undercoord) parameter &Mo-S-Mo&Undercoordination term for valency energy of Mo-S-Mo compound\\
    27&p(val4)  &Valence angle energy parameter &Mo-S-Mo&Energy parameter for valence angle of Mo-S-Mo compound\\
    28&Thetao,o&180 – equilibrium angle &S-S-Mo&Equilibrium angle of S-S-Mo compound\\
    29&p(val1)  &Valency energy (force constant) &S-S-Mo&Force constant for valency energy of S-S-Mo compound\\
    30&p(val2)  &Valency energy (force constant) &S-S-Mo&Force constant for valency energy of S-S-Mo compound\\
    31&p(val7)  &Valence energy (undercoord) parameter &S-S-Mo&Undercoordination term for valency energy of S-S-Mo compound\\
    32&p(val4)  &Valence angle energy parameter &S-S-Mo&Energy parameter for valence angle of S-S-Mo compound\\
    33&Thetao,o&180 – equilibrium angle &S-Mo-Mo&Equilibrium angle of S-Mo-Mo compound\\
    34&p(val1)  &Valency energy (force constant) &S-Mo-Mo&Force constant for valency energy of S-Mo-Mo compound\\
    35&p(val2)  &Valency energy (force constant) &S-Mo-Mo&Force constant for valency energy of S-Mo-Mo compound\\
    36&p(val7)  &Valence energy (undercoord) parameter &S-Mo-Mo&Undercoordination term for valency energy of S-Mo-Mo compound\\
    37&p(val4)  &Valence angle energy parameter &S-Mo-Mo&Energy parameter for valence angle of S-Mo-Mo compound\\
    38&Thetao,o&180 – equilibrium angle &H-S-Mo&Equilibrium angle of H-S-Mo compound\\
    39&p(val1)  &Valency energy (force constant) &H-S-Mo&Force constant for valency energy of H-S-Mo compound\\
    40&p(val2)  &Valency energy (force constant) &H-S-Mo&Force constant for valency energy of H-S-Mo compound\\
    41&p(val4)  &Valence angle energy parameter &H-S-Mo&Energy parameter for valence angle of H-S-Mo compound\\
    42&Thetao,o&180 – equilibrium angle &H-Mo-S&Equilibrium angle of H-Mo-S compound\\
    43&p(val1)  &Valency energy (force constant) &H-Mo-S&Force constant for valency energy of H-Mo-S compound\\
    44&p(val2)  &Valency energy (force constant) &H-Mo-S&Force constant for valency energy of H-Mo-S compound\\
    45&p(val4)  &Valence angle energy parameter &H-Mo-S&Energy parameter for valence angle of H-Mo-S compound \\
    \caption{Parameter Descriptions for $\mathrm{Mo-S}$ system}
    \label{tbl_param}
\end{tabularx}}

\pagebreak
\section{Discontinuity}\label{app:discontinuity}

\begin{table}[!htb]
    \centering
    \begin{tabular}{|c|c|c|c|c|c|}
    \hline
        parameter.0& $0.0156$ &parameter.15& $-0.1863$ &parameter.30& $-0.0685$ \\
        parameter.1& $0.2072$ &parameter.16& $0.0390$ &parameter.31& $0.0276$ \\
        parameter.2& $-0.1492$ &parameter.17& $-2.2476$ &parameter.32& $1.4114$ \\
        parameter.3& $2.1239$ &parameter.18& $-0.4673$ &parameter.33& $1.9474$ \\
        parameter.4& $-0.0328$ &parameter.19& $0.0270$ &parameter.34& $-0.0026$ \\
        parameter.5& $0.0116$ &parameter.20& $-0.0406$ &parameter.35& $-0.0909$ \\
        parameter.6& $-0.0173$ &parameter.21& $0.1514$ &parameter.36& $0.1456$ \\
        parameter.7& $-0.0219$ &parameter.22& $-0.0265$ &parameter.37& $3.2821$ \\
        parameter.8& $0.0147$ &parameter.23& $0.5671$ &parameter.38& $-0.0032$ \\
        parameter.9& $0.1544$ &parameter.24& $0.0473$ &parameter.39& $0.1504$ \\
        parameter.10& $0$ &parameter.25& $-0.0738$ &parameter.40& $-0.0816$ \\
        parameter.11& $-0.3914$ &parameter.26& $-0.0022$ &parameter.41& $2.8565$ \\
        parameter.12& $-0.0139$ &parameter.27& $-0.0418$ &parameter.42& $0.9831$ \\
        parameter.13& $0.1512$ &parameter.28& $-0.0044$ &parameter.43& $-0.3408$ \\
        parameter.14& $-0.3433$ &parameter.29& $0.0904$ &parameter.44& $0.1082$ \\
        \hline
    \end{tabular}
    \caption{This table gives the known search direction $\bm v$ where the discontinuity was discovered as in Section~\ref{sec:Chap3Discontinuity}. Figure~\ref{fig:Chap3ReaxFFdiscontinuous} shows the total error evaluations ($\mathcal{E}(\bm x)$) of the $\mathrm{Mo-S}$ system for various input configurations given by $\bm x + d\bm v$ where $\bm x$ is one of the initial sample points with error$=96{,}216.417$, $\bm v$ is as in this table and $d$ is a scalar value ranging from $0$ to $2$.}
    \label{tab:my_label}
\end{table}

\clearpage
\section{Discrete jumps geometries}~\label{app:discretejumpgeo}

\begin{table}[!htb]
    \centering
    \begin{tabular}{|c|c|c|c|}
    \hline
         Location & Output & Reaction involved & Jump in error\\
    \hline
         $(0.55, 0.56]$ & $j=111$ & $\mathrm{MoS(SH)_2H - MoS_2SHH}$ &  $-2{,}270.0205 $ \\
         $(1.21, 1.22]$ & $j=103$ & $\mathrm{MoS_3H_2 - MoS_2HSH}$ &  ${\color{white}+}2{,}794.7903 $ \\
         $(1.52, 1.53]$ & $j=303$ & $\mathrm{MoS_4H_2 - MoS_4H_2ha}$ &  $-2{,}095.6661 $ \\
    \hline
    \end{tabular}
    \caption{Identifying which molecules are involved in the dicrete jumps at various points of discontinuity in Figure~\ref{fig:Chap3ReaxFFdiscontinuous}}
    \label{tab:chapt3discontpoints}
\end{table}

\section{Additive models for grid evaluations}\label{sec:Chap3_GAM}
\begin{table}[!htb]
    \begin{tabular}{p{4cm}|p{2cm}|p{2cm}|p{2cm}|p{2cm}}
    \multicolumn{4}{@{}l}{\em Parametric Coefficients:}\\
    \centering
         & Estimate & Std. Error & t-value & Pr(>|t|)  \\ \hline
         (Intercept)& $73586.6$ & $162.8$ & $452$ & $<2e^{-16}$
    \end{tabular}
    
    \medskip
    \begin{tabular}{p{4cm}|p{2cm}|p{2cm}|p{2cm}|p{2cm}}
    \multicolumn{4}{@{}l}{\em Approximate Significance of smooth terms:}\\
         & edf & Ref.df & F & p-value  \\ \hline
         s(parameter9)& $8.495$ & $8.928$ & $35.37$ & $<2e^{-16}$ \\
         s(parameter21)& $4.878$ & $5.958$ & $560.57$ & $<2e^{-16}$
    \end{tabular}
    
    \bigskip
    \begin{tabular}{p{4cm}|p{2cm}}
         Diagnostics&   \\ \hline
         R-sq (adj.)& $0.938$  \\
         Deviance explained& $94.1\%$ \\
         GCV & $6.8716e^{6}$ \\
         Scale est. & $6.4668e^{6}$ 
    \end{tabular}
    \caption{Model summary of an additive model using gam() function from the mgcv package for grid points in Figure~\ref{fig:Chap3Contours} for coordinates $p \in \{22,9\}$ with $n=244$}
    \label{tab:Chap3_GAM1}
\end{table}

\begin{table}
    \begin{tabular}{p{4cm}|p{2cm}|p{2cm}|p{2cm}|p{2cm}}
    \multicolumn{4}{@{}l}{\em Parametric Coefficients:}\\
    \centering
         & Estimate & Std. Error & t-value & Pr(>|t|)  \\ \hline
         (Intercept)& $65863.5$ & $38.8$ & $1698$ & $<2e^{-16}$
    \end{tabular}
    
    \medskip
    \begin{tabular}{p{4cm}|p{2cm}|p{2cm}|p{2cm}|p{2cm}}
    \multicolumn{4}{@{}l}{\em Approximate Significance of smooth terms:}\\
         & edf & Ref.df & F & p-value  \\ \hline
         s(parameter44)& $8.973$ & $9.000$ & $1484$ & $<2e^{-16}$ \\
         s(parameter21)& $8.418$ & $8.905$ & $6188$ & $<2e^{-16}$
    \end{tabular}
    
    \bigskip
    \begin{tabular}{p{4cm}|p{2cm}}
         Diagnostics&   \\ \hline
         R-sq (adj.)& $0.996$  \\
         Deviance explained& $99.7\%$ \\
         GCV & $4.1521e^{5}$ \\
         Scale est. & $3.8538e^{5}$ 
    \end{tabular}
    \caption{Model summary of an additive model using gam() function from the mgcv package for grid points in Figure~\ref{fig:Chap3Contours} for coordinates $p \in \{22,45\}$ with $n=256$}
    \label{tab:Chap3_GAM2}
\end{table}

\section{Procedure}\label{app:fullproc}

Then, our procedure is as given below. Note that $K_1, K_2, K_3$ are counters chosen such that the process can break out of each loop within some known number of iterations if error thresholds $\mathcal{E}_{th1},\mathcal{E}_{th2},\mathcal{E}_{th3}$ are not met easily within each loop. In our implementation runs for obtaining the results in the section~\ref{sec:Chap3Results}, we used $K_1 = 5, K_2 = 5, K_3 = 20$ and $\mathcal{E}_{th1} = 200{,}000,\mathcal{E}_{th2} = 0,\mathcal{E}_{th3} = 0$. As for other constants, $M=20$, $\delta = 0.1$, $\delta_{th1} = 0.001$, $L=p$, $\delta' = 1$, $\delta_{th2} = 0.01$. `` Shrink $\delta'$ '' in the procedure below was achieved by using a scalar multiplier $0 \le c < 1$ such that $c\delta'$ is the new step size. Specifically, $c=1/\sqrt{10}$ was used for obtaining the results in the section~\ref{sec:Chap3Results}.
\begin{tcolorbox}[breakable, enhanced]
    \begin{itemize}
        \item Define 
        \begin{itemize}
            \item $k=1$. 
            \item counter thresholds $K_1,K_2,K_3 > 1$. 
             \item Define error thresholds $\mathcal{E}_{th1} > \mathcal{E}_{th2} > \mathcal{E}_{th3} > 0$.
        \end{itemize}
        \item Given $\bm x_i$. Compute $\mathcal{E}(\bm x_i)$.
        \item while ($\mathcal{E}(\bm x_i) > \mathcal{E}_{th3}$ \& $k \le K_3)$ do:\textit{\color{gray} (running both inner loops until a smaller threshold if needed)}
        \begin{itemize}
            \item \textit{\color{gray} Run simultaneous perturbation based sampling until thresholds}
            
            while ($\mathcal{E}(\bm x_i) \ge \mathcal{E}_{th1}$ \& $k \le K_1)$ do:
            \begin{itemize}
                \item Generate $\forall m=1,\dots,M, \bm \Delta_m = (\delta_1,\dots,\delta_p)$ such that $\delta_j = \pm 1$ w.p. $0.5$. 
                \item Define $\delta >0$.
                \item for $m = 1:M$ do:
                \begin{itemize}
                    \item $\bm u_i^m = \bm x_i + \delta ||\bm x_i||\bm \Delta_m$
                    \item Compute $\mathcal{E}_m(\bm u_i^m)$.
                \end{itemize}
                \item If $\Big(\min_m \mathcal{E}_m(\bm u_i^m) < \mathcal{E}(\bm x_i)\Big)$ then do:
                \begin{itemize}
                    \item $x_i = \argmin_{u_i^m}\mathcal{E}_m(\bm u_i^m)$
                    \item $\mathcal{E}(\bm x_i) = \min_m \mathcal{E}_m(\bm  u_i^m) \quad $ \textit{\color{gray} run parallel if possible}
                \end{itemize} 
                else: 
                \begin{itemize}
                \item if ($\delta < \delta_{th1}$)  then  exit loop
                
                else : Shrink $\delta$
                \end{itemize}
            \end{itemize}
            \item \textit{\color{gray} Run brute search until thresholds}
            
            while ($\mathcal{E}(\bm x_i) \ge \mathcal{E}_{th2}$ \& $k \le K_2)$ do:
            \begin{itemize}
                \item Define $\delta' >0$.
                \item Generate $\forall l=1,\dots,L, \bm \Delta_l = (\delta_1,\dots,\delta_p)$ such that for $\delta_{j=l} = \delta'$ , $\delta_{j\ne l} = 0$ otherwise. 
                \item for $l = 1:L$ do:
                \begin{itemize}
                    \item $\bm u_i^l = \bm x_i + \bm \Delta_l$
                    \item $\bm u_i^{L+l} = \bm x_i - \bm \Delta_l$
                    \item Compute $\mathcal{E}_l(\bm u_i^l)$ and $\mathcal{E}_l(\bm u_i^{L+l})$.
                \end{itemize}
                \item If $\Big(\min_l \mathcal{E}_l(\bm u_i^l) < \mathcal{E}(\bm x_i)\Big)$ then do:
                \begin{itemize}
                    \item $x_i = \argmin_{u_i^l}\mathcal{E}_l(\bm u_i^m)$
                    \item $\mathcal{E}(\bm x_i) = \min_l \mathcal{E}_l(\bm  u_i^l) \quad $ \textit{\color{gray} run parallel if possible}
                \end{itemize} 
                else: 
                \begin{itemize}
                \item if ($\delta < \delta_{th2}$)  then \textit{\color{gray} exit loop}
                
                else : Shrink $\delta'$
                \end{itemize}
            \end{itemize}
        \end{itemize}
    \end{itemize}
\end{tcolorbox}

\pagebreak
\section{$\mathrm{W-S}$ system specifications}\label{app:W-S}
Refer to \citet{INDEEDOPT} Supplementary Table 3 for full details.
    \paragraph{Parameter Descriptions for $\mathrm{W-S}$ system}
{\small
\begin{tabularx}{\textwidth}{c|c|c|c|c}
Parameter&Section&Atom type&Lower Limit&Upper Limit\\
\hline
1&Bond&W-S&30&200\\
2&Bond&W-S&30&200\\
3&Bond&W-S&1&-1\\
4&Bond&W-S&0.01&1\\
5&Bond&W-S&0.01&1\\
6&Bond&W-S&-0.05&-0.4\\
7&Bond&W-S&7.5&25\\
8&Bond&W-S&-0.05&-0.3\\
9&Bond&W-S&4.5&10\\
10&Off-diagonal&W-S&0.1&0.4\\
11&Off-diagonal&W-S&2.3&2.5\\
12&Off-diagonal&W-S&9&13\\
13&Off-diagonal&W-S&1.7&2.7\\
14&Off-diagonal&W-S&1.3&2\\
15&Angle&S-W-S&0.01&90\\
16&Angle&S-W-S&0.01&50\\
17&Angle&S-W-S&0.01&8\\
18&Angle&S-W-S&-0.01&-15\\
19&Angle&S-W-S&0.01&8\\
20&Angle&S-W-S&1&4\\
21&Angle&W-S-W&0.01&90\\
22&Angle&W-S-W&0.01&50\\
23&Angle&W-S-W&0.01&8\\
24&Angle&W-S-W&0.01&8\\
25&Angle&W-S-W&1&4\\
26&Angle&S-S-W&0.01&90\\
27&Angle&S-S-W&0.01&50\\
28&Angle&S-S-W&0.01&8\\
29&Angle&S-S-W&0.01&8\\
30&Angle&S-S-W&1&4\\
31&Angle&S-W-W&0.01&90\\
32&Angle&S-W-W&0.01&50\\
33&Angle&S-W-W&0.01&8\\
34&Angle&S-W-W&0.01&8\\
35&Angle&S-W-W&1&4\\
36&Angle&H-S-W&0.01&90\\
37&Angle&H-S-W&0.01&50\\
38&Angle&H-S-W&0.01&8\\
39&Angle&H-S-W&0.01&8\\
40&Angle&H-S-W&1&4\\
41&Angle&H-W-S&0.01&90\\
42&Angle&H-W-S&0.01&50\\
43&Angle&H-W-S&0.01&8\\
44&Angle&H-W-S&0.01&8\\
45&Angle&H-W-S&1&4\\
46&Angle&S-H-W&0.01&50\\
47&Angle&S-H-W&0.01&8\\
48&Angle&S-H-W&0.01&8\\
49&Angle&S-H-W&1&4\\
50&Angle&C-W-S&0.01&90\\
51&Angle&C-W-S&0.01&50\\
52&Angle&C-W-S&0.01&8\\
53&Angle&S-C-W&0.01&8\\
54&Angle&C-W-S&1&4\\
55&Angle&S-C-W&0.01&90\\
56&Angle&S-C-W&0.01&50\\
57&Angle&S-C-W&0.01&8\\
58&Angle&S-C-W&0.01&8\\
59&Angle&S-C-W&1&4\\
60&Angle&C-S-W&0.01&90\\
61&Angle&C-S-W&0.01&50\\
62&Angle&C-S-W&0.01&8\\
63&Angle&C-S-W&0.01&8\\
64&Angle&C-S-W&1&4\\
65&Angle&C-C-S&0.01&50\\
66&Angle&C-C-S&0.01&8\\
67&Angle&C-C-S&0.01&8\\
68&Angle&C-C-S&1&4
\label{tbl_param}
\end{tabularx}}
\end{appendices}
\end{document}

%% file: figures/Chapter3/flowdiag.tex
\begin{tikzpicture}[node distance=1cm]

        \node (start) [startstop, align=center] {Input \\ parameters $(\bm x)$};
        \node (in1) [startstop, above right=of start] {$e_2^* = \min_{\bm r_2} f(\bm x, \bm r_2)$};
        \node (in2) [startstop, below right=of start] {$e_1^* = \min_{\bm r_1} f(\bm x, \bm r_1)$};
        \node (sub) [connector, below right=of in1] {-};
        \node (stop) [startstop,  right=of sub] {$y_{j'}(\bm x)$};
        \node (side1) [startstop, above = of in1, align=center] {Initial geometry \\$\mathrm{MoH_5}$ $(\bm r_2^0)$};
        \node (side2) [startstop, below = of in2, align=center] {Initial geometry \\$\mathrm{MoH_4}$ $(\bm r_1^0)$};

        \draw [arrow] (start) |- (in1);
        \draw [arrow] (start) |- (in2);
        \draw [arrow] (in1) -| (sub);
        \draw [arrow] (in2) -| (sub);
        \draw [arrow] (sub) -- (stop);
        \draw [arrow] (side1) -- (in1);
        \draw [arrow] (side2) -- (in2);

        \begin{scope}[on background layer]
            \node[draw, dotted, inner sep=10pt, fit=(in1) (in2) (sub),label={center, align=center:ReaxFF \\ gray-box}, fill=gray!10] {};
        \end{scope}
    \end{tikzpicture}

%% file: figures/Chapter3/discrete.tex
\begin{tikzpicture}[x=1pt,y=1pt]
\definecolor{fillColor}{RGB}{255,255,255}
\path[use as bounding box,fill=fillColor,fill opacity=0.00] (0,0) rectangle (397.48,289.08);
\begin{scope}
\path[clip] ( 49.20, 61.20) rectangle (372.28,239.88);
\definecolor{drawColor}{RGB}{0,0,0}

\path[draw=drawColor,line width= 0.4pt,line join=round,line cap=round] ( 61.17,229.34) circle (  2.25);

\path[draw=drawColor,line width= 0.4pt,line join=round,line cap=round] ( 69.48,230.06) circle (  2.25);

\path[draw=drawColor,line width= 0.4pt,line join=round,line cap=round] ( 77.79,233.26) circle (  2.25);

\path[draw=drawColor,line width= 0.4pt,line join=round,line cap=round] ( 86.10,232.83) circle (  2.25);

\path[draw=drawColor,line width= 0.4pt,line join=round,line cap=round] ( 94.41,233.12) circle (  2.25);

\path[draw=drawColor,line width= 0.4pt,line join=round,line cap=round] (102.72,232.19) circle (  2.25);

\path[draw=drawColor,line width= 0.4pt,line join=round,line cap=round] (111.02,232.66) circle (  2.25);

\path[draw=drawColor,line width= 0.4pt,line join=round,line cap=round] (119.33,232.54) circle (  2.25);

\path[draw=drawColor,line width= 0.4pt,line join=round,line cap=round] (127.64,232.13) circle (  2.25);

\path[draw=drawColor,line width= 0.4pt,line join=round,line cap=round] (144.26,232.37) circle (  2.25);

\path[draw=drawColor,line width= 0.4pt,line join=round,line cap=round] (145.93,232.10) circle (  2.25);

\path[draw=drawColor,line width= 0.4pt,line join=round,line cap=round] (147.59,232.34) circle (  2.25);

\path[draw=drawColor,line width= 0.4pt,line join=round,line cap=round] (149.25,232.11) circle (  2.25);

\path[draw=drawColor,line width= 0.4pt,line join=round,line cap=round] (150.91,232.26) circle (  2.25);

\path[draw=drawColor,line width= 0.4pt,line join=round,line cap=round] (152.57,231.71) circle (  2.25);

\path[draw=drawColor,line width= 0.4pt,line join=round,line cap=round] (154.24,195.58) circle (  2.25);

\path[draw=drawColor,line width= 0.4pt,line join=round,line cap=round] (155.90,194.84) circle (  2.25);

\path[draw=drawColor,line width= 0.4pt,line join=round,line cap=round] (157.56,195.39) circle (  2.25);

\path[draw=drawColor,line width= 0.4pt,line join=round,line cap=round] (159.22,194.87) circle (  2.25);

\path[draw=drawColor,line width= 0.4pt,line join=round,line cap=round] (160.88,195.23) circle (  2.25);

\path[draw=drawColor,line width= 0.4pt,line join=round,line cap=round] (210.74,197.50) circle (  2.25);

\path[draw=drawColor,line width= 0.4pt,line join=round,line cap=round] (212.40,198.27) circle (  2.25);

\path[draw=drawColor,line width= 0.4pt,line join=round,line cap=round] (214.07,198.77) circle (  2.25);

\path[draw=drawColor,line width= 0.4pt,line join=round,line cap=round] (215.73,198.34) circle (  2.25);

\path[draw=drawColor,line width= 0.4pt,line join=round,line cap=round] (217.39,198.20) circle (  2.25);

\path[draw=drawColor,line width= 0.4pt,line join=round,line cap=round] (219.05,197.72) circle (  2.25);

\path[draw=drawColor,line width= 0.4pt,line join=round,line cap=round] (220.71,197.61) circle (  2.25);

\path[draw=drawColor,line width= 0.4pt,line join=round,line cap=round] (222.38,198.33) circle (  2.25);

\path[draw=drawColor,line width= 0.4pt,line join=round,line cap=round] (224.04,198.35) circle (  2.25);

\path[draw=drawColor,line width= 0.4pt,line join=round,line cap=round] (225.70,198.50) circle (  2.25);

\path[draw=drawColor,line width= 0.4pt,line join=round,line cap=round] (227.36, 67.82) circle (  2.25);

\path[draw=drawColor,line width= 0.4pt,line join=round,line cap=round] (260.60, 70.68) circle (  2.25);

\path[draw=drawColor,line width= 0.4pt,line join=round,line cap=round] (262.26, 70.36) circle (  2.25);

\path[draw=drawColor,line width= 0.4pt,line join=round,line cap=round] (263.93,116.14) circle (  2.25);

\path[draw=drawColor,line width= 0.4pt,line join=round,line cap=round] (265.59,116.24) circle (  2.25);

\path[draw=drawColor,line width= 0.4pt,line join=round,line cap=round] (267.25,116.30) circle (  2.25);

\path[draw=drawColor,line width= 0.4pt,line join=round,line cap=round] (268.91,116.54) circle (  2.25);

\path[draw=drawColor,line width= 0.4pt,line join=round,line cap=round] (270.57,116.50) circle (  2.25);

\path[draw=drawColor,line width= 0.4pt,line join=round,line cap=round] (272.24,117.01) circle (  2.25);

\path[draw=drawColor,line width= 0.4pt,line join=round,line cap=round] (273.90,116.97) circle (  2.25);

\path[draw=drawColor,line width= 0.4pt,line join=round,line cap=round] (275.56,117.37) circle (  2.25);

\path[draw=drawColor,line width= 0.4pt,line join=round,line cap=round] (277.22,117.39) circle (  2.25);

\path[draw=drawColor,line width= 0.4pt,line join=round,line cap=round] (310.46,126.94) circle (  2.25);

\path[draw=drawColor,line width= 0.4pt,line join=round,line cap=round] (312.12,126.92) circle (  2.25);

\path[draw=drawColor,line width= 0.4pt,line join=round,line cap=round] (313.78,126.87) circle (  2.25);

\path[draw=drawColor,line width= 0.4pt,line join=round,line cap=round] (315.45, 93.95) circle (  2.25);

\path[draw=drawColor,line width= 0.4pt,line join=round,line cap=round] (317.11, 95.34) circle (  2.25);

\path[draw=drawColor,line width= 0.4pt,line join=round,line cap=round] (318.77, 95.44) circle (  2.25);

\path[draw=drawColor,line width= 0.4pt,line join=round,line cap=round] (320.43, 94.92) circle (  2.25);

\path[draw=drawColor,line width= 0.4pt,line join=round,line cap=round] (322.09, 96.08) circle (  2.25);

\path[draw=drawColor,line width= 0.4pt,line join=round,line cap=round] (323.76, 97.39) circle (  2.25);

\path[draw=drawColor,line width= 0.4pt,line join=round,line cap=round] (325.42, 98.52) circle (  2.25);

\path[draw=drawColor,line width= 0.4pt,line join=round,line cap=round] (327.08, 99.80) circle (  2.25);

\path[draw=drawColor,line width= 0.4pt,line join=round,line cap=round] (343.70,113.05) circle (  2.25);

\path[draw=drawColor,line width= 0.4pt,line join=round,line cap=round] (352.01,121.41) circle (  2.25);

\path[draw=drawColor,line width= 0.4pt,line join=round,line cap=round] (360.32,129.64) circle (  2.25);
\end{scope}
\begin{scope}
\path[clip] (  0.00,  0.00) rectangle (397.48,289.08);
\definecolor{drawColor}{RGB}{0,0,0}

\path[draw=drawColor,line width= 0.4pt,line join=round,line cap=round] ( 61.17, 61.20) -- (310.46, 61.20);

\path[draw=drawColor,line width= 0.4pt,line join=round,line cap=round] ( 61.17, 61.20) -- ( 61.17, 55.20);

\path[draw=drawColor,line width= 0.4pt,line join=round,line cap=round] (144.26, 61.20) -- (144.26, 55.20);

\path[draw=drawColor,line width= 0.4pt,line join=round,line cap=round] (227.36, 61.20) -- (227.36, 55.20);

\path[draw=drawColor,line width= 0.4pt,line join=round,line cap=round] (310.46, 61.20) -- (310.46, 55.20);

\node[text=drawColor,anchor=base,inner sep=0pt, outer sep=0pt, scale=  1.00] at ( 61.17, 39.60) {0.0};

\node[text=drawColor,anchor=base,inner sep=0pt, outer sep=0pt, scale=  1.00] at (144.26, 39.60) {0.5};

\node[text=drawColor,anchor=base,inner sep=0pt, outer sep=0pt, scale=  1.00] at (227.36, 39.60) {1.0};

\node[text=drawColor,anchor=base,inner sep=0pt, outer sep=0pt, scale=  1.00] at (310.46, 39.60) {1.5};

\path[draw=drawColor,line width= 0.4pt,line join=round,line cap=round] ( 49.20, 63.42) -- ( 49.20,225.82);

\path[draw=drawColor,line width= 0.4pt,line join=round,line cap=round] ( 49.20, 63.42) -- ( 43.20, 63.42);

\path[draw=drawColor,line width= 0.4pt,line join=round,line cap=round] ( 49.20, 95.90) -- ( 43.20, 95.90);

\path[draw=drawColor,line width= 0.4pt,line join=round,line cap=round] ( 49.20,128.38) -- ( 43.20,128.38);

\path[draw=drawColor,line width= 0.4pt,line join=round,line cap=round] ( 49.20,160.86) -- ( 43.20,160.86);

\path[draw=drawColor,line width= 0.4pt,line join=round,line cap=round] ( 49.20,193.34) -- ( 43.20,193.34);

\path[draw=drawColor,line width= 0.4pt,line join=round,line cap=round] ( 49.20,225.82) -- ( 43.20,225.82);

\node[text=drawColor,rotate= 90.00,anchor=base,inner sep=0pt, outer sep=0pt, scale=  1.00] at ( 34.80, 63.42) {86000};

\node[text=drawColor,rotate= 90.00,anchor=base,inner sep=0pt, outer sep=0pt, scale=  1.00] at ( 34.80,128.38) {90000};

\node[text=drawColor,rotate= 90.00,anchor=base,inner sep=0pt, outer sep=0pt, scale=  1.00] at ( 34.80,193.34) {94000};

\path[draw=drawColor,line width= 0.4pt,line join=round,line cap=round] ( 49.20, 61.20) --
	(372.28, 61.20) --
	(372.28,239.88) --
	( 49.20,239.88) --
	cycle;
\end{scope}
\begin{scope}
\path[clip] ( 49.20, 61.20) rectangle (372.28,239.88);
\definecolor{drawColor}{RGB}{0,0,0}

\path[draw=drawColor,line width= 0.4pt,dash pattern=on 1pt off 3pt ,line join=round,line cap=round] (152.57, 61.20) -- (152.57,239.88);

\path[draw=drawColor,line width= 0.4pt,dash pattern=on 1pt off 3pt ,line join=round,line cap=round] (225.70, 61.20) -- (225.70,239.88);

\path[draw=drawColor,line width= 0.4pt,dash pattern=on 1pt off 3pt ,line join=round,line cap=round] (262.26, 61.20) -- (262.26,239.88);

\path[draw=drawColor,line width= 0.4pt,dash pattern=on 1pt off 3pt ,line join=round,line cap=round] (313.78, 61.20) -- (313.78,239.88);
\end{scope}
\begin{scope}
\path[clip] (  0.00,  0.00) rectangle (397.48,289.08);
\definecolor{drawColor}{RGB}{0,0,0}

\node[text=drawColor,anchor=base,inner sep=0pt, outer sep=0pt, scale=  1.00] at (210.74, 15.60) {Distance ($d$)};

\node[text=drawColor,rotate= 90.00,anchor=base,inner sep=0pt, outer sep=0pt, scale=  1.00] at ( 10.80,150.54) {Total error  $\mathcal{E}(\bm x + d\bm v)$};
\end{scope}
\end{tikzpicture}

%% file: figures/Chapter3/SGDwithFew-v1.tex
\begin{tikzpicture}[x=1pt,y=1pt,scale=0.75]
\definecolor{fillColor}{RGB}{255,255,255}
\path[use as bounding box,fill=fillColor,fill opacity=0.00] (0,0) rectangle (397.48,289.08);
\begin{scope}
\path[clip] ( 49.20, 61.20) rectangle (372.28,239.88);
\definecolor{drawColor}{RGB}{0,0,0}

\path[draw=drawColor,line width= 0.4pt,line join=round,line cap=round] ( 61.17,116.20) --
	( 64.16,190.58) --
	( 67.15,190.43) --
	( 70.14,190.43) --
	( 73.13,227.50) --
	( 76.12,227.50) --
	( 79.12,115.62) --
	( 82.11,115.62) --
	( 85.10,231.54) --
	( 88.09,231.54) --
	( 91.08,109.11) --
	( 94.07,109.11) --
	( 97.06,190.33) --
	(100.06,190.33) --
	(103.05,112.26) --
	(106.04,112.26) --
	(109.03,216.44) --
	(112.02,216.44) --
	(115.01,186.90) --
	(118.01,186.90) --
	(121.00,125.09) --
	(123.99,125.09) --
	(126.98,129.04) --
	(129.97,129.04) --
	(132.96,182.66) --
	(135.95,182.66) --
	(138.95,222.46) --
	(141.94,222.46) --
	(144.93,184.67) --
	(147.92,184.67) --
	(150.91,190.85) --
	(153.90,190.85) --
	(156.89,224.22) --
	(159.89,224.22) --
	(162.88,225.20) --
	(165.87,225.20) --
	(168.86,227.16) --
	(171.85,227.16) --
	(174.84,190.33) --
	(177.84,190.33) --
	(180.83,227.76) --
	(183.82,227.76) --
	(186.81, 90.44) --
	(189.80, 90.44) --
	(192.79, 97.85) --
	(195.78, 97.85) --
	(198.78, 83.40) --
	(201.77, 83.40) --
	(204.76, 95.02) --
	(207.75, 95.02) --
	(210.74, 90.22) --
	(213.73, 90.22) --
	(216.73, 85.82) --
	(219.72, 85.82) --
	(222.71, 92.96) --
	(225.70, 92.96) --
	(228.69, 67.82) --
	(231.68, 67.82) --
	(234.67,233.26) --
	(237.67,233.26) --
	(240.66, 87.37) --
	(243.65, 87.37) --
	(246.64, 88.93) --
	(249.63, 88.93) --
	(252.62, 92.18) --
	(255.62, 92.18) --
	(258.61, 94.74) --
	(261.60, 94.74) --
	(264.59, 89.51) --
	(267.58, 89.51) --
	(270.57, 87.75) --
	(273.56, 87.75) --
	(276.56, 92.75) --
	(279.55, 92.75) --
	(282.54, 87.70) --
	(285.53, 87.70) --
	(288.52, 88.31) --
	(291.51, 88.31) --
	(294.51, 90.39) --
	(297.50, 90.39) --
	(300.49, 88.85) --
	(303.48, 88.85) --
	(306.47, 95.11) --
	(309.46, 95.11) --
	(312.45,102.94) --
	(315.45,102.94) --
	(318.44, 89.55) --
	(321.43, 89.55) --
	(324.42,102.45) --
	(327.41,102.45) --
	(330.40, 85.27) --
	(333.40, 85.27) --
	(336.39,102.94) --
	(339.38,102.94) --
	(342.37, 83.05) --
	(345.36, 83.05) --
	(348.35, 87.02) --
	(351.34, 87.02) --
	(354.34, 89.85) --
	(357.33, 89.85) --
	(360.32, 87.79);
\end{scope}
\begin{scope}
\path[clip] (  0.00,  0.00) rectangle (397.48,289.08);
\definecolor{drawColor}{RGB}{0,0,0}

\path[draw=drawColor,line width= 0.4pt,line join=round,line cap=round] ( 61.17, 61.20) -- (360.32, 61.20);

\path[draw=drawColor,line width= 0.4pt,line join=round,line cap=round] ( 61.17, 61.20) -- ( 61.17, 55.20);

\path[draw=drawColor,line width= 0.4pt,line join=round,line cap=round] (121.00, 61.20) -- (121.00, 55.20);

\path[draw=drawColor,line width= 0.4pt,line join=round,line cap=round] (180.83, 61.20) -- (180.83, 55.20);

\path[draw=drawColor,line width= 0.4pt,line join=round,line cap=round] (240.66, 61.20) -- (240.66, 55.20);

\path[draw=drawColor,line width= 0.4pt,line join=round,line cap=round] (300.49, 61.20) -- (300.49, 55.20);

\path[draw=drawColor,line width= 0.4pt,line join=round,line cap=round] (360.32, 61.20) -- (360.32, 55.20);

\node[text=drawColor,anchor=base,inner sep=0pt, outer sep=0pt, scale=  1.00] at ( 61.17, 39.60) {0};

\node[text=drawColor,anchor=base,inner sep=0pt, outer sep=0pt, scale=  1.00] at (121.00, 39.60) {20};

\node[text=drawColor,anchor=base,inner sep=0pt, outer sep=0pt, scale=  1.00] at (180.83, 39.60) {40};

\node[text=drawColor,anchor=base,inner sep=0pt, outer sep=0pt, scale=  1.00] at (240.66, 39.60) {60};

\node[text=drawColor,anchor=base,inner sep=0pt, outer sep=0pt, scale=  1.00] at (300.49, 39.60) {80};

\node[text=drawColor,anchor=base,inner sep=0pt, outer sep=0pt, scale=  1.00] at (360.32, 39.60) {100};

\path[draw=drawColor,line width= 0.4pt,line join=round,line cap=round] ( 49.20, 61.20) --
	(372.28, 61.20) --
	(372.28,239.88) --
	( 49.20,239.88) --
	cycle;
\end{scope}
\begin{scope}
\path[clip] (  0.00,  0.00) rectangle (397.48,289.08);
\definecolor{drawColor}{RGB}{0,0,0}

\node[text=drawColor,anchor=base,inner sep=0pt, outer sep=0pt, scale=  1.00] at (210.74, 15.60) { };

\node[text=drawColor,rotate= 90.00,anchor=base,inner sep=0pt, outer sep=0pt, scale=  1.00] at ( 10.80,150.54) { };
\end{scope}
\begin{scope}
\path[clip] ( 49.20, 61.20) rectangle (372.28,239.88);
\definecolor{drawColor}{RGB}{255,0,0}

\path[draw=drawColor,line width= 0.4pt,dash pattern=on 1pt off 3pt ,line join=round,line cap=round] ( 49.20,116.20) -- (372.28,116.20);
\end{scope}
\begin{scope}
\path[clip] (  0.00,  0.00) rectangle (397.48,289.08);
\definecolor{drawColor}{RGB}{0,0,0}

\path[draw=drawColor,line width= 0.4pt,line join=round,line cap=round] ( 49.20, 61.20) -- ( 49.20,239.88);

\path[draw=drawColor,line width= 0.4pt,line join=round,line cap=round] ( 49.20, 79.82) -- ( 43.20, 79.82);

\path[draw=drawColor,line width= 0.4pt,line join=round,line cap=round] ( 49.20,129.89) -- ( 43.20,129.89);

\path[draw=drawColor,line width= 0.4pt,line join=round,line cap=round] ( 49.20,179.97) -- ( 43.20,179.97);

\path[draw=drawColor,line width= 0.4pt,line join=round,line cap=round] ( 49.20,230.04) -- ( 43.20,230.04);

\node[text=drawColor,rotate= 90.00,anchor=base,inner sep=0pt, outer sep=0pt, scale=  1.00] at ( 34.80, 79.82) {1e+04};

\node[text=drawColor,rotate= 90.00,anchor=base,inner sep=0pt, outer sep=0pt, scale=  1.00] at ( 34.80,129.89) {1e+05};

\node[text=drawColor,rotate= 90.00,anchor=base,inner sep=0pt, outer sep=0pt, scale=  1.00] at ( 34.80,179.97) {1e+06};

\node[text=drawColor,rotate= 90.00,anchor=base,inner sep=0pt, outer sep=0pt, scale=  1.00] at ( 34.80,230.04) {1e+07};

\node[text=drawColor,rotate= 90.00,anchor=base,inner sep=0pt, outer sep=0pt, scale= 1] at (10.80,150.54) {Total error = $\mathcal{E}_{\bm Q_1}(\bm x_i^{k})$};
\end{scope}
\end{tikzpicture}

%% file: figures/Chapter3/SGDwithFew-v2.tex
\begin{tikzpicture}[x=1pt,y=1pt,scale=0.75]
\definecolor{fillColor}{RGB}{255,255,255}
\path[use as bounding box,fill=fillColor,fill opacity=0.00] (0,0) rectangle (397.48,289.08);
\begin{scope}
\path[clip] ( 49.20, 61.20) rectangle (372.28,239.88);
\definecolor{drawColor}{RGB}{0,0,0}

\path[draw=drawColor,line width= 0.4pt,line join=round,line cap=round] ( 61.17,160.27) --
	( 64.16,225.02) --
	( 67.15,225.02) --
	( 70.14,230.62) --
	( 73.13,230.62) --
	( 76.12,231.91) --
	( 79.12,231.91) --
	( 82.11,232.22) --
	( 85.10,232.22) --
	( 88.09,168.58) --
	( 91.08,168.58) --
	( 94.07,183.26) --
	( 97.06,183.26) --
	(100.06,166.38) --
	(103.05,166.38) --
	(106.04,222.10) --
	(109.03,222.10) --
	(112.02,180.32) --
	(115.01,180.32) --
	(118.01,171.39) --
	(121.00,171.39) --
	(123.99,175.86) --
	(126.98,175.86) --
	(129.97,174.54) --
	(132.96,174.54) --
	(135.95,225.44) --
	(138.95,225.44) --
	(141.94,177.68) --
	(144.93,177.68) --
	(147.92,226.16) --
	(150.91,226.16) --
	(153.90,226.34) --
	(156.89,226.34) --
	(159.89,226.59) --
	(162.88,226.59) --
	(165.87,227.75) --
	(168.86,227.75) --
	(171.85,225.61) --
	(174.84,225.61) --
	(177.84,230.41) --
	(180.83,230.41) --
	(183.82,155.52) --
	(186.81,155.52) --
	(189.80,101.88) --
	(192.79,101.88) --
	(195.78,136.75) --
	(198.78,136.75) --
	(201.77,130.73) --
	(204.76,130.73) --
	(207.75,114.70) --
	(210.74,114.70) --
	(213.73,108.57) --
	(216.73,108.57) --
	(219.72,122.63) --
	(222.71,122.63) --
	(225.70,140.73) --
	(228.69,140.73) --
	(231.68,233.26) --
	(234.67,233.26) --
	(237.67,145.50) --
	(240.66,145.50) --
	(243.65,128.32) --
	(246.64,128.32) --
	(249.63, 95.69) --
	(252.62, 95.69) --
	(255.62,122.85) --
	(258.61,122.85) --
	(261.60, 89.92) --
	(264.59, 89.92) --
	(267.58, 67.82) --
	(270.57, 67.82) --
	(273.56,106.71) --
	(276.56,106.71) --
	(279.55,134.41) --
	(282.54,134.41) --
	(285.53,143.14) --
	(288.52,143.14) --
	(291.51,134.92) --
	(294.51,134.92) --
	(297.50,150.70) --
	(300.49,150.70) --
	(303.48,120.56) --
	(306.47,120.56) --
	(309.46,135.40) --
	(312.45,135.40) --
	(315.45,136.83) --
	(318.44,136.83) --
	(321.43,117.83) --
	(324.42,117.83) --
	(327.41,153.31) --
	(330.40,153.31) --
	(333.40,139.62) --
	(336.39,139.62) --
	(339.38,108.13) --
	(342.37,108.13) --
	(345.36,117.51) --
	(348.35,117.51) --
	(351.34,125.54) --
	(354.34,125.54) --
	(357.33,148.01);
\end{scope}
\begin{scope}
\path[clip] (  0.00,  0.00) rectangle (397.48,289.08);
\definecolor{drawColor}{RGB}{0,0,0}

\path[draw=drawColor,line width= 0.4pt,line join=round,line cap=round] ( 61.17, 61.20) -- (360.32, 61.20);

\path[draw=drawColor,line width= 0.4pt,line join=round,line cap=round] ( 61.17, 61.20) -- ( 61.17, 55.20);

\path[draw=drawColor,line width= 0.4pt,line join=round,line cap=round] (121.00, 61.20) -- (121.00, 55.20);

\path[draw=drawColor,line width= 0.4pt,line join=round,line cap=round] (180.83, 61.20) -- (180.83, 55.20);

\path[draw=drawColor,line width= 0.4pt,line join=round,line cap=round] (240.66, 61.20) -- (240.66, 55.20);

\path[draw=drawColor,line width= 0.4pt,line join=round,line cap=round] (300.49, 61.20) -- (300.49, 55.20);

\path[draw=drawColor,line width= 0.4pt,line join=round,line cap=round] (360.32, 61.20) -- (360.32, 55.20);

\node[text=drawColor,anchor=base,inner sep=0pt, outer sep=0pt, scale=  1.00] at ( 61.17, 39.60) {0};

\node[text=drawColor,anchor=base,inner sep=0pt, outer sep=0pt, scale=  1.00] at (121.00, 39.60) {20};

\node[text=drawColor,anchor=base,inner sep=0pt, outer sep=0pt, scale=  1.00] at (180.83, 39.60) {40};

\node[text=drawColor,anchor=base,inner sep=0pt, outer sep=0pt, scale=  1.00] at (240.66, 39.60) {60};

\node[text=drawColor,anchor=base,inner sep=0pt, outer sep=0pt, scale=  1.00] at (300.49, 39.60) {80};

\node[text=drawColor,anchor=base,inner sep=0pt, outer sep=0pt, scale=  1.00] at (360.32, 39.60) {100};

\path[draw=drawColor,line width= 0.4pt,line join=round,line cap=round] ( 49.20, 61.20) --
	(372.28, 61.20) --
	(372.28,239.88) --
	( 49.20,239.88) --
	cycle;
\end{scope}
\begin{scope}
\path[clip] (  0.00,  0.00) rectangle (397.48,289.08);
\definecolor{drawColor}{RGB}{0,0,0}

\node[text=drawColor,anchor=base,inner sep=0pt, outer sep=0pt, scale=  1.00] at (210.74, 15.60) { };

\node[text=drawColor,rotate= 90.00,anchor=base,inner sep=0pt, outer sep=0pt, scale=  1.00] at ( 10.80,150.54) { };
\end{scope}
\begin{scope}
\path[clip] ( 49.20, 61.20) rectangle (372.28,239.88);
\definecolor{drawColor}{RGB}{255,0,0}

\path[draw=drawColor,line width= 0.4pt,dash pattern=on 1pt off 3pt ,line join=round,line cap=round] ( 49.20,160.27) -- (372.28,160.27);
\end{scope}
\begin{scope}
\path[clip] (  0.00,  0.00) rectangle (397.48,289.08);
\definecolor{drawColor}{RGB}{0,0,0}

\path[draw=drawColor,line width= 0.4pt,line join=round,line cap=round] ( 49.20, 61.20) -- ( 49.20,239.57);

\path[draw=drawColor,line width= 0.4pt,line join=round,line cap=round] ( 49.20, 82.59) -- ( 43.20, 82.59);

\path[draw=drawColor,line width= 0.4pt,line join=round,line cap=round] ( 49.20,108.75) -- ( 43.20,108.75);

\path[draw=drawColor,line width= 0.4pt,line join=round,line cap=round] ( 49.20,134.91) -- ( 43.20,134.91);

\path[draw=drawColor,line width= 0.4pt,line join=round,line cap=round] ( 49.20,161.08) -- ( 43.20,161.08);

\path[draw=drawColor,line width= 0.4pt,line join=round,line cap=round] ( 49.20,187.24) -- ( 43.20,187.24);

\path[draw=drawColor,line width= 0.4pt,line join=round,line cap=round] ( 49.20,213.41) -- ( 43.20,213.41);

\path[draw=drawColor,line width= 0.4pt,line join=round,line cap=round] ( 49.20,239.57) -- ( 43.20,239.57);

\node[text=drawColor,rotate= 90.00,anchor=base,inner sep=0pt, outer sep=0pt, scale=  1.00] at ( 34.80, 82.59) {1e+02};

\node[text=drawColor,rotate= 90.00,anchor=base,inner sep=0pt, outer sep=0pt, scale=  1.00] at ( 34.80,134.91) {1e+04};

\node[text=drawColor,rotate= 90.00,anchor=base,inner sep=0pt, outer sep=0pt, scale=  1.00] at ( 34.80,187.24) {1e+06};

\node[text=drawColor,rotate= 90.00,anchor=base,inner sep=0pt, outer sep=0pt, scale=  1.00] at ( 34.80,239.57) {1e+08};

\node[text=drawColor,anchor=base,inner sep=0pt, outer sep=0pt, scale=  1.00] at (210.74, 15.60) {Iteration $k$};

\node[text=drawColor,rotate= 90.00,anchor=base,inner sep=0pt, outer sep=0pt, scale= 1] at (10.80,150.54) {Total error = $\mathcal{E}_{\bm Q_2}(\bm x_i^{k})$};
\end{scope}
\end{tikzpicture}

%% file: figures/Chapter3/RandomCDv2.tex
\begin{tikzpicture}[x=1pt,y=1pt]
\definecolor{fillColor}{RGB}{255,255,255}
\path[use as bounding box,fill=fillColor,fill opacity=0.00] (0,0) rectangle (397.48,289.08);
\begin{scope}
\path[clip] ( 49.20, 61.20) rectangle (372.28,239.88);
\definecolor{drawColor}{RGB}{0,0,0}

\path[draw=drawColor,line width= 0.4pt,line join=round,line cap=round] ( 61.17,233.26) --
	( 64.04,230.71) --
	( 66.92,230.71) --
	( 69.80,228.82) --
	( 72.67,225.23) --
	( 75.55,189.23) --
	( 78.42,189.23) --
	( 81.30,161.35) --
	( 84.18,161.27) --
	( 87.05,161.27) --
	( 89.93, 90.69) --
	( 92.81, 90.68) --
	( 95.68, 90.68) --
	( 98.56, 89.88) --
	(101.44, 89.88) --
	(104.31, 89.88) --
	(107.19, 89.88) --
	(110.07, 89.88) --
	(112.94, 89.88) --
	(115.82, 89.88) --
	(118.70, 89.77) --
	(121.57, 89.77) --
	(124.45, 88.48) --
	(127.32, 88.48) --
	(130.20, 88.40) --
	(133.08, 88.40) --
	(135.95, 88.40) --
	(138.83, 88.40) --
	(141.71, 88.40) --
	(144.58, 88.40) --
	(147.46, 88.40) --
	(150.34, 88.40) --
	(153.21, 88.40) --
	(156.09, 88.40) --
	(158.97, 88.31) --
	(161.84, 88.31) --
	(164.72, 88.31) --
	(167.60, 88.31) --
	(170.47, 88.16) --
	(173.35, 88.14) --
	(176.22, 88.14) --
	(179.10, 88.14) --
	(181.98, 88.14) --
	(184.85, 87.93) --
	(187.73, 87.93) --
	(190.61, 86.18) --
	(193.48, 86.13) --
	(196.36, 85.77) --
	(199.24, 85.41) --
	(202.11, 85.41) --
	(204.99, 85.41) --
	(207.87, 84.72) --
	(210.74, 84.52) --
	(213.62, 84.51) --
	(216.50, 84.51) --
	(219.37, 84.51) --
	(222.25, 84.51) --
	(225.12, 84.51) --
	(228.00, 84.51) --
	(230.88, 84.51) --
	(233.75, 84.51) --
	(236.63, 84.51) --
	(239.51, 84.51) --
	(242.38, 84.51) --
	(245.26, 84.51) --
	(248.14, 84.51) --
	(251.01, 84.49) --
	(253.89, 84.49) --
	(256.77, 84.49) --
	(259.64, 84.49) --
	(262.52, 84.49) --
	(265.40, 84.49) --
	(268.27, 84.49) --
	(271.15, 84.49) --
	(274.02, 84.49) --
	(276.90, 84.49) --
	(279.78, 84.49) --
	(282.65, 84.49) --
	(285.53, 84.49) --
	(288.41, 83.49) --
	(291.28, 83.49) --
	(294.16, 83.49) --
	(297.04, 82.63) --
	(299.91, 82.62) --
	(302.79, 82.61) --
	(305.67, 82.60) --
	(308.54, 82.60) --
	(311.42, 82.58) --
	(314.30, 82.58) --
	(317.17, 82.57) --
	(320.05, 82.57) --
	(322.92, 82.57) --
	(325.80, 82.57) --
	(328.68, 82.57) --
	(331.55, 82.55) --
	(334.43, 82.55) --
	(337.31, 82.54) --
	(340.18, 82.54) --
	(343.06, 82.53) --
	(345.94, 82.45) --
	(348.81, 82.45) --
	(351.69, 67.82) --
	(354.57, 67.82) --
	(357.44, 67.82) --
	(360.32, 67.82);
\end{scope}
\begin{scope}
\path[clip] (  0.00,  0.00) rectangle (397.48,289.08);
\definecolor{drawColor}{RGB}{0,0,0}

\path[draw=drawColor,line width= 0.4pt,line join=round,line cap=round] ( 58.29, 61.20) -- (345.94, 61.20);

\path[draw=drawColor,line width= 0.4pt,line join=round,line cap=round] ( 58.29, 61.20) -- ( 58.29, 55.20);

\path[draw=drawColor,line width= 0.4pt,line join=round,line cap=round] (115.82, 61.20) -- (115.82, 55.20);

\path[draw=drawColor,line width= 0.4pt,line join=round,line cap=round] (173.35, 61.20) -- (173.35, 55.20);

\path[draw=drawColor,line width= 0.4pt,line join=round,line cap=round] (230.88, 61.20) -- (230.88, 55.20);

\path[draw=drawColor,line width= 0.4pt,line join=round,line cap=round] (288.41, 61.20) -- (288.41, 55.20);

\path[draw=drawColor,line width= 0.4pt,line join=round,line cap=round] (345.94, 61.20) -- (345.94, 55.20);

\node[text=drawColor,anchor=base,inner sep=0pt, outer sep=0pt, scale=  1.00] at ( 58.29, 39.60) {0};

\node[text=drawColor,anchor=base,inner sep=0pt, outer sep=0pt, scale=  1.00] at (115.82, 39.60) {20};

\node[text=drawColor,anchor=base,inner sep=0pt, outer sep=0pt, scale=  1.00] at (173.35, 39.60) {40};

\node[text=drawColor,anchor=base,inner sep=0pt, outer sep=0pt, scale=  1.00] at (230.88, 39.60) {60};

\node[text=drawColor,anchor=base,inner sep=0pt, outer sep=0pt, scale=  1.00] at (288.41, 39.60) {80};

\node[text=drawColor,anchor=base,inner sep=0pt, outer sep=0pt, scale=  1.00] at (345.94, 39.60) {100};

\path[draw=drawColor,line width= 0.4pt,line join=round,line cap=round] ( 49.20, 67.82) -- ( 49.20,239.26);

\path[draw=drawColor,line width= 0.4pt,line join=round,line cap=round] ( 49.20, 67.82) -- ( 43.20, 67.82);

\path[draw=drawColor,line width= 0.4pt,line join=round,line cap=round] ( 49.20, 92.31) -- ( 43.20, 92.31);

\path[draw=drawColor,line width= 0.4pt,line join=round,line cap=round] ( 49.20,116.80) -- ( 43.20,116.80);

\path[draw=drawColor,line width= 0.4pt,line join=round,line cap=round] ( 49.20,141.29) -- ( 43.20,141.29);

\path[draw=drawColor,line width= 0.4pt,line join=round,line cap=round] ( 49.20,165.79) -- ( 43.20,165.79);

\path[draw=drawColor,line width= 0.4pt,line join=round,line cap=round] ( 49.20,190.28) -- ( 43.20,190.28);

\path[draw=drawColor,line width= 0.4pt,line join=round,line cap=round] ( 49.20,214.77) -- ( 43.20,214.77);

\path[draw=drawColor,line width= 0.4pt,line join=round,line cap=round] ( 49.20,239.26) -- ( 43.20,239.26);

\node[text=drawColor,rotate= 90.00,anchor=base,inner sep=0pt, outer sep=0pt, scale=  1.00] at ( 34.80, 67.82) {0};

\node[text=drawColor,rotate= 90.00,anchor=base,inner sep=0pt, outer sep=0pt, scale=  1.00] at ( 34.80, 92.31) {1000};

\node[text=drawColor,rotate= 90.00,anchor=base,inner sep=0pt, outer sep=0pt, scale=  1.00] at ( 34.80,141.29) {3000};

\node[text=drawColor,rotate= 90.00,anchor=base,inner sep=0pt, outer sep=0pt, scale=  1.00] at ( 34.80,190.28) {5000};

\node[text=drawColor,rotate= 90.00,anchor=base,inner sep=0pt, outer sep=0pt, scale=  1.00] at ( 34.80,239.26) {7000};

\path[draw=drawColor,line width= 0.4pt,line join=round,line cap=round] ( 49.20, 61.20) --
	(372.28, 61.20) --
	(372.28,239.88) --
	( 49.20,239.88) --
	cycle;
\end{scope}
\begin{scope}
\path[clip] (  0.00,  0.00) rectangle (397.48,289.08);
\definecolor{drawColor}{RGB}{0,0,0}

\node[text=drawColor,anchor=base,inner sep=0pt, outer sep=0pt, scale=  1.00] at (210.74, 15.60) {Iteration $k$};

\node[text=drawColor,rotate= 90.00,anchor=base,inner sep=0pt, outer sep=0pt, scale= 1] at (10.80,150.54) {Total error = $\mathcal{E}_{\bm Q}(\bm x_i^{k})$};
\end{scope}
\end{tikzpicture}

%% file: figures/Chapter3/BruteSearch-Box.tex
\begin{tikzpicture}[x=1pt,y=1pt]
\definecolor{fillColor}{RGB}{255,255,255}
\path[use as bounding box,fill=fillColor,fill opacity=0.00] (0,0) rectangle (397.48,289.08);
\begin{scope}
\path[clip] ( 49.20, 61.20) rectangle (372.28,239.88);
\definecolor{drawColor}{RGB}{0,0,0}

\path[draw=drawColor,line width= 0.4pt,line join=round,line cap=round] ( 58.92,175.41) -- ( 63.42,179.91);

\path[draw=drawColor,line width= 0.4pt,line join=round,line cap=round] ( 58.92,179.91) -- ( 63.42,175.41);

\path[draw=drawColor,line width= 0.4pt,line join=round,line cap=round] ( 65.72,164.85) -- ( 70.22,169.35);

\path[draw=drawColor,line width= 0.4pt,line join=round,line cap=round] ( 65.72,169.35) -- ( 70.22,164.85);

\path[draw=drawColor,line width= 0.4pt,line join=round,line cap=round] ( 72.51,172.42) -- ( 77.01,176.92);

\path[draw=drawColor,line width= 0.4pt,line join=round,line cap=round] ( 72.51,176.92) -- ( 77.01,172.42);

\path[draw=drawColor,line width= 0.4pt,line join=round,line cap=round] ( 79.31,165.87) -- ( 83.81,170.37);

\path[draw=drawColor,line width= 0.4pt,line join=round,line cap=round] ( 79.31,170.37) -- ( 83.81,165.87);

\path[draw=drawColor,line width= 0.4pt,line join=round,line cap=round] ( 86.11,171.88) -- ( 90.61,176.38);

\path[draw=drawColor,line width= 0.4pt,line join=round,line cap=round] ( 86.11,176.38) -- ( 90.61,171.88);

\path[draw=drawColor,line width= 0.4pt,line join=round,line cap=round] ( 92.91,174.63) -- ( 97.41,179.13);

\path[draw=drawColor,line width= 0.4pt,line join=round,line cap=round] ( 92.91,179.13) -- ( 97.41,174.63);

\path[draw=drawColor,line width= 0.4pt,line join=round,line cap=round] ( 99.71,155.03) -- (104.21,159.53);

\path[draw=drawColor,line width= 0.4pt,line join=round,line cap=round] ( 99.71,159.53) -- (104.21,155.03);

\path[draw=drawColor,line width= 0.4pt,line join=round,line cap=round] (106.51,170.74) -- (111.01,175.24);

\path[draw=drawColor,line width= 0.4pt,line join=round,line cap=round] (106.51,175.24) -- (111.01,170.74);

\path[draw=drawColor,line width= 0.4pt,line join=round,line cap=round] (113.31,172.92) -- (117.81,177.42);

\path[draw=drawColor,line width= 0.4pt,line join=round,line cap=round] (113.31,177.42) -- (117.81,172.92);

\path[draw=drawColor,line width= 0.4pt,line join=round,line cap=round] (120.11,120.05) -- (124.61,124.55);

\path[draw=drawColor,line width= 0.4pt,line join=round,line cap=round] (120.11,124.55) -- (124.61,120.05);

\path[draw=drawColor,line width= 0.4pt,line join=round,line cap=round] (126.91,173.78) -- (131.41,178.28);

\path[draw=drawColor,line width= 0.4pt,line join=round,line cap=round] (126.91,178.28) -- (131.41,173.78);

\path[draw=drawColor,line width= 0.4pt,line join=round,line cap=round] (133.70,168.94) -- (138.20,173.44);

\path[draw=drawColor,line width= 0.4pt,line join=round,line cap=round] (133.70,173.44) -- (138.20,168.94);

\path[draw=drawColor,line width= 0.4pt,line join=round,line cap=round] (140.50,149.89) -- (145.00,154.39);

\path[draw=drawColor,line width= 0.4pt,line join=round,line cap=round] (140.50,154.39) -- (145.00,149.89);

\path[draw=drawColor,line width= 0.4pt,line join=round,line cap=round] (147.30,169.13) -- (151.80,173.63);

\path[draw=drawColor,line width= 0.4pt,line join=round,line cap=round] (147.30,173.63) -- (151.80,169.13);

\path[draw=drawColor,line width= 0.4pt,line join=round,line cap=round] (154.10,151.52) -- (158.60,156.02);

\path[draw=drawColor,line width= 0.4pt,line join=round,line cap=round] (154.10,156.02) -- (158.60,151.52);

\path[draw=drawColor,line width= 0.4pt,line join=round,line cap=round] (160.90,126.60) -- (165.40,131.10);

\path[draw=drawColor,line width= 0.4pt,line join=round,line cap=round] (160.90,131.10) -- (165.40,126.60);

\path[draw=drawColor,line width= 0.4pt,line join=round,line cap=round] (167.70,172.64) -- (172.20,177.14);

\path[draw=drawColor,line width= 0.4pt,line join=round,line cap=round] (167.70,177.14) -- (172.20,172.64);

\path[draw=drawColor,line width= 0.4pt,line join=round,line cap=round] (174.50,173.06) -- (179.00,177.56);

\path[draw=drawColor,line width= 0.4pt,line join=round,line cap=round] (174.50,177.56) -- (179.00,173.06);

\path[draw=drawColor,line width= 0.4pt,line join=round,line cap=round] (181.30,168.07) -- (185.80,172.57);

\path[draw=drawColor,line width= 0.4pt,line join=round,line cap=round] (181.30,172.57) -- (185.80,168.07);

\path[draw=drawColor,line width= 0.4pt,line join=round,line cap=round] (188.10,165.59) -- (192.60,170.09);

\path[draw=drawColor,line width= 0.4pt,line join=round,line cap=round] (188.10,170.09) -- (192.60,165.59);

\path[draw=drawColor,line width= 0.4pt,line join=round,line cap=round] (194.89,161.28) -- (199.39,165.78);

\path[draw=drawColor,line width= 0.4pt,line join=round,line cap=round] (194.89,165.78) -- (199.39,161.28);

\path[draw=drawColor,line width= 0.4pt,line join=round,line cap=round] (201.69,170.48) -- (206.19,174.98);

\path[draw=drawColor,line width= 0.4pt,line join=round,line cap=round] (201.69,174.98) -- (206.19,170.48);

\path[draw=drawColor,line width= 0.4pt,line join=round,line cap=round] (208.49,175.73) -- (212.99,180.23);

\path[draw=drawColor,line width= 0.4pt,line join=round,line cap=round] (208.49,180.23) -- (212.99,175.73);

\path[draw=drawColor,line width= 0.4pt,line join=round,line cap=round] (215.29,171.45) -- (219.79,175.95);

\path[draw=drawColor,line width= 0.4pt,line join=round,line cap=round] (215.29,175.95) -- (219.79,171.45);

\path[draw=drawColor,line width= 0.4pt,line join=round,line cap=round] (222.09,172.66) -- (226.59,177.16);

\path[draw=drawColor,line width= 0.4pt,line join=round,line cap=round] (222.09,177.16) -- (226.59,172.66);

\path[draw=drawColor,line width= 0.4pt,line join=round,line cap=round] (228.89,176.08) -- (233.39,180.58);

\path[draw=drawColor,line width= 0.4pt,line join=round,line cap=round] (228.89,180.58) -- (233.39,176.08);

\path[draw=drawColor,line width= 0.4pt,line join=round,line cap=round] (235.69,168.29) -- (240.19,172.79);

\path[draw=drawColor,line width= 0.4pt,line join=round,line cap=round] (235.69,172.79) -- (240.19,168.29);

\path[draw=drawColor,line width= 0.4pt,line join=round,line cap=round] (242.49,174.89) -- (246.99,179.39);

\path[draw=drawColor,line width= 0.4pt,line join=round,line cap=round] (242.49,179.39) -- (246.99,174.89);

\path[draw=drawColor,line width= 0.4pt,line join=round,line cap=round] (249.29,166.48) -- (253.79,170.98);

\path[draw=drawColor,line width= 0.4pt,line join=round,line cap=round] (249.29,170.98) -- (253.79,166.48);

\path[draw=drawColor,line width= 0.4pt,line join=round,line cap=round] (256.08,168.72) -- (260.58,173.22);

\path[draw=drawColor,line width= 0.4pt,line join=round,line cap=round] (256.08,173.22) -- (260.58,168.72);

\path[draw=drawColor,line width= 0.4pt,line join=round,line cap=round] (262.88,170.94) -- (267.38,175.44);

\path[draw=drawColor,line width= 0.4pt,line join=round,line cap=round] (262.88,175.44) -- (267.38,170.94);

\path[draw=drawColor,line width= 0.4pt,line join=round,line cap=round] (269.68,166.69) -- (274.18,171.19);

\path[draw=drawColor,line width= 0.4pt,line join=round,line cap=round] (269.68,171.19) -- (274.18,166.69);

\path[draw=drawColor,line width= 0.4pt,line join=round,line cap=round] (276.48,175.98) -- (280.98,180.48);

\path[draw=drawColor,line width= 0.4pt,line join=round,line cap=round] (276.48,180.48) -- (280.98,175.98);

\path[draw=drawColor,line width= 0.4pt,line join=round,line cap=round] (283.28,175.35) -- (287.78,179.85);

\path[draw=drawColor,line width= 0.4pt,line join=round,line cap=round] (283.28,179.85) -- (287.78,175.35);

\path[draw=drawColor,line width= 0.4pt,line join=round,line cap=round] (290.08,174.28) -- (294.58,178.78);

\path[draw=drawColor,line width= 0.4pt,line join=round,line cap=round] (290.08,178.78) -- (294.58,174.28);

\path[draw=drawColor,line width= 0.4pt,line join=round,line cap=round] (296.88,176.09) -- (301.38,180.59);

\path[draw=drawColor,line width= 0.4pt,line join=round,line cap=round] (296.88,180.59) -- (301.38,176.09);

\path[draw=drawColor,line width= 0.4pt,line join=round,line cap=round] (303.68,173.03) -- (308.18,177.53);

\path[draw=drawColor,line width= 0.4pt,line join=round,line cap=round] (303.68,177.53) -- (308.18,173.03);

\path[draw=drawColor,line width= 0.4pt,line join=round,line cap=round] (310.48,170.57) -- (314.98,175.07);

\path[draw=drawColor,line width= 0.4pt,line join=round,line cap=round] (310.48,175.07) -- (314.98,170.57);

\path[draw=drawColor,line width= 0.4pt,line join=round,line cap=round] (317.28,170.35) -- (321.78,174.85);

\path[draw=drawColor,line width= 0.4pt,line join=round,line cap=round] (317.28,174.85) -- (321.78,170.35);

\path[draw=drawColor,line width= 0.4pt,line join=round,line cap=round] (324.07,174.14) -- (328.57,178.64);

\path[draw=drawColor,line width= 0.4pt,line join=round,line cap=round] (324.07,178.64) -- (328.57,174.14);

\path[draw=drawColor,line width= 0.4pt,line join=round,line cap=round] (330.87,171.82) -- (335.37,176.32);

\path[draw=drawColor,line width= 0.4pt,line join=round,line cap=round] (330.87,176.32) -- (335.37,171.82);

\path[draw=drawColor,line width= 0.4pt,line join=round,line cap=round] (337.67,176.04) -- (342.17,180.54);

\path[draw=drawColor,line width= 0.4pt,line join=round,line cap=round] (337.67,180.54) -- (342.17,176.04);

\path[draw=drawColor,line width= 0.4pt,line join=round,line cap=round] (344.47,171.26) -- (348.97,175.76);

\path[draw=drawColor,line width= 0.4pt,line join=round,line cap=round] (344.47,175.76) -- (348.97,171.26);

\path[draw=drawColor,line width= 0.4pt,line join=round,line cap=round] (351.27,173.86) -- (355.77,178.36);

\path[draw=drawColor,line width= 0.4pt,line join=round,line cap=round] (351.27,178.36) -- (355.77,173.86);

\path[draw=drawColor,line width= 0.4pt,line join=round,line cap=round] (358.07, 94.98) -- (362.57, 99.48);

\path[draw=drawColor,line width= 0.4pt,line join=round,line cap=round] (358.07, 99.48) -- (362.57, 94.98);
\end{scope}
\begin{scope}
\path[clip] (  0.00,  0.00) rectangle (397.48,289.08);
\definecolor{drawColor}{RGB}{0,0,0}

\path[draw=drawColor,line width= 0.4pt,line join=round,line cap=round] ( 61.17, 61.20) -- (360.32, 61.20);

\path[draw=drawColor,line width= 0.4pt,line join=round,line cap=round] ( 61.17, 61.20) -- ( 61.17, 55.20);

\path[draw=drawColor,line width= 0.4pt,line join=round,line cap=round] ( 67.97, 61.20) -- ( 67.97, 55.20);

\path[draw=drawColor,line width= 0.4pt,line join=round,line cap=round] ( 74.76, 61.20) -- ( 74.76, 55.20);

\path[draw=drawColor,line width= 0.4pt,line join=round,line cap=round] ( 81.56, 61.20) -- ( 81.56, 55.20);

\path[draw=drawColor,line width= 0.4pt,line join=round,line cap=round] ( 88.36, 61.20) -- ( 88.36, 55.20);

\path[draw=drawColor,line width= 0.4pt,line join=round,line cap=round] ( 95.16, 61.20) -- ( 95.16, 55.20);

\path[draw=drawColor,line width= 0.4pt,line join=round,line cap=round] (101.96, 61.20) -- (101.96, 55.20);

\path[draw=drawColor,line width= 0.4pt,line join=round,line cap=round] (108.76, 61.20) -- (108.76, 55.20);

\path[draw=drawColor,line width= 0.4pt,line join=round,line cap=round] (115.56, 61.20) -- (115.56, 55.20);

\path[draw=drawColor,line width= 0.4pt,line join=round,line cap=round] (122.36, 61.20) -- (122.36, 55.20);

\path[draw=drawColor,line width= 0.4pt,line join=round,line cap=round] (129.16, 61.20) -- (129.16, 55.20);

\path[draw=drawColor,line width= 0.4pt,line join=round,line cap=round] (135.95, 61.20) -- (135.95, 55.20);

\path[draw=drawColor,line width= 0.4pt,line join=round,line cap=round] (142.75, 61.20) -- (142.75, 55.20);

\path[draw=drawColor,line width= 0.4pt,line join=round,line cap=round] (149.55, 61.20) -- (149.55, 55.20);

\path[draw=drawColor,line width= 0.4pt,line join=round,line cap=round] (156.35, 61.20) -- (156.35, 55.20);

\path[draw=drawColor,line width= 0.4pt,line join=round,line cap=round] (163.15, 61.20) -- (163.15, 55.20);

\path[draw=drawColor,line width= 0.4pt,line join=round,line cap=round] (169.95, 61.20) -- (169.95, 55.20);

\path[draw=drawColor,line width= 0.4pt,line join=round,line cap=round] (176.75, 61.20) -- (176.75, 55.20);

\path[draw=drawColor,line width= 0.4pt,line join=round,line cap=round] (183.55, 61.20) -- (183.55, 55.20);

\path[draw=drawColor,line width= 0.4pt,line join=round,line cap=round] (190.35, 61.20) -- (190.35, 55.20);

\path[draw=drawColor,line width= 0.4pt,line join=round,line cap=round] (197.14, 61.20) -- (197.14, 55.20);

\path[draw=drawColor,line width= 0.4pt,line join=round,line cap=round] (203.94, 61.20) -- (203.94, 55.20);

\path[draw=drawColor,line width= 0.4pt,line join=round,line cap=round] (210.74, 61.20) -- (210.74, 55.20);

\path[draw=drawColor,line width= 0.4pt,line join=round,line cap=round] (217.54, 61.20) -- (217.54, 55.20);

\path[draw=drawColor,line width= 0.4pt,line join=round,line cap=round] (224.34, 61.20) -- (224.34, 55.20);

\path[draw=drawColor,line width= 0.4pt,line join=round,line cap=round] (231.14, 61.20) -- (231.14, 55.20);

\path[draw=drawColor,line width= 0.4pt,line join=round,line cap=round] (237.94, 61.20) -- (237.94, 55.20);

\path[draw=drawColor,line width= 0.4pt,line join=round,line cap=round] (244.74, 61.20) -- (244.74, 55.20);

\path[draw=drawColor,line width= 0.4pt,line join=round,line cap=round] (251.54, 61.20) -- (251.54, 55.20);

\path[draw=drawColor,line width= 0.4pt,line join=round,line cap=round] (258.33, 61.20) -- (258.33, 55.20);

\path[draw=drawColor,line width= 0.4pt,line join=round,line cap=round] (265.13, 61.20) -- (265.13, 55.20);

\path[draw=drawColor,line width= 0.4pt,line join=round,line cap=round] (271.93, 61.20) -- (271.93, 55.20);

\path[draw=drawColor,line width= 0.4pt,line join=round,line cap=round] (278.73, 61.20) -- (278.73, 55.20);

\path[draw=drawColor,line width= 0.4pt,line join=round,line cap=round] (285.53, 61.20) -- (285.53, 55.20);

\path[draw=drawColor,line width= 0.4pt,line join=round,line cap=round] (292.33, 61.20) -- (292.33, 55.20);

\path[draw=drawColor,line width= 0.4pt,line join=round,line cap=round] (299.13, 61.20) -- (299.13, 55.20);

\path[draw=drawColor,line width= 0.4pt,line join=round,line cap=round] (305.93, 61.20) -- (305.93, 55.20);

\path[draw=drawColor,line width= 0.4pt,line join=round,line cap=round] (312.73, 61.20) -- (312.73, 55.20);

\path[draw=drawColor,line width= 0.4pt,line join=round,line cap=round] (319.53, 61.20) -- (319.53, 55.20);

\path[draw=drawColor,line width= 0.4pt,line join=round,line cap=round] (326.32, 61.20) -- (326.32, 55.20);

\path[draw=drawColor,line width= 0.4pt,line join=round,line cap=round] (333.12, 61.20) -- (333.12, 55.20);

\path[draw=drawColor,line width= 0.4pt,line join=round,line cap=round] (339.92, 61.20) -- (339.92, 55.20);

\path[draw=drawColor,line width= 0.4pt,line join=round,line cap=round] (346.72, 61.20) -- (346.72, 55.20);

\path[draw=drawColor,line width= 0.4pt,line join=round,line cap=round] (353.52, 61.20) -- (353.52, 55.20);

\path[draw=drawColor,line width= 0.4pt,line join=round,line cap=round] (360.32, 61.20) -- (360.32, 55.20);

\node[text=drawColor,anchor=base,inner sep=0pt, outer sep=0pt, scale=  1.00] at ( 61.17, 39.60) {1};

\node[text=drawColor,anchor=base,inner sep=0pt, outer sep=0pt, scale=  1.00] at ( 74.76, 39.60) {3};

\node[text=drawColor,anchor=base,inner sep=0pt, outer sep=0pt, scale=  1.00] at ( 88.36, 39.60) {5};

\node[text=drawColor,anchor=base,inner sep=0pt, outer sep=0pt, scale=  1.00] at (101.96, 39.60) {7};

\node[text=drawColor,anchor=base,inner sep=0pt, outer sep=0pt, scale=  1.00] at (115.56, 39.60) {9};

\node[text=drawColor,anchor=base,inner sep=0pt, outer sep=0pt, scale=  1.00] at (135.95, 39.60) {12};

\node[text=drawColor,anchor=base,inner sep=0pt, outer sep=0pt, scale=  1.00] at (156.35, 39.60) {15};

\node[text=drawColor,anchor=base,inner sep=0pt, outer sep=0pt, scale=  1.00] at (176.75, 39.60) {18};

\node[text=drawColor,anchor=base,inner sep=0pt, outer sep=0pt, scale=  1.00] at (197.14, 39.60) {21};

\node[text=drawColor,anchor=base,inner sep=0pt, outer sep=0pt, scale=  1.00] at (217.54, 39.60) {24};

\node[text=drawColor,anchor=base,inner sep=0pt, outer sep=0pt, scale=  1.00] at (237.94, 39.60) {27};

\node[text=drawColor,anchor=base,inner sep=0pt, outer sep=0pt, scale=  1.00] at (258.33, 39.60) {30};

\node[text=drawColor,anchor=base,inner sep=0pt, outer sep=0pt, scale=  1.00] at (278.73, 39.60) {33};

\node[text=drawColor,anchor=base,inner sep=0pt, outer sep=0pt, scale=  1.00] at (299.13, 39.60) {36};

\node[text=drawColor,anchor=base,inner sep=0pt, outer sep=0pt, scale=  1.00] at (319.53, 39.60) {39};

\node[text=drawColor,anchor=base,inner sep=0pt, outer sep=0pt, scale=  1.00] at (339.92, 39.60) {42};

\node[text=drawColor,anchor=base,inner sep=0pt, outer sep=0pt, scale=  1.00] at (360.32, 39.60) {45};

\path[draw=drawColor,line width= 0.4pt,line join=round,line cap=round] ( 49.20, 67.82) -- ( 49.20,233.26);

\path[draw=drawColor,line width= 0.4pt,line join=round,line cap=round] ( 49.20, 67.82) -- ( 43.20, 67.82);

\path[draw=drawColor,line width= 0.4pt,line join=round,line cap=round] ( 49.20, 95.39) -- ( 43.20, 95.39);

\path[draw=drawColor,line width= 0.4pt,line join=round,line cap=round] ( 49.20,122.97) -- ( 43.20,122.97);

\path[draw=drawColor,line width= 0.4pt,line join=round,line cap=round] ( 49.20,150.54) -- ( 43.20,150.54);

\path[draw=drawColor,line width= 0.4pt,line join=round,line cap=round] ( 49.20,178.11) -- ( 43.20,178.11);

\path[draw=drawColor,line width= 0.4pt,line join=round,line cap=round] ( 49.20,205.69) -- ( 43.20,205.69);

\path[draw=drawColor,line width= 0.4pt,line join=round,line cap=round] ( 49.20,233.26) -- ( 43.20,233.26);

\node[text=drawColor,rotate= 90.00,anchor=base,inner sep=0pt, outer sep=0pt, scale=  1.00] at ( 34.80, 67.82) {50000};

\node[text=drawColor,rotate= 90.00,anchor=base,inner sep=0pt, outer sep=0pt, scale=  1.00] at ( 34.80,122.97) {60000};

\node[text=drawColor,rotate= 90.00,anchor=base,inner sep=0pt, outer sep=0pt, scale=  1.00] at ( 34.80,178.11) {70000};

\node[text=drawColor,rotate= 90.00,anchor=base,inner sep=0pt, outer sep=0pt, scale=  1.00] at ( 34.80,233.26) {80000};

\path[draw=drawColor,line width= 0.4pt,line join=round,line cap=round] ( 49.20, 61.20) --
	(372.28, 61.20) --
	(372.28,239.88) --
	( 49.20,239.88) --
	cycle;
\end{scope}
\begin{scope}
\path[clip] (  0.00,  0.00) rectangle (397.48,289.08);
\definecolor{drawColor}{RGB}{0,0,0}

\node[text=drawColor,anchor=base,inner sep=0pt, outer sep=0pt, scale=  1.00] at (210.74, 15.60) { };

\node[text=drawColor,rotate= 90.00,anchor=base,inner sep=0pt, outer sep=0pt, scale=  1.00] at ( 10.80,150.54) { };
\end{scope}
\begin{scope}
\path[clip] ( 49.20, 61.20) rectangle (372.28,239.88);
\definecolor{drawColor}{RGB}{255,0,0}

\path[draw=drawColor,line width= 0.4pt,dash pattern=on 1pt off 3pt ,line join=round,line cap=round] ( 49.20,178.36) -- (372.28,178.36);
\end{scope}
\begin{scope}
\path[clip] (  0.00,  0.00) rectangle (397.48,289.08);
\definecolor{drawColor}{RGB}{0,0,0}

\path[draw=drawColor,line width= 0.4pt,line join=round,line cap=round] (360.32, 61.20) -- (360.32, 61.20);

\path[draw=drawColor,line width= 0.4pt,line join=round,line cap=round] (360.32, 61.20) -- (360.32, 55.20);

\node[text=drawColor,anchor=base,inner sep=0pt, outer sep=0pt, scale=  1.00] at (210.74, 15.60) {Coordinate $p$};

\node[text=drawColor,rotate= 90.00,anchor=base,inner sep=0pt, outer sep=0pt, scale=  1.00] at ( 10.80,150.54) {Minimum total error};
\end{scope}
\end{tikzpicture}

%% file: figures/Chapter3/discrete2.tex
\begin{tikzpicture}[x=1pt,y=1pt]
\definecolor{fillColor}{RGB}{255,255,255}
\path[use as bounding box,fill=fillColor,fill opacity=0.00] (0,0) rectangle (397.48,289.08);
\begin{scope}
\path[clip] ( 66.00, 66.00) rectangle (343.48,223.08);
\definecolor{drawColor}{RGB}{0,0,0}

\path[draw=drawColor,line width= 0.4pt,line join=round,line cap=round] ( 76.28,213.81) circle (  2.25);

\path[draw=drawColor,line width= 0.4pt,line join=round,line cap=round] ( 83.41,214.44) circle (  2.25);

\path[draw=drawColor,line width= 0.4pt,line join=round,line cap=round] ( 90.55,217.26) circle (  2.25);

\path[draw=drawColor,line width= 0.4pt,line join=round,line cap=round] ( 97.69,216.88) circle (  2.25);

\path[draw=drawColor,line width= 0.4pt,line join=round,line cap=round] (104.83,217.14) circle (  2.25);

\path[draw=drawColor,line width= 0.4pt,line join=round,line cap=round] (111.96,216.32) circle (  2.25);

\path[draw=drawColor,line width= 0.4pt,line join=round,line cap=round] (119.10,216.73) circle (  2.25);

\path[draw=drawColor,line width= 0.4pt,line join=round,line cap=round] (126.24,216.63) circle (  2.25);

\path[draw=drawColor,line width= 0.4pt,line join=round,line cap=round] (133.37,216.27) circle (  2.25);

\path[draw=drawColor,line width= 0.4pt,line join=round,line cap=round] (147.65,216.47) circle (  2.25);

\path[draw=drawColor,line width= 0.4pt,line join=round,line cap=round] (149.07,216.24) circle (  2.25);

\path[draw=drawColor,line width= 0.4pt,line join=round,line cap=round] (150.50,216.45) circle (  2.25);

\path[draw=drawColor,line width= 0.4pt,line join=round,line cap=round] (151.93,216.25) circle (  2.25);

\path[draw=drawColor,line width= 0.4pt,line join=round,line cap=round] (153.36,216.38) circle (  2.25);

\path[draw=drawColor,line width= 0.4pt,line join=round,line cap=round] (154.78,215.90) circle (  2.25);

\path[draw=drawColor,line width= 0.4pt,line join=round,line cap=round] (156.21,184.14) circle (  2.25);

\path[draw=drawColor,line width= 0.4pt,line join=round,line cap=round] (157.64,183.49) circle (  2.25);

\path[draw=drawColor,line width= 0.4pt,line join=round,line cap=round] (159.07,183.96) circle (  2.25);

\path[draw=drawColor,line width= 0.4pt,line join=round,line cap=round] (160.49,183.51) circle (  2.25);

\path[draw=drawColor,line width= 0.4pt,line join=round,line cap=round] (161.92,183.83) circle (  2.25);

\path[draw=drawColor,line width= 0.4pt,line join=round,line cap=round] (204.74,185.82) circle (  2.25);

\path[draw=drawColor,line width= 0.4pt,line join=round,line cap=round] (206.17,186.50) circle (  2.25);

\path[draw=drawColor,line width= 0.4pt,line join=round,line cap=round] (207.60,186.94) circle (  2.25);

\path[draw=drawColor,line width= 0.4pt,line join=round,line cap=round] (209.02,186.56) circle (  2.25);

\path[draw=drawColor,line width= 0.4pt,line join=round,line cap=round] (210.45,186.44) circle (  2.25);

\path[draw=drawColor,line width= 0.4pt,line join=round,line cap=round] (211.88,186.02) circle (  2.25);

\path[draw=drawColor,line width= 0.4pt,line join=round,line cap=round] (213.31,185.92) circle (  2.25);

\path[draw=drawColor,line width= 0.4pt,line join=round,line cap=round] (214.73,186.55) circle (  2.25);

\path[draw=drawColor,line width= 0.4pt,line join=round,line cap=round] (216.16,186.57) circle (  2.25);

\path[draw=drawColor,line width= 0.4pt,line join=round,line cap=round] (217.59,186.70) circle (  2.25);

\path[draw=drawColor,line width= 0.4pt,line join=round,line cap=round] (219.02, 71.82) circle (  2.25);

\path[draw=drawColor,line width= 0.4pt,line join=round,line cap=round] (247.56, 74.34) circle (  2.25);

\path[draw=drawColor,line width= 0.4pt,line join=round,line cap=round] (248.99, 74.05) circle (  2.25);

\path[draw=drawColor,line width= 0.4pt,line join=round,line cap=round] (250.42,114.30) circle (  2.25);

\path[draw=drawColor,line width= 0.4pt,line join=round,line cap=round] (251.85,114.39) circle (  2.25);

\path[draw=drawColor,line width= 0.4pt,line join=round,line cap=round] (253.27,114.44) circle (  2.25);

\path[draw=drawColor,line width= 0.4pt,line join=round,line cap=round] (254.70,114.65) circle (  2.25);

\path[draw=drawColor,line width= 0.4pt,line join=round,line cap=round] (256.13,114.62) circle (  2.25);

\path[draw=drawColor,line width= 0.4pt,line join=round,line cap=round] (257.56,115.06) circle (  2.25);

\path[draw=drawColor,line width= 0.4pt,line join=round,line cap=round] (258.98,115.03) circle (  2.25);

\path[draw=drawColor,line width= 0.4pt,line join=round,line cap=round] (260.41,115.38) circle (  2.25);

\path[draw=drawColor,line width= 0.4pt,line join=round,line cap=round] (261.84,115.40) circle (  2.25);

\path[draw=drawColor,line width= 0.4pt,line join=round,line cap=round] (290.39,123.79) circle (  2.25);

\path[draw=drawColor,line width= 0.4pt,line join=round,line cap=round] (291.81,123.78) circle (  2.25);

\path[draw=drawColor,line width= 0.4pt,line join=round,line cap=round] (293.24,123.73) circle (  2.25);

\path[draw=drawColor,line width= 0.4pt,line join=round,line cap=round] (294.67, 94.79) circle (  2.25);

\path[draw=drawColor,line width= 0.4pt,line join=round,line cap=round] (296.10, 96.01) circle (  2.25);

\path[draw=drawColor,line width= 0.4pt,line join=round,line cap=round] (297.52, 96.10) circle (  2.25);

\path[draw=drawColor,line width= 0.4pt,line join=round,line cap=round] (298.95, 95.65) circle (  2.25);

\path[draw=drawColor,line width= 0.4pt,line join=round,line cap=round] (300.38, 96.66) circle (  2.25);

\path[draw=drawColor,line width= 0.4pt,line join=round,line cap=round] (301.81, 97.81) circle (  2.25);

\path[draw=drawColor,line width= 0.4pt,line join=round,line cap=round] (303.23, 98.81) circle (  2.25);

\path[draw=drawColor,line width= 0.4pt,line join=round,line cap=round] (304.66, 99.93) circle (  2.25);

\path[draw=drawColor,line width= 0.4pt,line join=round,line cap=round] (318.93,111.58) circle (  2.25);

\path[draw=drawColor,line width= 0.4pt,line join=round,line cap=round] (326.07,118.93) circle (  2.25);

\path[draw=drawColor,line width= 0.4pt,line join=round,line cap=round] (333.21,126.17) circle (  2.25);
\end{scope}
\begin{scope}
\path[clip] (  0.00,  0.00) rectangle (397.48,289.08);
\definecolor{drawColor}{RGB}{0,0,0}

\path[draw=drawColor,line width= 0.4pt,line join=round,line cap=round] ( 76.28, 66.00) -- (290.39, 66.00);

\path[draw=drawColor,line width= 0.4pt,line join=round,line cap=round] ( 76.28, 66.00) -- ( 76.28, 60.00);

\path[draw=drawColor,line width= 0.4pt,line join=round,line cap=round] (147.65, 66.00) -- (147.65, 60.00);

\path[draw=drawColor,line width= 0.4pt,line join=round,line cap=round] (219.02, 66.00) -- (219.02, 60.00);

\path[draw=drawColor,line width= 0.4pt,line join=round,line cap=round] (290.39, 66.00) -- (290.39, 60.00);

\node[text=drawColor,anchor=base,inner sep=0pt, outer sep=0pt, scale=  1.00] at ( 76.28, 44.40) {0.0};

\node[text=drawColor,anchor=base,inner sep=0pt, outer sep=0pt, scale=  1.00] at (147.65, 44.40) {0.5};

\node[text=drawColor,anchor=base,inner sep=0pt, outer sep=0pt, scale=  1.00] at (219.02, 44.40) {1.0};

\node[text=drawColor,anchor=base,inner sep=0pt, outer sep=0pt, scale=  1.00] at (290.39, 44.40) {1.5};

\path[draw=drawColor,line width= 0.4pt,line join=round,line cap=round] ( 66.00, 67.95) -- ( 66.00,210.72);

\path[draw=drawColor,line width= 0.4pt,line join=round,line cap=round] ( 66.00, 67.95) -- ( 60.00, 67.95);

\path[draw=drawColor,line width= 0.4pt,line join=round,line cap=round] ( 66.00, 96.50) -- ( 60.00, 96.50);

\path[draw=drawColor,line width= 0.4pt,line join=round,line cap=round] ( 66.00,125.06) -- ( 60.00,125.06);

\path[draw=drawColor,line width= 0.4pt,line join=round,line cap=round] ( 66.00,153.61) -- ( 60.00,153.61);

\path[draw=drawColor,line width= 0.4pt,line join=round,line cap=round] ( 66.00,182.17) -- ( 60.00,182.17);

\path[draw=drawColor,line width= 0.4pt,line join=round,line cap=round] ( 66.00,210.72) -- ( 60.00,210.72);

\node[text=drawColor,rotate= 90.00,anchor=base,inner sep=0pt, outer sep=0pt, scale=  1.00] at ( 51.60, 67.95) {86000};

\node[text=drawColor,rotate= 90.00,anchor=base,inner sep=0pt, outer sep=0pt, scale=  1.00] at ( 51.60,125.06) {90000};

\node[text=drawColor,rotate= 90.00,anchor=base,inner sep=0pt, outer sep=0pt, scale=  1.00] at ( 51.60,182.17) {94000};

\path[draw=drawColor,line width= 0.4pt,line join=round,line cap=round] ( 66.00, 66.00) --
	(343.48, 66.00) --
	(343.48,223.08) --
	( 66.00,223.08) --
	cycle;
\end{scope}
\begin{scope}
\path[clip] ( 66.00, 66.00) rectangle (343.48,223.08);
\definecolor{drawColor}{RGB}{223,83,107}

\path[draw=drawColor,line width= 0.4pt,line join=round,line cap=round] ( 74.03, 69.57) -- ( 78.53, 74.07);

\path[draw=drawColor,line width= 0.4pt,line join=round,line cap=round] ( 74.03, 74.07) -- ( 78.53, 69.57);

\path[draw=drawColor,line width= 0.4pt,line join=round,line cap=round] ( 81.16, 72.75) -- ( 85.66, 77.25);

\path[draw=drawColor,line width= 0.4pt,line join=round,line cap=round] ( 81.16, 77.25) -- ( 85.66, 72.75);

\path[draw=drawColor,line width= 0.4pt,line join=round,line cap=round] ( 88.30, 77.48) -- ( 92.80, 81.98);

\path[draw=drawColor,line width= 0.4pt,line join=round,line cap=round] ( 88.30, 81.98) -- ( 92.80, 77.48);

\path[draw=drawColor,line width= 0.4pt,line join=round,line cap=round] ( 95.44, 81.64) -- ( 99.94, 86.14);

\path[draw=drawColor,line width= 0.4pt,line join=round,line cap=round] ( 95.44, 86.14) -- ( 99.94, 81.64);

\path[draw=drawColor,line width= 0.4pt,line join=round,line cap=round] (102.58, 86.84) -- (107.08, 91.34);

\path[draw=drawColor,line width= 0.4pt,line join=round,line cap=round] (102.58, 91.34) -- (107.08, 86.84);

\path[draw=drawColor,line width= 0.4pt,line join=round,line cap=round] (109.71, 92.16) -- (114.21, 96.66);

\path[draw=drawColor,line width= 0.4pt,line join=round,line cap=round] (109.71, 96.66) -- (114.21, 92.16);

\path[draw=drawColor,line width= 0.4pt,line join=round,line cap=round] (116.85, 97.77) -- (121.35,102.27);

\path[draw=drawColor,line width= 0.4pt,line join=round,line cap=round] (116.85,102.27) -- (121.35, 97.77);

\path[draw=drawColor,line width= 0.4pt,line join=round,line cap=round] (123.99,102.84) -- (128.49,107.34);

\path[draw=drawColor,line width= 0.4pt,line join=round,line cap=round] (123.99,107.34) -- (128.49,102.84);

\path[draw=drawColor,line width= 0.4pt,line join=round,line cap=round] (131.12,109.00) -- (135.62,113.50);

\path[draw=drawColor,line width= 0.4pt,line join=round,line cap=round] (131.12,113.50) -- (135.62,109.00);

\path[draw=drawColor,line width= 0.4pt,line join=round,line cap=round] (145.40,119.96) -- (149.90,124.46);

\path[draw=drawColor,line width= 0.4pt,line join=round,line cap=round] (145.40,124.46) -- (149.90,119.96);

\path[draw=drawColor,line width= 0.4pt,line join=round,line cap=round] (146.82,120.79) -- (151.32,125.29);

\path[draw=drawColor,line width= 0.4pt,line join=round,line cap=round] (146.82,125.29) -- (151.32,120.79);

\path[draw=drawColor,line width= 0.4pt,line join=round,line cap=round] (148.25,121.59) -- (152.75,126.09);

\path[draw=drawColor,line width= 0.4pt,line join=round,line cap=round] (148.25,126.09) -- (152.75,121.59);

\path[draw=drawColor,line width= 0.4pt,line join=round,line cap=round] (149.68,123.02) -- (154.18,127.52);

\path[draw=drawColor,line width= 0.4pt,line join=round,line cap=round] (149.68,127.52) -- (154.18,123.02);

\path[draw=drawColor,line width= 0.4pt,line join=round,line cap=round] (151.11,123.88) -- (155.61,128.38);

\path[draw=drawColor,line width= 0.4pt,line join=round,line cap=round] (151.11,128.38) -- (155.61,123.88);

\path[draw=drawColor,line width= 0.4pt,line join=round,line cap=round] (152.53,124.44) -- (157.03,128.94);

\path[draw=drawColor,line width= 0.4pt,line join=round,line cap=round] (152.53,128.94) -- (157.03,124.44);

\path[draw=drawColor,line width= 0.4pt,line join=round,line cap=round] (153.96,125.26) -- (158.46,129.76);

\path[draw=drawColor,line width= 0.4pt,line join=round,line cap=round] (153.96,129.76) -- (158.46,125.26);

\path[draw=drawColor,line width= 0.4pt,line join=round,line cap=round] (155.39,126.56) -- (159.89,131.06);

\path[draw=drawColor,line width= 0.4pt,line join=round,line cap=round] (155.39,131.06) -- (159.89,126.56);

\path[draw=drawColor,line width= 0.4pt,line join=round,line cap=round] (156.82,127.35) -- (161.32,131.85);

\path[draw=drawColor,line width= 0.4pt,line join=round,line cap=round] (156.82,131.85) -- (161.32,127.35);

\path[draw=drawColor,line width= 0.4pt,line join=round,line cap=round] (158.24,128.72) -- (162.74,133.22);

\path[draw=drawColor,line width= 0.4pt,line join=round,line cap=round] (158.24,133.22) -- (162.74,128.72);

\path[draw=drawColor,line width= 0.4pt,line join=round,line cap=round] (159.67,129.49) -- (164.17,133.99);

\path[draw=drawColor,line width= 0.4pt,line join=round,line cap=round] (159.67,133.99) -- (164.17,129.49);

\path[draw=drawColor,line width= 0.4pt,line join=round,line cap=round] (202.49,152.41) -- (206.99,156.91);

\path[draw=drawColor,line width= 0.4pt,line join=round,line cap=round] (202.49,156.91) -- (206.99,152.41);

\path[draw=drawColor,line width= 0.4pt,line join=round,line cap=round] (203.92,153.58) -- (208.42,158.08);

\path[draw=drawColor,line width= 0.4pt,line join=round,line cap=round] (203.92,158.08) -- (208.42,153.58);

\path[draw=drawColor,line width= 0.4pt,line join=round,line cap=round] (205.35,153.90) -- (209.85,158.40);

\path[draw=drawColor,line width= 0.4pt,line join=round,line cap=round] (205.35,158.40) -- (209.85,153.90);

\path[draw=drawColor,line width= 0.4pt,line join=round,line cap=round] (206.77,154.76) -- (211.27,159.26);

\path[draw=drawColor,line width= 0.4pt,line join=round,line cap=round] (206.77,159.26) -- (211.27,154.76);

\path[draw=drawColor,line width= 0.4pt,line join=round,line cap=round] (208.20,155.19) -- (212.70,159.69);

\path[draw=drawColor,line width= 0.4pt,line join=round,line cap=round] (208.20,159.69) -- (212.70,155.19);

\path[draw=drawColor,line width= 0.4pt,line join=round,line cap=round] (209.63,156.08) -- (214.13,160.58);

\path[draw=drawColor,line width= 0.4pt,line join=round,line cap=round] (209.63,160.58) -- (214.13,156.08);

\path[draw=drawColor,line width= 0.4pt,line join=round,line cap=round] (211.06,155.56) -- (215.56,160.06);

\path[draw=drawColor,line width= 0.4pt,line join=round,line cap=round] (211.06,160.06) -- (215.56,155.56);

\path[draw=drawColor,line width= 0.4pt,line join=round,line cap=round] (212.48,156.47) -- (216.98,160.97);

\path[draw=drawColor,line width= 0.4pt,line join=round,line cap=round] (212.48,160.97) -- (216.98,156.47);

\path[draw=drawColor,line width= 0.4pt,line join=round,line cap=round] (213.91,156.67) -- (218.41,161.17);

\path[draw=drawColor,line width= 0.4pt,line join=round,line cap=round] (213.91,161.17) -- (218.41,156.67);

\path[draw=drawColor,line width= 0.4pt,line join=round,line cap=round] (215.34,157.70) -- (219.84,162.20);

\path[draw=drawColor,line width= 0.4pt,line join=round,line cap=round] (215.34,162.20) -- (219.84,157.70);

\path[draw=drawColor,line width= 0.4pt,line join=round,line cap=round] (216.77,157.87) -- (221.27,162.37);

\path[draw=drawColor,line width= 0.4pt,line join=round,line cap=round] (216.77,162.37) -- (221.27,157.87);

\path[draw=drawColor,line width= 0.4pt,line join=round,line cap=round] (245.31,165.08) -- (249.81,169.58);

\path[draw=drawColor,line width= 0.4pt,line join=round,line cap=round] (245.31,169.58) -- (249.81,165.08);

\path[draw=drawColor,line width= 0.4pt,line join=round,line cap=round] (246.74,165.94) -- (251.24,170.44);

\path[draw=drawColor,line width= 0.4pt,line join=round,line cap=round] (246.74,170.44) -- (251.24,165.94);

\path[draw=drawColor,line width= 0.4pt,line join=round,line cap=round] (248.17,166.01) -- (252.67,170.51);

\path[draw=drawColor,line width= 0.4pt,line join=round,line cap=round] (248.17,170.51) -- (252.67,166.01);

\path[draw=drawColor,line width= 0.4pt,line join=round,line cap=round] (249.60,166.86) -- (254.10,171.36);

\path[draw=drawColor,line width= 0.4pt,line join=round,line cap=round] (249.60,171.36) -- (254.10,166.86);

\path[draw=drawColor,line width= 0.4pt,line join=round,line cap=round] (251.02,166.82) -- (255.52,171.32);

\path[draw=drawColor,line width= 0.4pt,line join=round,line cap=round] (251.02,171.32) -- (255.52,166.82);

\path[draw=drawColor,line width= 0.4pt,line join=round,line cap=round] (252.45,166.66) -- (256.95,171.16);

\path[draw=drawColor,line width= 0.4pt,line join=round,line cap=round] (252.45,171.16) -- (256.95,166.66);

\path[draw=drawColor,line width= 0.4pt,line join=round,line cap=round] (253.88,166.84) -- (258.38,171.34);

\path[draw=drawColor,line width= 0.4pt,line join=round,line cap=round] (253.88,171.34) -- (258.38,166.84);

\path[draw=drawColor,line width= 0.4pt,line join=round,line cap=round] (255.31,167.50) -- (259.81,172.00);

\path[draw=drawColor,line width= 0.4pt,line join=round,line cap=round] (255.31,172.00) -- (259.81,167.50);

\path[draw=drawColor,line width= 0.4pt,line join=round,line cap=round] (256.73,167.54) -- (261.23,172.04);

\path[draw=drawColor,line width= 0.4pt,line join=round,line cap=round] (256.73,172.04) -- (261.23,167.54);

\path[draw=drawColor,line width= 0.4pt,line join=round,line cap=round] (258.16,168.40) -- (262.66,172.90);

\path[draw=drawColor,line width= 0.4pt,line join=round,line cap=round] (258.16,172.90) -- (262.66,168.40);

\path[draw=drawColor,line width= 0.4pt,line join=round,line cap=round] (259.59,168.41) -- (264.09,172.91);

\path[draw=drawColor,line width= 0.4pt,line join=round,line cap=round] (259.59,172.91) -- (264.09,168.41);

\path[draw=drawColor,line width= 0.4pt,line join=round,line cap=round] (288.14,176.09) -- (292.64,180.59);

\path[draw=drawColor,line width= 0.4pt,line join=round,line cap=round] (288.14,180.59) -- (292.64,176.09);

\path[draw=drawColor,line width= 0.4pt,line join=round,line cap=round] (289.56,176.93) -- (294.06,181.43);

\path[draw=drawColor,line width= 0.4pt,line join=round,line cap=round] (289.56,181.43) -- (294.06,176.93);

\path[draw=drawColor,line width= 0.4pt,line join=round,line cap=round] (290.99,177.22) -- (295.49,181.72);

\path[draw=drawColor,line width= 0.4pt,line join=round,line cap=round] (290.99,181.72) -- (295.49,177.22);

\path[draw=drawColor,line width= 0.4pt,line join=round,line cap=round] (292.42,178.32) -- (296.92,182.82);

\path[draw=drawColor,line width= 0.4pt,line join=round,line cap=round] (292.42,182.82) -- (296.92,178.32);

\path[draw=drawColor,line width= 0.4pt,line join=round,line cap=round] (293.85,178.78) -- (298.35,183.28);

\path[draw=drawColor,line width= 0.4pt,line join=round,line cap=round] (293.85,183.28) -- (298.35,178.78);

\path[draw=drawColor,line width= 0.4pt,line join=round,line cap=round] (295.27,178.98) -- (299.77,183.48);

\path[draw=drawColor,line width= 0.4pt,line join=round,line cap=round] (295.27,183.48) -- (299.77,178.98);

\path[draw=drawColor,line width= 0.4pt,line join=round,line cap=round] (296.70,179.57) -- (301.20,184.07);

\path[draw=drawColor,line width= 0.4pt,line join=round,line cap=round] (296.70,184.07) -- (301.20,179.57);

\path[draw=drawColor,line width= 0.4pt,line join=round,line cap=round] (298.13,180.59) -- (302.63,185.09);

\path[draw=drawColor,line width= 0.4pt,line join=round,line cap=round] (298.13,185.09) -- (302.63,180.59);

\path[draw=drawColor,line width= 0.4pt,line join=round,line cap=round] (299.56,181.18) -- (304.06,185.68);

\path[draw=drawColor,line width= 0.4pt,line join=round,line cap=round] (299.56,185.68) -- (304.06,181.18);

\path[draw=drawColor,line width= 0.4pt,line join=round,line cap=round] (300.98,182.40) -- (305.48,186.90);

\path[draw=drawColor,line width= 0.4pt,line join=round,line cap=round] (300.98,186.90) -- (305.48,182.40);

\path[draw=drawColor,line width= 0.4pt,line join=round,line cap=round] (302.41,183.11) -- (306.91,187.61);

\path[draw=drawColor,line width= 0.4pt,line join=round,line cap=round] (302.41,187.61) -- (306.91,183.11);

\path[draw=drawColor,line width= 0.4pt,line join=round,line cap=round] (316.68,194.95) -- (321.18,199.45);

\path[draw=drawColor,line width= 0.4pt,line join=round,line cap=round] (316.68,199.45) -- (321.18,194.95);

\path[draw=drawColor,line width= 0.4pt,line join=round,line cap=round] (323.82,203.37) -- (328.32,207.87);

\path[draw=drawColor,line width= 0.4pt,line join=round,line cap=round] (323.82,207.87) -- (328.32,203.37);

\path[draw=drawColor,line width= 0.4pt,line join=round,line cap=round] (330.96,215.01) -- (335.46,219.51);

\path[draw=drawColor,line width= 0.4pt,line join=round,line cap=round] (330.96,219.51) -- (335.46,215.01);
\end{scope}
\begin{scope}
\path[clip] ( 66.00, 66.00) rectangle (343.48,223.08);
\definecolor{drawColor}{RGB}{0,0,0}

\path[draw=drawColor,line width= 0.4pt,dash pattern=on 1pt off 3pt ,line join=round,line cap=round] (154.78, 66.00) -- (154.78,223.08);

\path[draw=drawColor,line width= 0.4pt,dash pattern=on 1pt off 3pt ,line join=round,line cap=round] (217.59, 66.00) -- (217.59,223.08);

\path[draw=drawColor,line width= 0.4pt,dash pattern=on 1pt off 3pt ,line join=round,line cap=round] (248.99, 66.00) -- (248.99,223.08);

\path[draw=drawColor,line width= 0.4pt,dash pattern=on 1pt off 3pt ,line join=round,line cap=round] (293.24, 66.00) -- (293.24,223.08);
\end{scope}
\begin{scope}
\path[clip] (  0.00,  0.00) rectangle (397.48,289.08);
\definecolor{drawColor}{RGB}{0,0,0}

\path[draw=drawColor,line width= 0.4pt,line join=round,line cap=round] (343.48, 66.00) -- (343.48,223.08);

\path[draw=drawColor,line width= 0.4pt,line join=round,line cap=round] (343.48, 74.86) -- (349.48, 74.86);

\path[draw=drawColor,line width= 0.4pt,line join=round,line cap=round] (343.48,102.46) -- (349.48,102.46);

\path[draw=drawColor,line width= 0.4pt,line join=round,line cap=round] (343.48,130.05) -- (349.48,130.05);

\path[draw=drawColor,line width= 0.4pt,line join=round,line cap=round] (343.48,157.65) -- (349.48,157.65);

\path[draw=drawColor,line width= 0.4pt,line join=round,line cap=round] (343.48,185.24) -- (349.48,185.24);

\path[draw=drawColor,line width= 0.4pt,line join=round,line cap=round] (343.48,212.84) -- (349.48,212.84);

\node[text=drawColor,rotate= 90.00,anchor=base,inner sep=0pt, outer sep=0pt, scale=  1.00] at (365.08, 74.86) {210000};

\node[text=drawColor,rotate= 90.00,anchor=base,inner sep=0pt, outer sep=0pt, scale=  1.00] at (365.08,130.05) {230000};

\node[text=drawColor,rotate= 90.00,anchor=base,inner sep=0pt, outer sep=0pt, scale=  1.00] at (365.08,185.24) {250000};

\node[text=drawColor,rotate= 90.00,anchor=base,inner sep=0pt, outer sep=0pt, scale=  1.00] at ( 27.60,144.54) {Total error $\mathcal{E}(\bm x + d\bm v)^{T=5000}$};

\node[text=drawColor,rotate= 90.00,anchor=base,inner sep=0pt, outer sep=0pt, scale=  1.00] at (389.08,144.54) {Total error  $\mathcal{E}(\bm x + d\bm v)^{T=0}$ };

\path[draw=drawColor,line width= 0.4pt,line join=round,line cap=round] (265.04,246.59) circle (  1.80);
\definecolor{drawColor}{RGB}{255,0,0}

\path[draw=drawColor,line width= 0.4pt,line join=round,line cap=round] (263.24,235.19) -- (266.84,238.79);

\path[draw=drawColor,line width= 0.4pt,line join=round,line cap=round] (263.24,238.79) -- (266.84,235.19);
\definecolor{drawColor}{RGB}{0,0,0}

\node[text=drawColor,anchor=base west,inner sep=0pt, outer sep=0pt, scale=  0.80] at (272.24,243.84) {Total error  at $T=5000$};

\node[text=drawColor,anchor=base west,inner sep=0pt, outer sep=0pt, scale=  0.80] at (272.24,234.24) {Total error at $T=0$};

\node[text=drawColor,anchor=base,inner sep=0pt, outer sep=0pt, scale=  1.00] at (210.74, 15.60) {Distance ($d$)};
\end{scope}
\end{tikzpicture}

%% file: figures/Chapter3/Results1.tex
\begin{tikzpicture}[x=1pt,y=1pt]
\definecolor{fillColor}{RGB}{255,255,255}
\path[use as bounding box,fill=fillColor,fill opacity=0.00] (0,0) rectangle (397.48,289.08);
\begin{scope}
\path[clip] ( 49.20, 61.20) rectangle (372.28,239.88);
\definecolor{drawColor}{RGB}{255,0,0}

\path[draw=drawColor,line width= 0.4pt,line join=round,line cap=round] ( 61.17,233.26) --
	( 71.48,211.46) --
	( 81.80,176.28) --
	( 92.11,124.59) --
	(102.43,124.59) --
	(112.74,115.39) --
	(123.06,108.03) --
	(133.38,101.30) --
	(143.69, 99.97) --
	(154.01, 97.93) --
	(164.32, 93.98) --
	(174.64, 90.56) --
	(184.95, 88.06) --
	(195.27, 85.67) --
	(205.58, 83.62) --
	(215.90, 81.93) --
	(226.22, 80.34) --
	(236.53, 78.78) --
	(246.85, 77.13) --
	(257.16, 75.51) --
	(267.48, 73.99) --
	(277.79, 72.60) --
	(288.11, 70.97) --
	(298.43, 69.87) --
	(308.74, 68.85) --
	(319.06, 67.82);
\end{scope}
\begin{scope}
\path[clip] (  0.00,  0.00) rectangle (397.48,289.08);
\definecolor{drawColor}{RGB}{0,0,0}

\path[draw=drawColor,line width= 0.4pt,line join=round,line cap=round] ( 49.20, 90.13) -- ( 49.20,236.65);

\path[draw=drawColor,line width= 0.4pt,line join=round,line cap=round] ( 49.20, 90.13) -- ( 43.20, 90.13);

\path[draw=drawColor,line width= 0.4pt,line join=round,line cap=round] ( 49.20,126.76) -- ( 43.20,126.76);

\path[draw=drawColor,line width= 0.4pt,line join=round,line cap=round] ( 49.20,163.39) -- ( 43.20,163.39);

\path[draw=drawColor,line width= 0.4pt,line join=round,line cap=round] ( 49.20,200.02) -- ( 43.20,200.02);

\path[draw=drawColor,line width= 0.4pt,line join=round,line cap=round] ( 49.20,236.65) -- ( 43.20,236.65);

\node[text=drawColor,rotate= 90.00,anchor=base,inner sep=0pt, outer sep=0pt, scale=  1.00] at ( 34.80, 90.13) {5.0};

\node[text=drawColor,rotate= 90.00,anchor=base,inner sep=0pt, outer sep=0pt, scale=  1.00] at ( 34.80,126.76) {5.5};

\node[text=drawColor,rotate= 90.00,anchor=base,inner sep=0pt, outer sep=0pt, scale=  1.00] at ( 34.80,163.39) {6.0};

\node[text=drawColor,rotate= 90.00,anchor=base,inner sep=0pt, outer sep=0pt, scale=  1.00] at ( 34.80,200.02) {6.5};

\node[text=drawColor,rotate= 90.00,anchor=base,inner sep=0pt, outer sep=0pt, scale=  1.00] at ( 34.80,236.65) {7.0};

\path[draw=drawColor,line width= 0.4pt,line join=round,line cap=round] ( 49.20, 61.20) --
	(372.28, 61.20) --
	(372.28,239.88) --
	( 49.20,239.88) --
	cycle;
\end{scope}
\begin{scope}
\path[clip] ( 49.20, 61.20) rectangle (372.28,239.88);
\definecolor{drawColor}{RGB}{255,0,0}

\path[draw=drawColor,line width= 0.4pt,line join=round,line cap=round] ( 61.17,111.49) --
	( 71.48, 99.19) --
	( 81.80, 92.94) --
	( 92.11, 84.75) --
	(102.43, 80.02) --
	(112.74, 71.01) --
	(123.06, 67.69);

\path[draw=drawColor,line width= 0.4pt,line join=round,line cap=round] ( 61.17,110.05) --
	( 71.48,102.11) --
	( 81.80, 93.41) --
	( 92.11, 88.89) --
	(102.43, 82.87) --
	(112.74, 81.90) --
	(123.06, 80.88) --
	(133.38, 80.04) --
	(143.69, 79.42) --
	(154.01, 78.98) --
	(164.32, 78.45) --
	(174.64, 77.76) --
	(184.95, 77.32) --
	(195.27, 77.01) --
	(205.58, 76.80) --
	(215.90, 76.27);

\path[draw=drawColor,line width= 0.4pt,line join=round,line cap=round] ( 61.17,146.24) --
	( 71.48,135.67) --
	( 81.80,134.58) --
	( 92.11,118.93) --
	(102.43,114.44) --
	(112.74,109.95) --
	(123.06,107.66) --
	(133.38,105.74) --
	(143.69,102.18) --
	(154.01,100.63) --
	(164.32, 99.41) --
	(174.64, 98.31) --
	(184.95, 96.83) --
	(195.27, 95.77) --
	(205.58, 94.69) --
	(215.90, 92.51) --
	(226.22, 89.48) --
	(236.53, 88.50) --
	(246.85, 87.52) --
	(257.16, 86.77) --
	(267.48, 86.04) --
	(277.79, 85.25) --
	(288.11, 84.61) --
	(298.43, 84.10) --
	(308.74, 83.62);

\path[draw=drawColor,line width= 0.4pt,line join=round,line cap=round] ( 61.17,231.40) --
	( 71.48,206.21) --
	( 81.80,179.76) --
	( 92.11,147.76) --
	(102.43,120.25) --
	(112.74,117.45) --
	(123.06,109.99) --
	(133.38,104.72) --
	(143.69,101.75) --
	(154.01, 99.37) --
	(164.32, 97.52) --
	(174.64, 95.97) --
	(184.95, 94.78) --
	(195.27, 93.70) --
	(205.58, 92.81) --
	(215.90, 92.13) --
	(226.22, 84.88) --
	(236.53, 77.01) --
	(246.85, 65.85);

\path[draw=drawColor,line width= 0.4pt,line join=round,line cap=round] ( 61.17,109.71) --
	( 71.48,101.05) --
	( 81.80, 92.84) --
	( 92.11, 90.10) --
	(102.43, 87.32) --
	(112.74, 82.49) --
	(123.06, 80.08) --
	(133.38, 78.07) --
	(143.69, 76.62) --
	(154.01, 75.21) --
	(164.32, 73.79) --
	(174.64, 71.95) --
	(184.95, 68.22) --
	(195.27, 66.03);

\path[draw=drawColor,line width= 0.4pt,line join=round,line cap=round] ( 61.17,101.20) --
	( 71.48, 95.73) --
	( 81.80, 92.85) --
	( 92.11, 87.89) --
	(102.43, 84.05) --
	(112.74, 79.07) --
	(123.06, 78.59);
\definecolor{drawColor}{RGB}{0,0,0}

\path[draw=drawColor,line width= 0.4pt,dash pattern=on 1pt off 3pt ,line join=round,line cap=round] ( 49.20, 68.08) -- (372.29, 68.08);
\end{scope}
\begin{scope}
\path[clip] (  0.00,  0.00) rectangle (397.48,289.08);
\definecolor{drawColor}{RGB}{0,0,0}

\path[draw=drawColor,line width= 0.4pt,line join=round,line cap=round] ( 61.17, 61.20) -- (360.32, 61.20);

\path[draw=drawColor,line width= 0.4pt,line join=round,line cap=round] ( 61.17, 61.20) -- ( 61.17, 55.20);

\path[draw=drawColor,line width= 0.4pt,line join=round,line cap=round] ( 71.48, 61.20) -- ( 71.48, 55.20);

\path[draw=drawColor,line width= 0.4pt,line join=round,line cap=round] ( 81.80, 61.20) -- ( 81.80, 55.20);

\path[draw=drawColor,line width= 0.4pt,line join=round,line cap=round] ( 92.11, 61.20) -- ( 92.11, 55.20);

\path[draw=drawColor,line width= 0.4pt,line join=round,line cap=round] (102.43, 61.20) -- (102.43, 55.20);

\path[draw=drawColor,line width= 0.4pt,line join=round,line cap=round] (112.74, 61.20) -- (112.74, 55.20);

\path[draw=drawColor,line width= 0.4pt,line join=round,line cap=round] (123.06, 61.20) -- (123.06, 55.20);

\path[draw=drawColor,line width= 0.4pt,line join=round,line cap=round] (133.38, 61.20) -- (133.38, 55.20);

\path[draw=drawColor,line width= 0.4pt,line join=round,line cap=round] (143.69, 61.20) -- (143.69, 55.20);

\path[draw=drawColor,line width= 0.4pt,line join=round,line cap=round] (154.01, 61.20) -- (154.01, 55.20);

\path[draw=drawColor,line width= 0.4pt,line join=round,line cap=round] (164.32, 61.20) -- (164.32, 55.20);

\path[draw=drawColor,line width= 0.4pt,line join=round,line cap=round] (174.64, 61.20) -- (174.64, 55.20);

\path[draw=drawColor,line width= 0.4pt,line join=round,line cap=round] (184.95, 61.20) -- (184.95, 55.20);

\path[draw=drawColor,line width= 0.4pt,line join=round,line cap=round] (195.27, 61.20) -- (195.27, 55.20);

\path[draw=drawColor,line width= 0.4pt,line join=round,line cap=round] (205.58, 61.20) -- (205.58, 55.20);

\path[draw=drawColor,line width= 0.4pt,line join=round,line cap=round] (215.90, 61.20) -- (215.90, 55.20);

\path[draw=drawColor,line width= 0.4pt,line join=round,line cap=round] (226.22, 61.20) -- (226.22, 55.20);

\path[draw=drawColor,line width= 0.4pt,line join=round,line cap=round] (236.53, 61.20) -- (236.53, 55.20);

\path[draw=drawColor,line width= 0.4pt,line join=round,line cap=round] (246.85, 61.20) -- (246.85, 55.20);

\path[draw=drawColor,line width= 0.4pt,line join=round,line cap=round] (257.16, 61.20) -- (257.16, 55.20);

\path[draw=drawColor,line width= 0.4pt,line join=round,line cap=round] (267.48, 61.20) -- (267.48, 55.20);

\path[draw=drawColor,line width= 0.4pt,line join=round,line cap=round] (277.79, 61.20) -- (277.79, 55.20);

\path[draw=drawColor,line width= 0.4pt,line join=round,line cap=round] (288.11, 61.20) -- (288.11, 55.20);

\path[draw=drawColor,line width= 0.4pt,line join=round,line cap=round] (298.43, 61.20) -- (298.43, 55.20);

\path[draw=drawColor,line width= 0.4pt,line join=round,line cap=round] (308.74, 61.20) -- (308.74, 55.20);

\path[draw=drawColor,line width= 0.4pt,line join=round,line cap=round] (319.06, 61.20) -- (319.06, 55.20);

\path[draw=drawColor,line width= 0.4pt,line join=round,line cap=round] (329.37, 61.20) -- (329.37, 55.20);

\path[draw=drawColor,line width= 0.4pt,line join=round,line cap=round] (339.69, 61.20) -- (339.69, 55.20);

\path[draw=drawColor,line width= 0.4pt,line join=round,line cap=round] (350.00, 61.20) -- (350.00, 55.20);

\path[draw=drawColor,line width= 0.4pt,line join=round,line cap=round] (360.32, 61.20) -- (360.32, 55.20);

\node[text=drawColor,anchor=base,inner sep=0pt, outer sep=0pt, scale=  1.00] at ( 61.17, 39.60) {1};

\node[text=drawColor,anchor=base,inner sep=0pt, outer sep=0pt, scale=  1.00] at ( 81.80, 39.60) {3};

\node[text=drawColor,anchor=base,inner sep=0pt, outer sep=0pt, scale=  1.00] at (102.43, 39.60) {5};

\node[text=drawColor,anchor=base,inner sep=0pt, outer sep=0pt, scale=  1.00] at (123.06, 39.60) {7};

\node[text=drawColor,anchor=base,inner sep=0pt, outer sep=0pt, scale=  1.00] at (143.69, 39.60) {9};

\node[text=drawColor,anchor=base,inner sep=0pt, outer sep=0pt, scale=  1.00] at (164.32, 39.60) {11};

\node[text=drawColor,anchor=base,inner sep=0pt, outer sep=0pt, scale=  1.00] at (184.95, 39.60) {13};

\node[text=drawColor,anchor=base,inner sep=0pt, outer sep=0pt, scale=  1.00] at (205.58, 39.60) {15};

\node[text=drawColor,anchor=base,inner sep=0pt, outer sep=0pt, scale=  1.00] at (226.22, 39.60) {17};

\node[text=drawColor,anchor=base,inner sep=0pt, outer sep=0pt, scale=  1.00] at (246.85, 39.60) {19};

\node[text=drawColor,anchor=base,inner sep=0pt, outer sep=0pt, scale=  1.00] at (267.48, 39.60) {21};

\node[text=drawColor,anchor=base,inner sep=0pt, outer sep=0pt, scale=  1.00] at (288.11, 39.60) {23};

\node[text=drawColor,anchor=base,inner sep=0pt, outer sep=0pt, scale=  1.00] at (308.74, 39.60) {25};

\node[text=drawColor,anchor=base,inner sep=0pt, outer sep=0pt, scale=  1.00] at (329.37, 39.60) {27};

\node[text=drawColor,anchor=base,inner sep=0pt, outer sep=0pt, scale=  1.00] at (350.00, 39.60) {29};

\node[text=drawColor,anchor=base,inner sep=0pt, outer sep=0pt, scale=  1.00] at (210.74, 15.60) {Iteration};

\node[text=drawColor,rotate= 90.00,anchor=base,inner sep=0pt, outer sep=0pt, scale=  1.00] at ( 10.80,150.54) {$\log_{10} \mathcal{E}(\bm x_i)$};
\end{scope}
\end{tikzpicture}